\documentclass[aps,superscriptaddress,showpacs,showkeys,nofootinbib,floatfix]{revtex4}
\usepackage{graphicx,epsfig,wrapfig,amssymb}
\usepackage{color}
\usepackage{amsfonts}
\usepackage{amsmath}
\usepackage{latexsym}      
\definecolor{darkgreen}{rgb}{0,0.65,0}

\newcommand{\be}{\begin{equation}}
\newcommand{\ee}{\end{equation}}
\newcommand{\ba}{\begin{eqnarray}}
\newcommand{\ea}{\end{eqnarray}}
\newcommand{\la}{\langle}
\newcommand{\ra}{\rangle}
\newcommand{\di}{ {\rm d} }

\newcommand{\tvec}[1]{\mbox{\boldmath{$#1$}}}

\begin{document}
%===================  TITLE, AUTHORS, AFFILIATIONS ===================
\newcommand*{\Pavia}{Dipartimento di Fisica Nucleare e Teorica, 
Universit\`a degli Studi di Pavia, Pavia, Italy}\affiliation{\Pavia}
\newcommand*{\INFN}{Istituto Nazionale di Fisica Nucleare, 
Sezione di Pavia, Pavia, Italy}\affiliation{\INFN}
\newcommand*{\Dubna}{Joint Institute for Nuclear Research, Dubna,
141980 Russia}\affiliation{\Dubna}
\newcommand*{\UConn}{Department of Physics, University of Connecticut, 
Storrs, CT 06269, USA}\affiliation{\UConn}

\title{Azimuthal spin asymmetries in light-cone constituent quark models}

\author{S.~Boffi}\affiliation{\Pavia}\affiliation{\INFN}
\author{A.~V.~Efremov}\affiliation{\Dubna}
\author{B.~Pasquini}\affiliation{\Pavia}\affiliation{\INFN}
\author{P.~Schweitzer}\affiliation{\UConn}%\affiliation{\Bochum}

\date{March 2009}
%===================  PREPRINT NUMBER, JOURNAL =======================
%\preprint{\red{\large version 01b }}
%===================  ABSTRACT =======================================
\begin{abstract}
  We present results for all leading-twist
  azimuthal spin asymmetries in semi-inclusive lepton-nucleon 
  deep-inelastic scattering 
  due to T-even transverse-momentum dependent parton 
  distribution functions on the basis of a light-cone constituent quark model.
  Attention is paid to discuss the range of applicability 
  of the model, especially with regard to the scale dependence of the
  observables and the transverse-momentum dependence of the distributions.
  We find good agreement with available experimental data and 
  present predictions to be further tested by future CLAS, 
  COMPASS and HERMES data.
 \end{abstract}
 % Hadron-induced inclusive production with identified leptons,
                % photons, or other nonhadronic particles (energy > 10 GeV)
\maketitle

%\tableofcontents

%===================  SECTION 1: INTRODUCTION ========================
\section{Introduction}
\label{Sec-1:introduction}

The composite nature of the nucleon has been explored for a long time by means
of deep inelastic scattering (DIS) of a lepton beam in the Bjorken regime, 
i.e., when $q$ and $P$ denote the four-momentum transfer and the nucleon 
momentum, in the limit of $P\cdot q$ and $Q^2=-q^2\to\infty$
while $x=Q^2/(2P\cdot q)$ is fixed.
As a consequence of the high scale $Q$, scattering occurs  in a collinear 
configuration between the incident lepton and a single 'parton' in 
the nucleon. The 
factorization theorem  allows the inclusive DIS cross section to be expressed 
as a convolution of two contributions: one corresponds to the hard process 
occurring at short distance between probe and parton; the other accounts 
for the coherent long-distance interactions between parton and target,
and is described in terms of parton distributions. 
At leading order (leading twist)  $x$ can be interpreted as the
fraction of the longitudinal momentum of the parent (fast-moving) nucleon 
carried by the active parton, and one may distinguish three kinds of parton 
distributions. Two of them are well-known from measurements of structure 
functions in DIS and other processes: $f_1^a(x)$ is
 the number density of unpolarized partons with longitudinal 
momentum fraction $x$ in an unpolarized nucleon, and $g_1(x)$ 
gives the net helicity of partons in a longitudinally polarized
nucleon. The third one, the
(chiral-odd) transversity $h_1(x)$ describing the number density of partons
with polarization parallel to that of a transversely polarized nucleon 
minus the number density of partons with antiparallel polarization, requires 
a quark helicity flip that cannot be achieved in the inclusive DIS. 
Other processes have to be explored for that.

However, in addition to the information on the longitudinal behaviour in
 momentum space along the direction in which the nucleon is moving, a complete 
three-dimensional picture of the nucleon also requires knowledge of the 
transverse motion of partons \cite{Collins:2003fm,Collins:2007ph}.
A full account of the orbital motion, which 
is also an important issue to understand the spin structure of the nucleon,
can be given in terms of transverse-momentum dependent parton 
distribution functions (TMDs). There are eight leading-twist TMDs 
$f_1(x,p_T)$, $f_{1T}^{\perp}(x,p_T)$, $g_{1L}(x,p_T)$, 
$g_{1T}(x, p_T)$, $h_1(x,p_T)$, $h_{1L}^{\perp}(x,p_T)$, 
$h_{1T}^{\perp}(x,p_T)$, $h_1^{\perp}(x,p_T)$~\cite{Boer:1997nt}.
Two of them, the 
Boer-Mulders and Sivers functions $h_1^{\perp}(x,p_T)$ and
$f_{1T}^{\perp}(x, p_T)$ \cite{Sivers:1989cc,Boer:1997nt}, 
are T-odd, i.e. 
they change sign under na\"{\i}ve time reversal, which is defined as 
usual time reversal, but without interchange of initial and final states.
The other six leading-twist TMDs are T-even. 

In order to be sensitive to intrinsic transverse parton momenta it is 
necessary to measure adequate transverse momenta of the produced hadrons 
in the final state, e.g., in processes like semi-inclusive lepton-nucleon 
DIS (SIDIS), hadron production in $e^+e^-$ annihilation or the Drell-Yan
processes in hadron-hadron collisions~\cite{Collins:2003fm,Collins:2007ph,Cahn:1978se,Collins:1984kg,Sivers:1989cc,Efremov:1992pe,Collins:1992kk,Collins:1993kq,Kotzinian:1994dv,Mulders:1995dh,Boer:1997nt,Boer:1997mf,Boer:1999mm,Brodsky:2002cx,Collins:2002kn,Belitsky:2002sm,Bacchetta:2004zf,Cherednikov:2007tw,D'Alesio:2007jt,Barone:2001sp,Goeke:2005hb}.

Here, factorization has been 
proved at leading twist~\cite{Collins:1981uk,Ji:2004wu,Collins:2004nx} 
allowing to access information on TMDs as 
well as on fragmentation functions (FFs) describing the hadronization process 
of the hit quark decaying into the detected hadrons. At leading twist, 
the fragmentation of unpolarized hadrons is described in terms of two
 fragmentation functions, $D_1(z,K_T)$ and $H_1^{\perp}(z,K_T)$,
 where $z$ is the energy fraction taken out by the detected hadron and 
$K_T=|\tvec K_T|$ its transverse momentum. The function $D_1(z,K_T)$ 
describes the decay of  an unpolarized quark, whereas the Collins function 
$H_1^{\perp}(z,K_T)$ describes a left-right asymmetry in the decay of 
a transversely polarized quark
\cite{Efremov:1992pe,Collins:1992kk,Collins:1993kq}.

By measuring the angular distribution of produced hadrons, in SIDIS it is 
possible to access information on all eight leading-twist TMDs in combinations
 with the two leading-twist FFs. Restricting ourselves to the 
one-photon-exchange approximation and considering spin degrees of freedom
 such as the beam helicity and the target spin, the contraction between the 
lepton and hadron tensors in the SIDIS lepton-nucleon cross section can be 
decomposed in a model-independent way in terms of eighteen structure functions,
 thus exhibiting a non trivial azimuthal dependence of the detected hadron 
around the (space-like) direction defined by the virtual 
photon~\cite{Diehl:2005pc,Dombey:1969wk,Gourdin:1973qx,Kotzinian:1994dv,Boffi:1993gs,Bacchetta:2006tn}.
According to factorization each 
of the leading-twist structure functions
 can be conceived as a convolution between one TMD and one FF. Since
 structure functions enter the cross section with a defined angular 
coefficient, they can be accessed by looking at specific azimuthal SIDIS
 asymmetries.  This has become now a powerful tool for studying the
 three-dimensional structure of the nucleon~\cite{Arneodo:1986cf,Airapetian:1999tv,Airapetian:2001eg,Airapetian:2002mf,Avakian:2003pk,Airapetian:2004tw,Alexakhin:2005iw,Diefenthaler:2005gx,Gregor:2005qv,Ageev:2006da,Avakian:2005ps,Airapetian:2005jc,Airapetian:2006rx,Abe:2005zx,Diefenthaler:2007rj,Kotzinian:2007uv,Alekseev:2008dn,Seidl:2008xc}, 
and many more data are expected to come in the future.
The remarkable experimental progress was accompanied by and motivated 
numerous theoretical and phenomenological studies in literature
\cite{Bacchetta:2002tk,Yuan:2003wk,Jakob:1997wg,Pasquini:2008ax,Bacchetta:2008af,Efremov:2008mp,She:2009jq,Meissner:2007rx,Courtoy:2008vi,Efremov:2002qh,Efremov:2003eq,D'Alesio:2004up,Anselmino:2004ki,Efremov:2004qs,Pasquini:2006iv,Efremov:2004tp,Collins:2005ie,Collins:2005rq,Anselmino:2005nn,Vogelsang:2005cs,Efremov:2006qm,Anselmino:2007fs,Arnold:2008ap,Anselmino:2008sg,Anselmino:2008jk,Gamberg:2007wm,Bacchetta:2007wc,Kotzinian:2006dw,Avakian:2007mv,Brodsky:2006hj,Burkardt:2007rv,Pobylitsa:2003ty,Avakian:2007xa,Bacchetta:1999kz,Metz:2008ib,Avakian:2009nj,Avakian:2008dz}.

In this paper we compute azimuthal spin asymmetries due to the 
T-even transverse-momentum dependent parton distributions functions.
For that we use the predictions from the light-cone 
constituent quark model (CQM) of Ref.~\cite{Pasquini:2008ax}, which has been 
successfully applied also in the calculation of electroweak properties of the 
nucleon~\cite{Pasquini:2007iz} and generalized parton 
distributions~\cite{Boffi:2007yc}.
Such a model, based on the light-cone wave function (LCWF) overlap 
representation of TMDs, is well suited to 
illustrate the relevance of the different 
orbital angular momentum components of the nucleon wave function 
for the respective observables.
To best of our knowledge, this is the first attempt to describe all
leading-twist spin asymmetries in SIDIS due to T-even TMDs in a single
approach. We include also studies of the far better known collinear double 
spin asymmetries $A_1$ and $A_{LL}$ --- not only for sake of completeness,
but also to demonstrate the capability of the approach to describe reliably
the gross features of spin effects in the nucleon.

The article is organized as follows. 
In Sec.~\ref{Sec-2:TMDs-and-SIDIS} the relevant definitions of azimuthal 
asymmetries in SIDIS are recalled. 
In Sec.~\ref{Sec-3:model} the main ingredients of the light-cone CQM of 
Ref.~\cite{Pasquini:2008ax} are reviewed, and the 
results for T-even TMDs discussed. 
In Sec.~\ref{Sec-4:A1-and-ALL} we first study the collinear double spin
asymmetries $A_1$ and $A_{LL}$. 
This and the following Sec.~\ref{Sec-5:pT-and-Gauss} devoted to a discussion
of the transverse-momentum dependence of the TMDs, help to assert the range
of applicability of the approach.
In Secs.~\ref{Sec-6:A_LT}--\ref{Sec-9:pretzelosity} we evaluate the 
leading-twist azimuthal double and single spin asymmetries due to T-even
TMDs --- focusing on their $x$-dependence.
Sec.~\ref{Sec-10:Ph-dep} exemplifies how the approach can be applied
to make predictions for the transverse hadron momentum dependence of 
spin asymmetries.
Concluding remarks are given in Sec.~\ref{sect:conclusions}.
Finally, a more detailed discussion about the model predictions for the 
sensitivity of the azimuthal asymmetries on different orbital 
angular momentum components are discussed in the Appendix.

%===================  SECTION 2: TMDs and SIDIS ======================
\section{Spin and azimuthal asymmetries in SIDIS}
\label{Sec-2:TMDs-and-SIDIS}

%------ BEGIN FIGURE 1: Kinematics of SIDIS ---------------------------
	\begin{wrapfigure}{RD}{8cm}
	\centering
        \includegraphics[width=8cm]{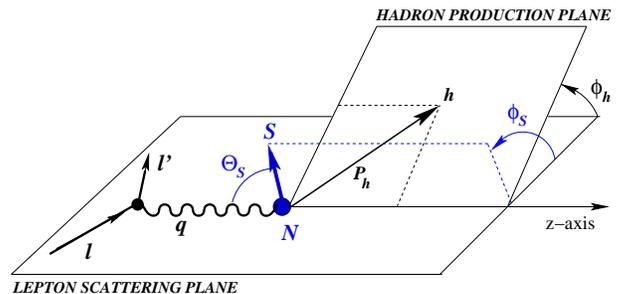}
        \caption{\label{Fig1-SIDIS-kinematics}
    	Kinematics of the SIDIS process $lN\to l^\prime h X$
    	and the definitions of azimuthal angles in the lab frame.}
\end{wrapfigure}
%------ END FIGURE 1 -------------------------------------------------

The SIDIS process is sketched in Fig.~\ref{Fig1-SIDIS-kinematics}.
Let us denote the momenta of the target, incoming and outgoing lepton
by $P$, $l$ and $l'$ and introduce the four-momentum transfer 
$q= l-l'$ with $Q^2= - q^2$. Then the relevant SIDIS
variables are defined as $x = Q^2/(2P\cdot q)$, $y = (P\cdot q)/(P\cdot l)$ and
$z = (P\cdot P_h)/(P\cdot q)$.
The component of the momentum of the produced hadron transverse with 
respect to the virtual photon is denoted by ${\tvec P}_{\!h\perp}$ and 
$P_{h\perp}=|{\tvec P}_{\!h\perp}|$.

The SIDIS cross section (differential in $x$, $y$, $z$ and the azimuthal angle 
$\phi_h$ of the produced hadron defined in Fig.~\ref{Fig1-SIDIS-kinematics})
has the following general decomposition \cite{Kotzinian:1994dv,Diehl:2005pc},
where $\sigma_0$ is the spin- and $\phi$-independent part of the cross section,
and where the dots indicate power suppressed ('subleading-twist') terms,
\ba
    \frac{\di^4\sigma}{\di x\,\di y\,\di z\,\di\phi_h}
    &=&	\frac{\di^4\sigma_0}{\di x\,\di y\,\di z\,\di\phi_h}\Biggl\{ 1
     +             \cos(2\phi_h)\,       p_1(y)\,A_{UU}^{\cos(2\phi_h)}
     +          S_L\sin(2\phi_h)\,       p_1(y)\,A_{UL}^{\sin(2\phi_h)} \nonumber\\
    &+&\lambda\,S_L\,                    p_2(y)\,A_{LL}
     + \lambda\,S_T\cos( \phi_h-\phi_S)\,p_2(y)\,A_{LT}^{\cos( \phi_h-\phi_S)}
     +          S_T\sin( \phi_h-\phi_S)\,        A_{UT}^{\sin( \phi_h-\phi_S)} \nonumber\\     
    &+&         S_T\sin( \phi_h+\phi_S)\,p_1(y)\,A_{UT}^{\sin( \phi_h+\phi_S)}
     +          S_T\sin(3\phi_h-\phi_S)\,p_1(y)\,A_{UT}^{\sin(3\phi_h-\phi_S)}\Biggr\}
     + \dots \;\;\; \label{Eq:azim-distr-in-SIDIS}
\ea
with
\be
	p_1(y) =  \frac{1-y}{1-y+\frac12\,y^2} \;,\;\;\;
	p_2(y) =  \frac{y(1-\frac12\,y)}{1-y+\frac12\,y^2} \;.\;\;\;
\ee

In $A_{XY}^{\rm weight}$ the index $X$ describes the beam polarization, 
which is unpolarized (U) or longitudinal (L, characterized then by the
beam helicity $\lambda$).
The index $Y$ denotes the target polarization, which is unpolarized (U),
longitudinal~(L) or transverse (T) with respect to the virtual photon.
In experiments the target is polarized with respect to the beam, of course,
but this is up to corrections of ${\cal O}(1/Q)$ the same.
As we shall deal with leading-twist observables, such corrections
will be neglected through out.
The superscript 'weight' reminds us of the kind of angular distribution of the 
produced hadrons with no index indicating an isotropic $\phi$-distribution, 
and
$\phi_S$ is the azimuthal angle of the target's transverse polarization 
vector, see Fig.~\ref{Fig1-SIDIS-kinematics}.
The asymmetries are defined in terms of structure functions,
$A_{XY}^{\rm weight}=F_{XY}^{\rm weight}/F_{UU}$, and the latter have
the following partonic (tree-level) description in the Bjorken-limit
\cite{Mulders:1995dh,Boer:1997nt}
\ba
 F_{UU} &=&\phantom{-}\,{\cal C}\biggl[\;f_1 D_1 \;\biggr],
        {\phantom{\Biggl|}} \label{FUU}\\
 F_{LL}	&=&\phantom{-}\,{\cal C}\biggl[\;g_{1L} D_1 \;\biggr],
        {\phantom{\Biggl|}} \label{FLL}\\
 F_{UT}^{\sin\left(\phi_h -\phi_S\right)} &=& -\,{\cal C}\biggl[
	\frac{{\tvec h}_\perp\cdot {\tvec p}_T^{ }}{M} f_{1T}^{\perp }  D_1\biggr],
	\label{Eq:FUTSiv}\\
 F_{LT}^{\cos(\phi_h -\phi_S)} &=& \phantom{-}\,{\cal C}\biggl[
	\frac{{\tvec h}_\perp\cdot{\tvec p}_T^{ }}{M} g_{1T}^\perp D_1 \biggr],
        \label{Eq:FLT}  \\
 F_{UT}^{\sin\left(\phi_h +\phi_S\right)} &=& \phantom{-}\,{\cal C}\biggl[
	\frac{{\tvec h}_\perp\cdot{\tvec K}_T^{ }}{z\,m_h}h_{1} H_1^{\perp }\biggr],
	\label{Eq:FUTCol}\\
 F_{UU}^{\cos(2\phi_h)} 	&=& \phantom{-}\,{\cal C}\biggl[
   	\frac{2\, \bigl({\tvec h}_\perp \cdot {\tvec K}_T^{ } \bigr)
   	\,\bigl({\tvec h}_\perp \cdot {\tvec p}_T^{ } \bigr)
    	-{\tvec K}_T^{ } \cdot {\tvec p}_T^{ }}{z\,m_h M}
    	h_{1}^{\perp } H_{1}^{\perp }\biggr],		\label{FUUcos2phi}\\
 F_{UL}^{\sin(2\phi_h)} 	&=& \phantom{-}\,{\cal C}\biggl[
   	\frac{2\,\bigl({\tvec h}_\perp \cdot {\tvec K}_T^{ } \bigr)
   	\,\bigl({\tvec h}_\perp \cdot {\tvec p}_T^{ } \bigr)
   	-{\tvec K}_T^{ } \cdot {\tvec p}_T^{ }}{z\, m_h M}
    	h_{1L}^{\perp } H_{1}^{\perp }\biggr], 		\label{FULsin2phi}\\
 F_{UT}^{\sin\left(3\phi_h -\phi_S\right)} &=& -\,{\cal C}\biggl[
   	\frac{2\, \bigl({\tvec h}_\perp \cdot {\tvec p}_T^{ } \bigr)\, 
        \bigl( {\tvec p}_T^{ }\cdot {\tvec K}_T^{ } \bigr)
   	+{\tvec p}_T^2\, \bigl({\tvec h}_\perp \cdot {\tvec K}_T^{ } \bigr)
   	-4\, ({\tvec h}_\perp\cdot{\tvec p}_T^{ })^2 \, 
	({\tvec h}_\perp\cdot{\tvec K}_T^{ })}{2\, z\, m_h M^2}
    	\,h_{1T}^{\perp }   H_1^{\perp }\biggr],
	\label{Eq:FUTpretzel}
\ea
where ${\tvec h}_\perp={\tvec P}_{h\perp}/P_{h\perp}$ and $M$ ($m_h$) is the 
mass of the nucleon (produced hadron). The convolution is defined as 
\be\label{Eq:conv-integral}
   {\cal C}\biggl[\;w\;j\;J\biggr]=\int\di^2{\tvec p}_T^{ }\int\di^2{\tvec K}_T
   \;\delta^{(2)}(z\,{\tvec p}_T^{ }+{\tvec K}_T^{ }-{\tvec P}_{h\perp}^{ })
   \,w({\tvec p}_T,\,{\tvec K}_T)\,
   \sum_a e_a^2\;x\,j^a(x,p_T)\;J^a(z, K_T)\;,
\ee
where $p_T=|{\tvec p}_T|$.
These convolution integrals can be solved analytically only in the case
of the structure functions $F_{UU}$ and $F_{LL}$, in which case the 
weight function $w$ in Eq.~(\ref{Eq:conv-integral}) is simply unity.
The unpolarized cross section is given in terms of $F_{UU}$ by
\be\label{Eq:sigma0}
       \frac{\di^4\sigma_0}{\di x\,\di y\,\di z\,\di\phi_h}  =
       \frac{2\,\alpha^2 s}{Q^4}\,\biggl(1-y+\frac{y^2}{2}\biggr) \;
       F_{UU}(x,z) \;,\;\;
       F_{UU}(x,z) = \sum_a e_a^2 x\,f_1^a(x)\,D_1^a(z)\;.
\ee
Recalling the notation $g_1^a(x)=\int\di^2{\tvec p}_T\, g_{1L}^a(x,p_T)$,
the double spin asymmetry $A_{LL}$ is given as follows
\be\label{Eq:ALL}
       A_{LL} = \frac{F_{LL}}{F_{UU}} 
       = \frac{\sum_a e_a^2 \,x\,g_1^a(x)\,D_1^a(z)}
              {\sum_a e_a^2 \,x\,f_1^a(x)\,D_1^a(z)}\;.
\ee
If no hadron is observed in the final state, the inclusive version 
of the double spin asymmetry (\ref{Eq:ALL}) is commonly referred
to as $A_1$ and given by
\be\label{Eq:A1}
       A_1 = \frac{\sum_ae_a^2\,x\,g_1^a(x)}{\sum_a e_a^2\,x\,f_1^a(x)}\;.
\ee

For all other structure functions (\ref{Eq:FUTSiv}--\ref{Eq:FUTpretzel})
the convolution integrals cannot be solved analytically, unless one assumes models 
for the transverse parton momentum dependence of TMDs. A popular 
model is the Gaussian Ansatz, where one assumes
\ba\label{Eq:Gauss-ansatz}
    j^a(x,p_T) = j^a(x)\;
    \frac{\exp(-p_T^{\:2}/\la p_T^2(j)\ra)}{\pi\la p_T^2(j)\ra} \;,\;\;\;
    J^a(z,K_T) = J^a(z)\;
    \frac{\exp(-K_T^2/\la K_T^2(J)\ra)}{\pi\;\la K_T^2(J)\ra}
    \ea
for some generic transverse parton momentum dependent distribution
$j^a(x,p_T)$ and fragmentation $J^a(z,K_T)$ functions. This is, 
of course, a crude approximation. However, besides being convenient
\cite{Mulders:1995dh}, this Ansatz is also phenomenologically useful, provided
the transverse hadron momenta are small compared to the relevant hard scale,
$\la P_{h\perp}\ra\ll Q$ in SIDIS, and one is interested in catching the gross
features of the effects \cite{D'Alesio:2004up}.
A high precision description of $p_T$-effects requires methods along the
QCD-based formalism of \cite{Collins:1984kg}, see \cite{Landry:2002ix} and
references therein for examples.

Using this Ansatz we obtain the following results
\ba
\hspace{-15mm}&& 
F_{UT}^{\sin\left(\phi_h -\phi_S\right)\phantom{3}} = -
   B_0\sum_a e_a^2 \,x\,f_{1T}^{\perp(1)a}(x)\,D_1^a(z),\hspace{19mm}
   B_0=\frac{\sqrt{\pi}\,M}
   {\{\la p_T^2(f_{1T}^\perp)\ra+\la K_T^2(D_1)\ra/z^2\}^{1/2}},
   \label{Eq:GaussFUTSiv}\\ 
\hspace{-15mm}&&
F_{LT}^{\cos(\phi_h -\phi_S)\phantom{3}} = \phantom{-}
   B_0^\prime\sum_a e_a^2 \,x\,g_{1T}^{\perp(1)a}(x)\,D_1^a(z),\hspace{20mm}
   B_0^\prime = \frac{\sqrt{\pi}\,M}
   {\{\la p_T^2(g_{1T}^\perp)\ra+\la K_T^2(D_1)\ra/z^2\}^{1/2}},
   \label{Eq:GaussFLT}  \\ 
\hspace{-15mm}&&
F_{UT}^{\sin\left(\phi_h +\phi_S\right)\phantom{3}} 	=  \phantom{-} 
   B_1\sum_ae_a^2\,x\,h_1^a(x)\,H_1^{\perp(1/2)a}(z),\hspace{17mm}
   B_1=\frac{2}{\{1+R(h_1)\}^{1/2}},
   \label{Eq:GaussFUTCol}\\ 
\hspace{-15mm}&&
F_{UU}^{\cos(2\phi_h)\phantom{-\phi_S}} 	=  \phantom{-}
   B_2\sum_a e_a^2\,x\,h_{1}^{\perp(1)a}(x)H_1^{\perp(1/2)a}(z),
	\;\;\hspace{10mm}
   B_2 =\frac{8\,z\,M\, [\pi\la K_T^2(H_1^\perp)\ra]^{-1/2}}
   {1+ R(h_{1}^\perp)},
   \label{Eq:GaussFUUcos2phi}\\ 
\hspace{-15mm}&&
F_{UL}^{\sin(2\phi_h)\phantom{-\phi_S}} 	= \phantom{-}
   B_2^\prime\sum_ae_a^2\,x\,h_{1L}^{\perp(1)a}(x)H_1^{\perp(1/2)a}(z),
	\;\;\;\hspace{10mm}
   B_2^\prime =\frac{8\,z\,M\, [\pi\la K_T^2(H_1^\perp)\ra]^{-1/2}}
   {1+ R(h_{1L}^\perp)},
   \label{Eq:GaussFULsin2phi}\\ 
\hspace{-15mm}&&
F_{UT}^{\sin\left(3\phi_h -\phi_S\right)} = -
   B_3\sum_a e_a^2 \,x\,h_{1T}^{\perp(1)a}(x)H_1^{\perp(1/2)a}(z),
	\;\;\,\hspace{10mm}
   B_3=\frac{3}{\{R(h_{1T}^\perp)^{1/3}+R(h_{1T}^\perp)^{-1/3}\}^{3/2}},\;\;
   \label{Eq:GaussFUTpretzel}
\ea
where
\ba
   j^{a}(x)=\int\di^2{\tvec p}_T^{ }\,
   j^{a}(x,p_T),&\qquad&
   j^{(1)a}(x)=\int\di^2{\tvec p}_T^{ }\,\frac{p_T^2}{2M^2}\,
   j^{a}(x,p_T),
\label{Eq:integratedTMD}\\
H_1^{\perp(1/2)a}(z)=\int{\rm d}^2{\tvec K}_T^{ }\,\frac{K_T}{2zm_h}
H_1^{\perp a}(z,K_T),&\qquad&
   R(j) = \frac{z^2\la p_T^2(j)\ra}{\la K_T^2(H_1)\ra} \;.
\ea
Of course, other models of $p_T$-dependence can also be assumed.
In that case, however, one typically cannot solve the convolution
analytically, as in Eqs.~(\ref{Eq:GaussFUTSiv})--(\ref{Eq:GaussFUTpretzel}),
and has to use numerical integration.

Notice that one could avoid the model dependence at this point 
by including adequate powers of transverse hadron momentum in the
weights of the asymmetries \cite{Boer:1997nt}. 
The analysis of such $P_{h\perp}$-weighted asymmetries
is more involved, and so far only preliminary data not corrected
for acceptance effects have been shown \cite{Gregor:2005qv}.

%========= SECTION 3: LIGHT-CONE CONSTITUENT QUARK MODEL =============
\section{TMDs in the light-cone constituent 
quark model}
\label{Sec-3:model}

A convenient way to describe parton distributions is to 
use the representation in terms of overlaps of LCWFs.
This representation can be viewed as a generalization of the famous Drell-Yan 
formula for electromagnetic form factors~\cite{Drell:1970km}, 
and it was recently derived 
and 
applied in phenomenological calculations 
for 
the generalized parton distributions~\cite{Diehl:2000xz,Boffi:2002yy} 
and transverse-momentum dependent parton distributions~\cite{Ji:2002xn,Pasquini:2008ax}.
In practice, this representation becomes useful in 
phenomenological applications where one can reasonably truncate the
expansion of the hadron state to the Fock components with a few partons.
In our approach, we consider the minimum Fock sector with just three valence quarks. This truncation allows to describe the parton distributions in those kinematical regions where the valence degrees of freedom are effective, while
the contributions from quarks and gluons
are suppressed.

The three-quark component of the nucleon has been studied 
extensively in the literature~\cite{Lepage:1980fj,Chernyak:1983ej,King:1986wi,Chernyak:1987nv,Braun:1999te,Braun:2000kw,Stefanis:1999wy} in terms of quark distribution 
amplitudes defined as hadron-to-vacuum 
transition matrix elements of non-local gauge-invariant light-cone operators.
Unlike these works,
the authors of Refs.~\cite{Ji:2002xn,Ji:2003yj,Burkardt:2002uc} considered the 
wave-function amplitudes keeping full transverse-momentum dependence of 
partons and proposed a systematic way to enumerate independent amplitudes of 
a LCWF given a particular parton combination.
 Within this general classification scheme, 
one finds that the nucleon state with three valence quarks 
has six independent scalar amplitudes which serve to parametrize the 
contribution
from the four different orbital angular momentum components $L_z$ compatible
with total angular momentum conservation, i.e.  $L_z=0,\pm 1 , 2$.
An application of this method has been developed in 
Ref.~\cite{Pasquini:2008ax} 
for the calculation of the TMDs within a light-cone CQM
 which will be used here to make quantitative estimates of the azimuthal 
asymmetries.
The key ingredient of the model is to derive the LCWF by boosting
equal-time model wave function.
 The equal-time wave function is constructed as a product of a momentum 
wave function which is in a pure S-wave state and invariant under permutations,
 and a spin-isospin wave function which is uniquely determined by SU(6)
symmetry requirements.
The corresponding solution in light-cone dynamics is obtained through the
unitary  Melosh rotations acting
on the spin of the individual quarks.
By applying the Melosh rotations, the Pauli spinors of the quarks in the
 nucleon rest frame are converted into light-cone spinors.
The effects of the relativistic spin-dynamics are evident in the presence
of spin-flip terms in the Melosh rotations generating non-zero orbital 
angular momentum components which can be mapped out into six independent 
scalar amplitudes.
The explicit expressions of these light-cone amplitudes can be found in 
Ref.~\cite{Pasquini:2008ax}, while the corresponding results for the TMDs
are given by
\begin{eqnarray}
\label{eq:f1}
f^a_1(x,p_T)&=&
N^a \int{\rm d}[X]\
   \delta(x-x_3)\delta({\tvec p}_{T}-{\tvec p}_{\perp\,3})\
\vert \psi(\{x_i\},\{{\tvec p}_{\perp\,i}\})\vert^2,
\end{eqnarray}
\begin{eqnarray}
\label{eq:g1}
g^a_{1L}(x,p_T)&=&
P^a\int{\rm d}[X]\
   \delta(x-x_3)\delta({\tvec p}_{T}-{\tvec p}_{\perp\,3})\
\frac{(m+ x M_0)^2 -{\tvec p}^2_{T}}{(m+ xM_0)^2 + {\tvec p}^2_{T}}\;
\vert \psi(\{x_i\},\{{\tvec p}_{\perp\,i}\})\vert^2,
\end{eqnarray}
\begin{eqnarray}
\label{eq:g1T}
g^{a}_{1T}(x,p_T)&=&
P^a
\int{\rm d}[X]\
   \delta(x-x_3)\delta({\tvec p}_{T}-{\tvec p}_{\perp\,3})\
\frac{2M(m+ xM_0)}{(m+ xM_0)^2 + {\tvec p}^2_{T}}\;
\vert \psi(\{x_i\},\{{\tvec p}_{\perp\,i}\})\vert^2,
\end{eqnarray}
\begin{eqnarray}
\label{eq:h1}
h^a_1(x,p_T)&=&
P^a
\int{\rm d}[X]\
   \delta(x-x_3)\delta({\tvec p}_{T}-{\tvec p}_{\perp\,3})\
\frac{(m+ xM_0)^2}{(m+ xM_0)^2 + {\tvec p}^2_{T}}\;
\vert \psi(\{x_i\},\{{\tvec p}_{\perp\,i}\})\vert^2,
\end{eqnarray}
\begin{eqnarray}
\label{eq:h1T}
h^{\perp\,a}_{1T}(x,p_T)&=&-
P^a
\int{\rm d}[X]\
   \delta(x-x_3)\delta({\tvec p}_{T}-{\tvec p}_{\perp\,3})\
\frac{2M^2}{(m+ xM_0)^2 + {\tvec p}^2_{T}}\;
\vert \psi(\{x_i\},\{{\tvec p}_{\perp\,i}\})\vert^2,
\end{eqnarray}
\begin{eqnarray}
h^{\perp\, a}_{1L}(x,p_T)&=&
- P^a
\int{\rm d}[X]\
   \delta(x-x_3)\delta({\tvec p}_{T}-{\tvec p}_{\perp\,3})\
\frac{2M(m+ xM_0)}{(m+ xM_0)^2 + {\tvec p}^2_T}\;
\vert \psi(\{x_i\},\{{\tvec p}_{\perp\,i}\})\vert^2,
\label{eq:h1L}
\end{eqnarray}
where we introduced 
the integration measure

\vspace{-0.5cm}
\be
   {\rm d}[X] = 
   {\rm d}x_1{\rm d}x_2{\rm d}x_3
   \delta\left(1-\sum_{i=1}^3 x_i\right)
   \frac{{\rm d}^2 {\tvec p}_{\perp\,1}{\rm d}^2{\tvec p}_{\perp\,2}
     {\rm d}^2{\tvec p}_{\perp\,3}}{[2(2\pi^3)]^2}
   \delta\left(\sum_{i=1}^3 {\tvec p}_{\perp\,i}\right).
\ee

In Eqs.~(\ref{eq:f1})-(\ref{eq:h1L}),  $M_0$ is 
the mass of the non-interacting three-quark system, 
and $m$ the constituent quark mass.
Furthermore,
 the flavor dependence is given by 
the factors $N^u=2$, $N^d=1,$ and  $P^u={\frac{4}{3}}$,  
$P^d={-\frac{1}{3}}$, as dictated by SU(6) symmetry.
A further consequence of the assumed SU(6) symmetry is the factorization
in Eqs.~(\ref{eq:f1})-(\ref{eq:h1L}) of the momentum-dependent wave function
$\psi(\{x_i\},\{{\tvec p}_{\perp\,i}\})$  
from the spin-dependent factor arising from the Melosh rotations.
Thanks to this factorized form one finds the following relations
\begin{eqnarray}
\label{eq:61}
2h^a_1(x,p_T)
&=&g^a_{1L}(x,p_T)+\frac{P^a}{N^a}f^a_1(x,p_T),\\
\frac{P^a}{N^a}f^a_1(x,p_T)
&=&h_1^a(x,p_T) -\frac{p_T^2}{2M^2}h_{1T}^{\perp \,a}(x,p_T),
\label{eq:61a}\\
h_{1L}^{\perp q}(x,p_T)
&=&-g_{1T}^a(x,p_T).
\label{eq:61b}
\end{eqnarray}
Eq.~(\ref{eq:61}) is a generalization of analogous relations discussed in 
\cite{Pasquini:2006iv,Pasquini:2005dk} and was also rederived together 
with Eq.~(\ref{eq:61a}) in Ref.~\cite{Avakian:2008dz}.
Eq.~(\ref{eq:61b}) was already found in the diquark spectator model 
of Ref.~\cite{Jakob:1997wg}. 
In QCD the various TMDs are all independent of each other, and describe 
different aspects of the nucleon structure. However, it is natural 
to encounter relations among TMDs in simple models 
limiting to the valence-quark contribution and imposing SU(6) symmetry.
The specific form of the relations can be traced back to the Melosh rotations 
which relate longitudinal and transverse nucleon polarization states in 
a Lorentz-invariant way.
A similar situation occurs with the bag model~\cite{Avakian:2008dz}.
In the diquark spectator model of Ref.~\cite{Jakob:1997wg}
the relations (\ref{eq:61}) and 
(\ref{eq:61a}) hold only for the separate scalar and axial contributions, 
while Eq.~(\ref{eq:61b}) is verified more generally for both $u$ and $d$ 
flavors. 
Since only two out of the four functions $f_1$, $g_{1L}$, $h_1$,
$h_{1T}^\perp$ are linearly independent, there are numerous relations 
among them. For example, subtracting 
(\ref{eq:61}) and (\ref{eq:61a}) one gets a particularly interesting 
relation between pretzelosity, transversity and helicity 
distribution~\cite{Avakian:2008dz}
\begin{eqnarray}
\label{eq:62}
g_1^a(x,p_T)-h_1^a(x,p_T)=h_{1T}^{\perp(1)a}(x,p_T).
\end{eqnarray}
This relation was recently discussed also in connection with the quark orbital
angular momentum distribution
\cite{She:2009jq}.
In the version of the diquark spectator model of Ref.~\cite{Bacchetta:2008af}
the relation~(\ref{eq:62}) is not supported in the axial-vector diquark
sector, but it remains valid  for the scalar sector 
(see also~\cite{Meissner:2007rx,She:2009jq}).
Interestingly, in Ref.~\cite{Efremov:2008mp} the $h_{1T}^{\perp}(x,p_T)$
distribution was reconsidered also within a covariant parton model
with the remarkable finding that the model satisfies the relation 
(\ref{eq:62})  without assuming SU(6) symmetry.

The results in Eqs.~(\ref{eq:f1})-(\ref{eq:h1L}) are applied in the 
following to a specific CQM taking the form of the momentum wave function 
from Ref.~\cite{Schlumpf:1992ce}
\ba
\psi(\{x_i,\boldsymbol{ p}_{\perp i}\})=
2(2\pi)^3\bigg[\frac{1}{M_0}\frac{\omega_1\omega_2\omega_3}{x_1x_2x_3}\bigg]^{1/2}
\frac{N'}{(M_0^2+\beta^2)^\gamma},
\label{eq:30}
\ea 
where $\omega_i$ is the free-quark energy and $N'$ is a normalization factor 
such that 
${\int{\rm d}[X]\vert \psi(\{x_i\},\{{\tvec p}_{\perp\,i}\})\vert^2=1}$.
In Eq.~(\ref{eq:30}), the scale $\beta$, 
the parameter $\gamma$ for the power-law behaviour, and the quark mass $m$ 
are taken from Ref.~\cite{Schlumpf:1992ce}, i.e. $\beta=0.607 $ GeV, 
$\gamma=3.4$ and $m=0.267$ GeV. According to the analysis of 
Ref.~\cite{Schlumpf:1992pp} these values lead to a very good description 
of many baryonic properties.

The results  Eqs.~(\ref{eq:f1})-(\ref{eq:h1L}) are general and can be applied 
to any CQM adopting the appropriate nucleon wave function.
For example, we also considered the prediction in the hypercentral CQM 
model of Refs.~\cite{Faccioli:1998aq,Ferraris:1995ui}.
It has been observed that the 
description of nucleon properties using the model wave function either from 
\cite{Schlumpf:1992pp} 
or  from \cite{Faccioli:1998aq,Ferraris:1995ui}
agree typically within (10--20)$\,\%$, which might be considered
as an indication of the accuracy of the CQM approach. In the following we 
shall assume that such is also the accuracy of the T-even TMDs from CQM 
\cite{Schlumpf:1992pp}. 
The numerical results for T-even TMDs obtained 
in this way were discussed in detail in Ref.~\cite{Pasquini:2008ax}.
In order to compute T-odd TMDs it is necessary to go
beyond the mere CQM scenario, and introduce gauge-boson degrees of
freedom, which was beyond the scope of Ref.~\cite{Pasquini:2008ax} and 
this work, where we concentrate on asymmetries due to T-even TMDs.

%------ BEGIN FIGURE 2: Angular momentum decomposition of TMDs---------
\begin{figure}[b!]
 	\centering
	\vspace{-1cm}
        \includegraphics[width=12cm]{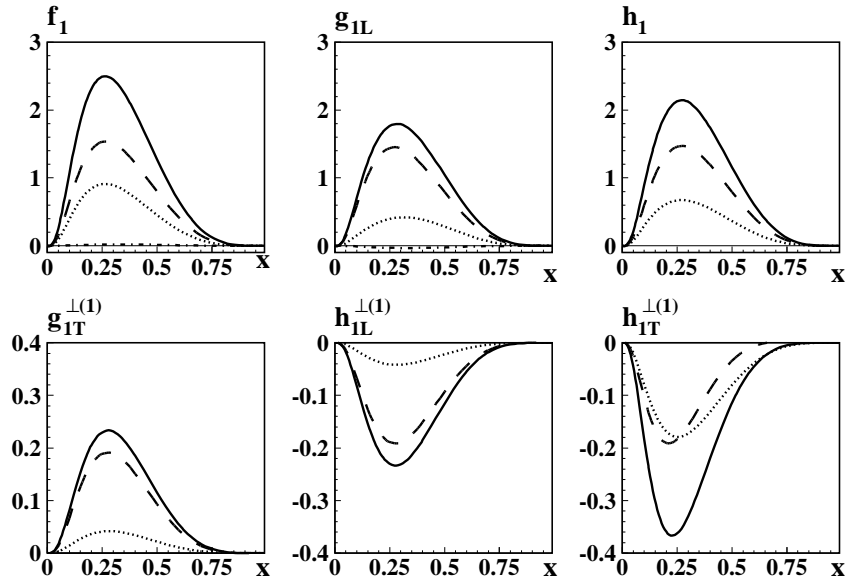}
	\caption{\label{Fig2:TMD}
    Parton distribution functions and transverse moments of TMDs 
    as functions of $x$ from the light-cone CQM \cite{Pasquini:2008ax}.
    In all panels the solid curves show the total results for the
    'flavour-less' TMDs, i.e.\ the TMDs of definite flavour follow from 
    multiplying by the spin-flavour factors $N^a$ or $P^a$, see 
    Eqs.~(\ref{eq:f1})-(\ref{eq:h1T}).
    The other curves show how much the different angular momentum components 
    of the nucleon wave function contribute to the total results: 
    In the case of $f_1(x)$, $g_1(x)$, $h_1(x)$
    the dashed, and dotted  curves 
    correspond to the contribution from the squares of the S- and P-wave 
    components of the nucleon wave function, respectively.
    The $D$-wave contribution is absent in $h_1$, while 
    for $f_1$ and $g_1$ it is quite small and corresponds to the
    hardly-visible dashed-dotted curves.
    In the case of $g_{1T}^{\perp(1)}(x)$, $h_{1L}^{(1)\perp}(x)$ 
    the dashed and dotted curves give
    the results from the S-P and P-D interference terms, respectively.
      In the case of $ h_{1T}^{(1)\perp}(x)$, the dashed curve is the 
      result from the P-wave interference, and the dotted curve
      is due to the interference of S and D waves.}
\end{figure}
%------ END FIGURE 2 --------------------------------------------------

In  Fig.~\ref{Fig2:TMD} we show the results for the
integrals in ${\tvec p}_T$ of the TMDs 
defined in Eq.~(\ref{Eq:integratedTMD}), omitting the flavour dependence
given by the SU(6) isospin factors $N^a$ and $P^a$ in Eqs.~(\ref{eq:h1L}).  
The solid curves correspond to the total results, obtained as the sum of the 
partial-wave contributions. The other
curves show the contributions of the different orbital
angular momentum components of the nucleon wave function.
The unpolarized distribution $f_1$, the helicity distribution $g_{1}$, and the transversity  
$h_{1}$  involve matrix elements which are all diagonal in the orbital angular 
momentum. 
In the plots of these functions, the dashed curves give the contribution from
the
S-wave component, and 
the dotted curves correspond to the P-wave contribution. 
The D-wave gives a negligible contribution to the $f_1$ and 
$g_{1}$ distributions (dashed-dotted curves),
while it is absent in the case of $h_1$.
Although all these three functions are dominated by the S waves,
they have a non-negligible contribution also from the P waves, 
with the largest (smallest) weight in the case of $f_1$ ($g_{1}$).
The functions $g_{1T}^{\perp(1)}$ and $h_{1L}^{\perp(1)}$
 involve a transfer of orbital angular 
momentum by one unit between the initial and final nucleon state.
In our model, they are simply related by Eq.~(\ref{eq:61b}).
For these functions, the dashed curves in  Fig.~\ref{Fig2:TMD} show
the contribution from the interference of S and P waves,
and the  dotted curves correspond to the
results from the P- and D-wave interference term.
The S-P interference term gives the largest contribution
in the full $x$ range, while the P- and D-wave interference  term
contributes at most by 20$\%$.
In the case of $h_{1T}^{\perp(1)}$ one has a  mismatch of  orbital 
angular momentum between the initial and final nucleon state equal to 
$\Delta L_z=2$. In the plot of this function in Fig.~\ref{Fig2:TMD},
the dashed curve gives the result for the interference of the
$L_z=1$ and $L_z=-1$ components, and the dotted 
curve refers to the contribution from the interference of the
the S- and D-wave components.
Thanks to the interference with the S wave, we note that here 
the contribution from the D wave is amplified.
Furthermore, at variance with the other distribution functions,
the different partial-wave contributions do not have 
the same $x$-dependence, and for $x\gtrsim 0.6$ 
the P waves are suppressed with respect to
 the S-D wave interference term.
This peculiar behaviour makes the  $h_{1T}^{\perp(1)}$ function 
interesting, especially in the study of the interplay between
 the different partial wave components in the azimuthal spin asymmetries,
as discussed in the Appendix.

%======== SECTION 4: A_1 and A_LL ====================================
\section{\boldmath 
Collinear double spin asymmetries $A_1$ and $A_{LL}$}
\label{Sec-4:A1-and-ALL}

Before discussing azimuthal asymmetries in SIDIS, we consider first 
the double spin asymmetry $A_{LL}$ and its inclusive analog $A_1$,
Eqs.~(\ref{Eq:ALL},~\ref{Eq:A1}).
The study of these observables in the model framework is instructive,
because in this case evolution equations (and fragmentation functions)
are known and complications due to $p_T$-dependence are avoided.
This allows us to test the model under 'controlled conditions'
in two respects.
First, in which $x$-range and with what accuracy is the model 
applicable? Of course, the performance of the model could vary
with observables. Nevertheless, this exercise will 
give us valuable insights in this respect.
Second, how stable are the results under evolution? 
In this case we can compare exact results, with results obtained making 
{\sl assumptions} on the evolution. 
The experience made here will be useful later, when dealing with azimuthal 
asymmetries whose evolution is practically not solved.

A related key question emerging not only here but in any nonperturbative 
calculation concerns the scale at which the model results for the parton 
distributions hold. From the point of view of
QCD where both quark and gluon degrees of freedom contribute, the role of the 
low-energy quark models is to provide initial conditions for the QCD evolution 
equations. Therefore, we assume 
the existence of a low scale $Q_0^2$ where glue and sea quark contributions 
are suppressed, and the dynamics inside the nucleon is described in terms 
of three valence (constituent) quarks confined by an effective long-range 
interaction.
In fact, glue and sea quark degrees of freedom might be thought of at this
low scale to be contained in the structure of the constituent quarks,
which are massive objects.
The actual value of $Q_0^2$ is fixed evolving back unpolarized data, 
until the  valence distribution matches the condition that the second moment,
i.e.  the momentum fraction carried by the valence quarks, is equal to 
one~\cite{Traini:1997jz}.

Using LO evolution equations, we find 
$Q^2_0=0.079$ GeV$^2$~\cite{Boffi:2003yj}.
Although there is no rigorous relation between the QCD quarks and the 
constituent quarks, and a more fundamental description of the transition from 
soft to hard regimes would be very helpful, this strategy reflects the present 
state of the art for quark model 
calculations~\cite{Scopetta:1997wk,Broniowski:2007si,Pasquini:2004gc},
and has been validated with a fair comparison to 
experiments~\cite{Traini:1997jz,Scopetta:1997wk}.

\begin{figure}[h!]
          \centering
	\vspace{-4mm}
          \includegraphics[height=4.1cm]{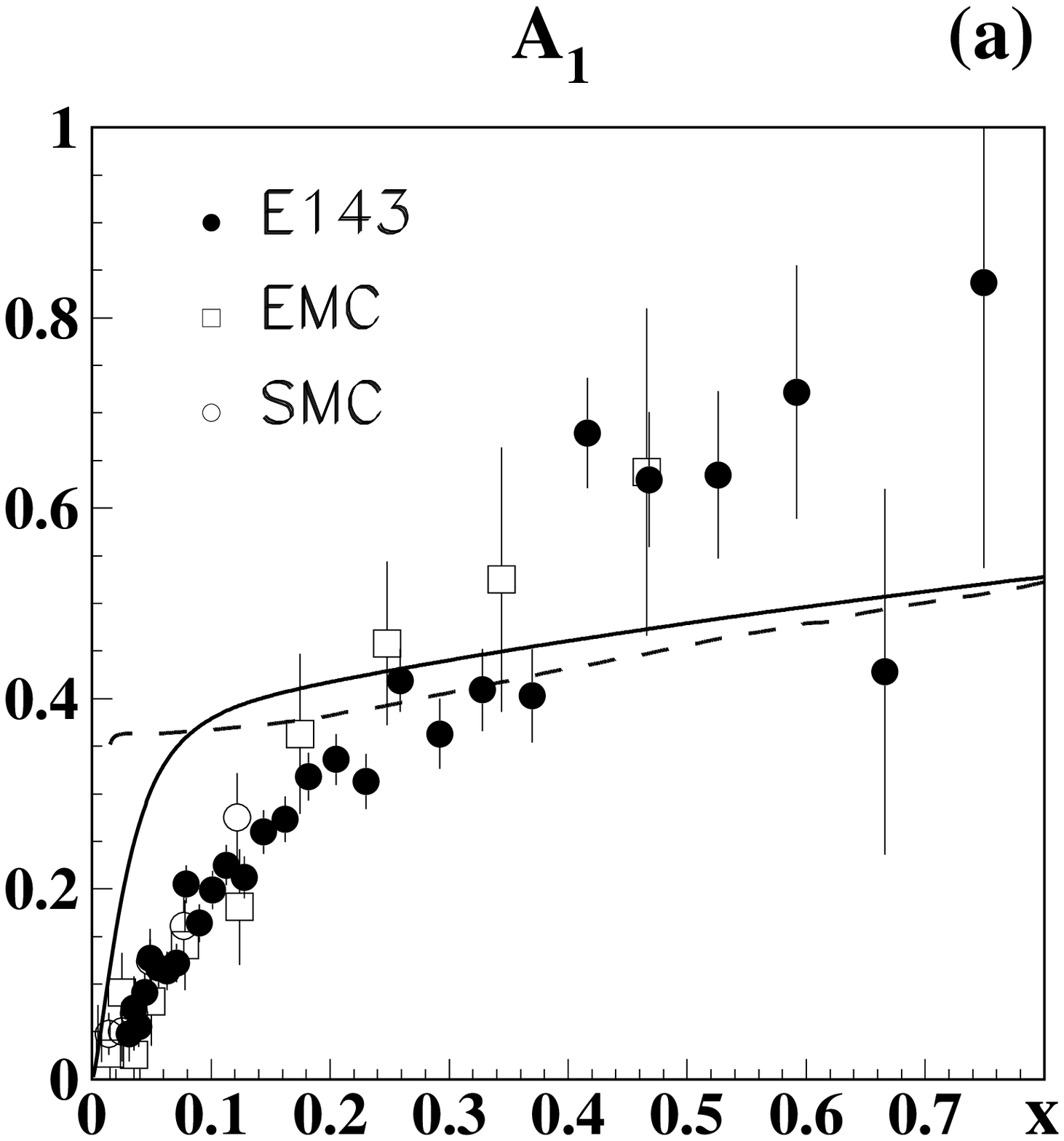}
          \hspace{-8mm}
          \includegraphics[height=4cm]{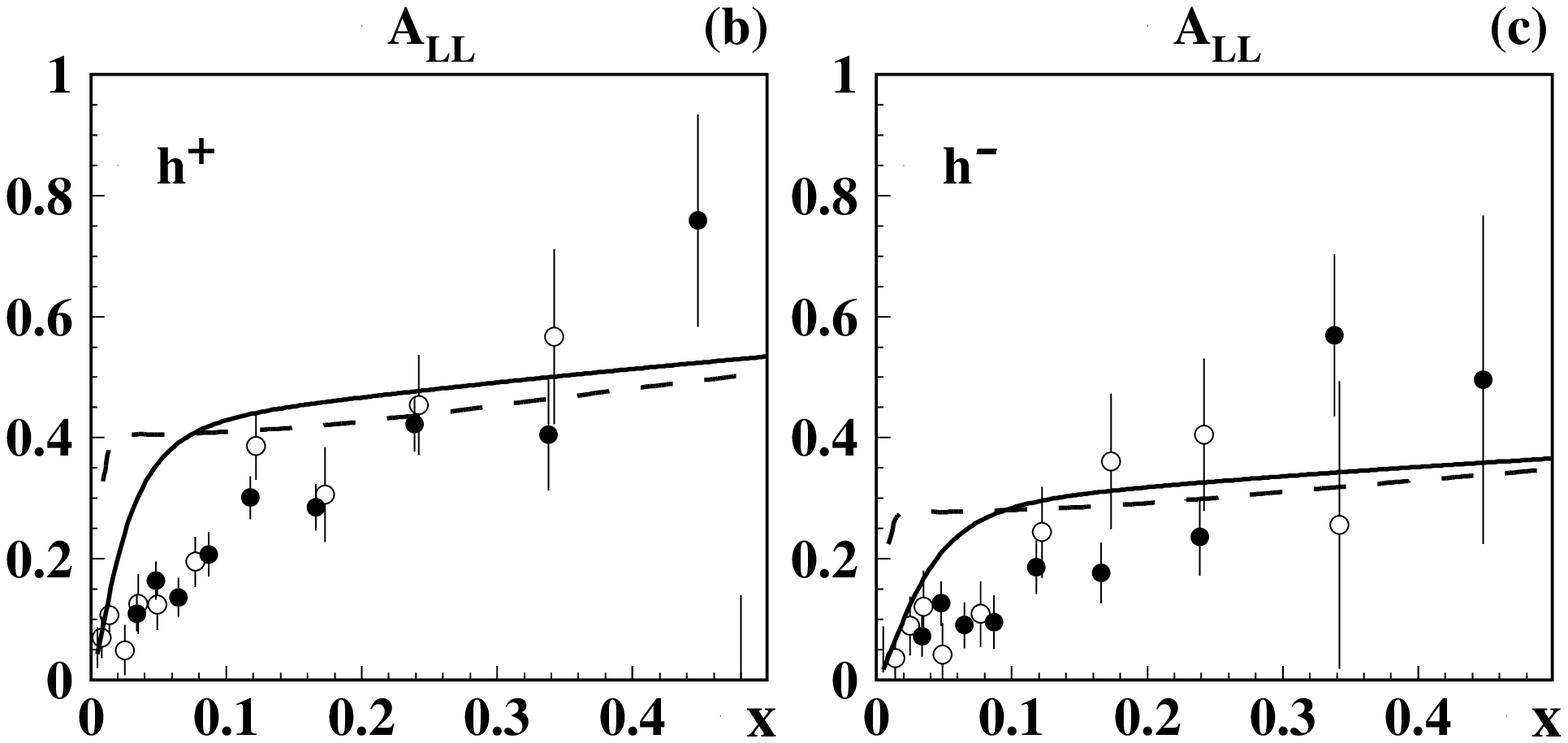}
          \caption{\label{Fig2:A1-ALL} 
	  The inclusive (a) and semi-inclusive (b, c) double spin asymmetries,
          $A_1$ and $A_{LL}$, defined in Eqs.~(\ref{Eq:ALL},~\ref{Eq:A1}), 
          in DIS off proton as functions of $x$. 
          The theoretical curves are obtained with $g_1^a(x)$ and
          $f_1^a(x)$ from the light-cone CQM \cite{Pasquini:2008ax} 
          as follows: 
          both functions LO-evolved to the $\la Q^2\ra$ of the experiments
          (solid~curves), 
          and both at the low scale of the model (dashed curves).
          In (b, c) we use always the parametrization
          \cite{Kretzer:2000yf} for $D_1^a$ at $Q^2=2.5$ GeV$^2$.
          The data in (a) are from 
          Refs.~\cite{Abe:1998wq,Ashman:1987hv,Adeva:1999pa},
          in (b,c) are from SMC (open circles) \cite{Adeva:1997qz}
          and HERMES (black squares) \cite{Airapetian:2004zf}.}
\end{figure}

Fig.~\ref{Fig2:A1-ALL}a shows the inclusive 
double spin asymmetry $A_1$ in DIS off proton, Eq.~(\ref{Eq:A1}).
The two theoretical curves are obtained using $g_1^a(x)$ and $f_1^a(x)$ 
from the light-cone CQM \cite{Pasquini:2008ax}. In one case both
distribution functions are LO-evolved from the low scale of the model $Q^2_0$
to $Q^2=3.0$ GeV$^2$ (solid curve) using the evolution codes of Refs.
\cite{Miyama:1995bd,Hirai:1997gb}, and in the other case both distribution 
functions
are taken at the low scale of the model (dashed curve).
The results differ moderately at $x\gtrsim 0.1$ reflecting the 
weak scale dependence of $A_1$ \cite{Kotikov:1997df}.

As can be seen in Fig.~\ref{Fig2:A1-ALL}a, the description of data from the 
E143, EMC and SMC experiments \cite{Abe:1998wq,Ashman:1987hv,Adeva:1999pa}
is reasonable. For $x\gtrsim 0.15$ the model describes the
$A_1$ data within an accuracy of about $30\,\%$.
The description improves in the valence-$x$ region of 
$x\gtrsim 0.2$ though the accuracy of the data at large $x$
is not sufficient to draw definite conclusions.
Since the model contains no antiquark- and gluon-degrees of freedom,
it is not surprising to observe that it does not work at small-$x$.
As an intermediate summary, it can be said that the results are weakly 
scale dependent, and the model well catches the main features of the 
observable $A_1$ in its range of applicability, namely in the valence-$x$ 
region.

%------ BEGIN FIGURE 3: A1 in DIS, ALL in SIDIS -----------------------
%------ END FIGURE 3 --------------------------------------------------

Since in the following we will deal with SIDIS, we repeat 
the exercise with the double spin asymmetry $A_{LL}$, Eq.~(\ref{Eq:ALL}).
Figs.~\ref{Fig2:A1-ALL}b, c show $A_{LL}$ in DIS production of charged 
hadrons from proton.
The theoretical curves are obtained using $g_1^a(x)$ and $f_1^a(x)$ 
from the light-cone CQM \cite{Pasquini:2008ax}, once LO-evolved to 
$Q^2=2.5$ GeV$^2$ (solid curve), and once left at the initial scale 
of the model (dashed curve).
For the fragmentation function $D_1^a(z)$ we use in both cases,
and throughout this work, the LO parametrization \cite{Kretzer:2000yf} 
at $Q^2=2.5$ GeV$^2$.
Again we observe a weak scale dependence, and a good description of data
in the valence $x$-region where 
the model describes the data within an accuracy of ${\cal O}(20\,\%)$.
Thus, in the SIDIS case we make comparably positive experience
as in the inclusive case. 
Notice that the result with $g_1^a(x)$ and $f_1^a(x)$ taken at the low scale 
is, strictly speaking, not the consistent result for $A_{LL}$ at such a low 
scale because we use the parametrization for $D_1^a$ at  $Q^2=2.5$ GeV$^2$.
 
We remark that we could have tried to describe the double spin
asymmetries $A_1$ and $A_{LL}$ with $g_1^a(x)$ from the model, and 
$f_1^a(x)$ from a parametrization, for example \cite{Gluck:1998xa}.
Such an approach would correspond to the strategy to use the model 
only as input for the part which is responsible for the spin effect, 
and to use for the well known denominator of the spin asymmetry
standard parametrizations,
which has the advantage that the model uncertainty is only in the
numerator. In the case of $A_1$ and $A_{LL}$, 
however, such an approach yields a bad description of the data.
This can be traced back to the fact that the $f_1^a(x)$ from the model 
\cite{Pasquini:2008ax} and from parametrizations \cite{Gluck:1998xa}
have different large-$x$ behaviour. Interestingly, it happens 
to be the case also in the case of $g_1^a(x)$ from the model 
\cite{Pasquini:2008ax} and from parametrizations \cite{Gluck:2000dy},
such that the uncertainties partly {\sl cancel} in the ratio,
leading to a better description of the data.
It is important to stress that here we deal with chiral-even functions,
where antiquark and gluon degrees of freedom are of importance.
In the case of chiral-odd TMDs the situation is different, and 
a different approach could be more successful.
We will come back to this point later on.

The above discussion allows to assign a 'typical accuracy' to the approach.
In this context it is of interest to make the following observation. 
The present version of the model uses SU(6) symmetry, such 
that $g_1^u(x)=-4g_1^d(x)$ at the low scale, 
and similarly for other polarized distribution functions or TMDs.  
Due to isospin symmetry the structure functions of the neutron 
are related to those of the proton by interchanging $u$ and $d$ flavour.
Therefore, at the low scale $A_1^n\propto\frac49\,g_1^{u/n}+\frac19\,g_1^{d/n}$
$=\frac49\,g_1^d+\frac19\,g_1^u=0$. At higher scales $A_1^n\neq 0$ 
due to evolution but the effect remains small.
Rather than claiming $A_1^n\approx 0$,  
it is more meaningful to state that SU(6) predicts $A_1^n$ to be small
compared to, say, the $A_1$ of proton. Thus, for the SU(6) symmetry concept 
to be a useful tool, we expect for the ratio
\be\label{Eq:A1-neutron}
       \biggl|\frac{A_1^n}{A_1^p}\biggr| \ll 1 \;.
\ee
Experimentally $A_1^n$ (extracted by subtracting deuteron 
and proton data, or from $^3$He data, modulo nuclear corrections)
is found clearly non-zero. However, in the valence-$x$ region the SU(6) 
expectation is supported by data \cite{Zheng:2003un,Dharmawardnane:2006zd,Anthony:1996mw,Abe:1997dp,Adams:1995ufa,Chen:2008ja}: 
the ratio  (\ref{Eq:A1-neutron}) is of the order of magnitude 20$\,\%$
--- which is indeed a 'zero' within the model accuracy.

Notice that in SIDIS the SU(6) symmetric TMDs are weighted with fragmentation 
functions, such that in general azimuthal asymmetries from the neutron are 
non-zero already at the low scale (with the exception of $\pi^0$ 
into which $u$ and $d$ quarks fragment with equal strength).
Nevertheless, also in SIDIS the results for a neutron target are highly
sensitive to SU(6) breaking effects, and we refrain from showing them here.
An adequate description of spin asymmetries from a neutron target requires
to account systematically for possible SU(6)-breaking effects.
This is similar to what one observes in the case of the electric form factor 
of the neutron, where S'-wave components in the nucleon wave function were 
found essential to reproduce the experimental data~\cite{Pasquini:2007iz,JuliaDiaz:2003gq}. 
Results for the TMDs with such SU(6) breaking terms 
will be discussed elsewhere.

Let us draw conclusions from the study presented in this Section.
The double spin asymmetries in inclusive DIS, $A_1$, and SIDIS, 
$A_{LL}$, are weakly scale-dependent. The model describes the data
on these observables within an accuracy of $\sim 20-30\%$
in the valence-$x$ region. This suggests that the approach could
also be useful for studies of azimuthal spin asymmetries. In fact,
at this point it is worth to stress that
(i)    azimuthal phenomena are expected to yield sizable effects
       especially in the valence-$x$ region,
(ii)   {\sl first} data on azimuthal asymmetries often have
       uncertainties comparable to the observed model accuracy,
(iii)  in proposals for future experiments predictions of new effects 
       of an accuracy of ${\cal O}(30\,\%)$ are useful enough.

%===================  SECTION 5: ASYMMETRIES AND GAUSS ===============
\section{\boldmath $P_T$-dependence, Gauss Ansatz, applicability of the model}
\label{Sec-5:pT-and-Gauss}

When dealing with azimuthal asymmetries in SIDIS it is very convenient 
to use the Gaussian model for the distribution of transverse parton momenta,
see Sec.~\ref{Sec-2:TMDs-and-SIDIS}.
If one {\sl assumes} the Gaussian Ansatz (\ref{Eq:Gauss-ansatz}) for 
$f_1^a(x,p_T)$ and $D_1^a(z,K_T)$, then a good description of the SIDIS 
data (more precisely, {\sl mean values} for $\la P_{h\perp}(z)\ra$ not 
corrected for acceptance effects) from HERMES \cite{Airapetian:2002mf}
is obtained with the following parameters \cite{Collins:2005ie} 
\be\label{Eq:fit-pT2-KT2}
        \la p_T^2(f_1)\ra = 0.33\,{\rm GeV}^2 \;,\;\;\;
        \la K^2_T(D_1)\ra = 0.16\,{\rm GeV}^2 \;.\ee
Numerically very similar results were obtained in \cite{Anselmino:2005nn} 
from a study of EMC data \cite{Arneodo:1986cf} on the Cahn effect
\cite{Cahn:1978se}. 

The $p_T$-dependence of TMDs in the model \cite{Pasquini:2008ax} is 
definitely not of Gaussian form. However, the essential question is: 
Can it be reasonably {\sl approximated} by a Gaussian form?

In order to discuss that let us make the following two exercises. 
First, we define the mean transverse 
momenta $(n=1)$ and the mean square transverse momenta $(n=2)$ in 
the TMD $j(x,p_T)$ as follows 
\be\label{Eq:define-mean-pT}
       \la p_{T,j}^n\ra = \frac{\int\di x\int\di^2p_T \;p_T^n\,j(x,p_T)}
                           {\int\di x\int\di^2p_T \;j(x,p_T)} \;.
\ee
In Table~\ref{Table:pT-model} we show results for these quantities
for T-even twist-2 TMDs from \cite{Pasquini:2008ax}. 

In order to see to which extent the results for the $p_T$-dependence 
of TMDs from \cite{Pasquini:2008ax} can be approximated by a Gaussian 
behaviour, we remind that in the Gaussian model the following relation holds 
\be\label{Eq:Gauss-relation-pT-pT2}
       \la p_T^2\ra \stackrel{\rm Gauss}{=} \frac{4}{\pi}\,\la p_T\ra^2\;.
\ee
In Table~\ref{Table:pT-model}  we show also the 
results for the ratio $\frac{4\la p_T\ra^2}{\pi\la p_T^2\ra}$ 
that would be unity for a Gaussian $p_T$-distribution.
Remarkably, the model results for this ratio from 
\cite{Pasquini:2008ax} deviate from unity by not more  than $10\,\%$.
Of course, although the Gaussian model relation 
(\ref{Eq:Gauss-relation-pT-pT2})
works within $10\,\%$, it does not necessarily imply that the $p_T$-dependence
in the model is Gaussian within such an accuracy. 
We make therefore the following second exercise.

\begin{table}[b!]
\begin{tabular}{cllll}
\hline
  \ \hspace{3cm} \ 
& \ \hspace{2cm} \ 
& \ \hspace{2cm} \ 
& \ \hspace{2cm} \ 
& \ \hspace{2cm} \  \cr
TMD $j$ & $\begin{array}{l}\la p_T\ra \cr {\rm in \; GeV}   \end{array}$
        & $\begin{array}{l}\la p_T^2\ra \cr {\rm in \; GeV}^2 \end{array}$
        & $\displaystyle\frac{4\la p_T\ra^2}{\pi\la p_T^2\ra}$ 
        & $\displaystyle\frac{\la p_T^2(j)\ra}{\la p_T^2(f_1)\ra}$ \cr
& & & & \cr
\hline
$f_1$          & 0.239 & 0.080 & 0.909 & 1    \cr
$g_1$          & 0.206 & 0.059 & 0.916 & 0.74 \cr
$h_1$          & 0.210 & 0.063 & 0.891 & 0.79 \cr
$g_{1T}^\perp$ & 0.206 & 0.059 & 0.916 & 0.74 \cr
$h_{1L}^\perp$ & 0.206 & 0.059 & 0.916 & 0.74 \cr
$h_{1T}^\perp$ & 0.190 & 0.050 & 0.919 & 0.63 \cr
& & & & \cr
\end{tabular}
\caption{\label{Table:pT-model}
The mean transverse momenta and the mean square transverse 
momenta of T-even TMDs, as defined in Eq.~(\ref{Eq:define-mean-pT}), 
from the light-cone CQM \cite{Pasquini:2008ax}.
If the transverse momenta in the TMDs were Gaussian, then the
result for the ratio in the fourth column would be unity, see text.
The last column shows the $\la p_T^2(j)\ra$ in units of 
$\la p_T^2(f_1)\ra$.}
\end{table}

We ask the question: what is the 
difference between computing in the model \cite{Pasquini:2008ax} an 
observable using exact model $p_T$-dependence of TMDs and computing 
it by approximating the true $p_T$-dependence by a Gaussian?
We can rephrase this question also as follows:
when integrating out the transverse momenta of produced hadrons
and focusing, for example, on the $x$-dependence of azimuthal asymmetries,
the $p_T$-model-dependence is weakened, but to what extent?

In order to answer that question we choose the double spin asymmetry
$A_{LT}^{\cos(\phi_h-\phi_S)}$ and use the model \cite{Bacchetta:2007wc}
for $D_1^a(z,K_T)$, which also refers to a low scale. (We stress that 
the results presented here are to be considered as an exploratory study 
of $p_T$-model effects. Our
predictions for this asymmetry
will be given in the subsequent Section.)

\begin{wrapfigure}[20]{RD}{7.5cm}
    	\centering
	\vspace{-8mm}
        \includegraphics[width=6.3cm]{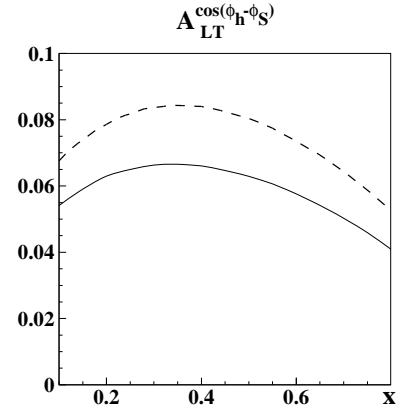}
	\vspace{-7mm}
        \caption{\label{Fig3:alt_gaussian}	  
          $A_{LT}^{\cos(\phi_h-\phi_S)}$ in $\pi^+$ production off proton, 
          as function of~$x$. 
	  Solid curve: exact result obtained using $g_{1T}(x,p_T)$ from 
          \cite{Pasquini:2008ax} and $D_1(z,K_T)$ from \cite{Bacchetta:2007wc}.
	  Dashed curve: an approximation obtained using the integrated
          functions $g_{1T}^{(1)}(x)$, $D_1(z)$ from  
          \cite{Pasquini:2008ax,Bacchetta:2007wc} and 'simulating' their
          $p_T$-dependence by means of the Gaussian Ansatz,
          as described in the text.}  
\end{wrapfigure}

In Fig.~\ref{Fig3:alt_gaussian} we see the results for 
$A_{LT}^{\cos(\phi_h-\phi_S)}$ obtained as follows.
The solid curve shows the result from solving numerically
the convolution integral in (\ref{Eq:FLT}) with 
$g_{1T}^{\perp}(x,p_T)$ from  \cite{Pasquini:2008ax} 
and $D_1^a(z,K_T)$ from \cite{Bacchetta:2007wc}.
The dotted curve shows the result for the asymmetry obtained from the 
Gaussian model, Eq.~(\ref{Eq:GaussFLT}), using $g_{1T}^{\perp(1)a}(x)$ from  
\cite{Pasquini:2008ax} and $D_1^a(z)$ from \cite{Bacchetta:2007wc},
assuming the Gaussian Ansatz (\ref{Eq:Gauss-ansatz}) for these functions,
and assigning the Gaussian widths according to Eq.~(\ref{Eq:define-mean-pT}).

%------ BEGIN FIGURE 4: Gaussian Ansatz for A_LT ---------------------
%------ END FIGURE 4 -------------------------------------------------

The different results agree within an accuracy of $20\,\%$, see
Fig.~\ref{Fig3:alt_gaussian}. Such an uncertainty is 'within the 
model accuracies' of Refs.~\cite{Pasquini:2008ax,Bacchetta:2007wc}.
Thus we conclude that the true transverse-momentum dependence 
in the models \cite{Pasquini:2008ax,Bacchetta:2007wc} can be 
{\sl approximated} by the Gaussian Ansatz with a satisfactory
precision for practical purposes.

Next we address the question how to use consistently the model predictions 
\cite{Pasquini:2008ax} for phenomenology --- in view of the fact that they 
refer to a very low hadronic scale. The fact that $\la p_T^2\ra$ 
of $f_1^{a}$ in that model is smaller compared to what is required by 
phenomenology,  Eq.~(\ref{Eq:fit-pT2-KT2}), is perfectly reasonable.
Sudakov effects make the $p_T$-distributions broader, i.e.\ $\la p_T^2\ra$ 
larger, when evolving to larger (experimentally relevant) scales.

This $p_T$-broadening is expected to be independent of the quark
polarization, in first approximation. Thus, what we can use for phenomenology
are the model results for $\la p_T^2\ra$ in units of the mean square 
transverse momenta of $f_1$, see last column in Table~\ref{Table:pT-model},
and take the 'unit' $\la p_T^2(f_1)\ra$ from phenomenology,
Eq.~(\ref{Eq:fit-pT2-KT2}).

On the basis of the considerations in this and in the previous Section  
we are in the position to establish our strategy to treat azimuthal 
asymmetries in the following. Let us summarize.

\begin{itemize}
\item We will mainly focus on the $x$-dependence of the asymmetries,
      especially in the valence-$x$ region (see Sec.~\ref{Sec-4:A1-and-ALL}).
\item We will assume the Gaussian model, which is a reasonable approximation
      (this Section, see above).
\item When information on a specific Gaussian width of a polarized TMD
      is needed, we will use the model prediction for the corresponding 
      ratio (see last column in Table~\ref{Table:pT-model}), and 
      the value from Eq.~(\ref{Eq:fit-pT2-KT2}) for the width of $f_1^{a}$ .
\item We will not discuss the $z$-dependence of the azimuthal asymmetries,
      because here integrals over the $x$-dependence enter which extend, 
      depending on the experiment, to low-$x$ regions where the model
      is not applicable.\footnote{\label{Footnote-1} 
      We recall that the numerators and denominators of the asymmetries 
      (\ref{Eq:FUTSiv})--(\ref{Eq:FUTpretzel}) are actually weighted 
      by $1/Q^4 \propto 1/x^2$ which strongly emphasizes the role 
      of the small-$x$ region, whose description is beyond the 
      range of applicability of the model.}
\item Similar warnings apply to the $P_{h\perp}$-dependence of the 
      asymmetries. We shall therefore address this point with particular
      care, see Sec.~\ref{Sec-10:Ph-dep} below.
\end{itemize}

%===================  SECTION 6: A_LT^sin(phi+phiS) ==================
\section{\boldmath The double-spin asymmetry $A_{LT}^{\cos(\phi_h-\phi_S)}$}
\label{Sec-6:A_LT}

We start the discussion of azimuthal asymmetries with the double-spin 
asymmetry $A_{LT}^{\cos(\phi_h-\phi_S)}=F_{LT}^{\cos(\phi_h-\phi_S)}/F_{UU}$
which is proportional to $\sum_ae_a^2\;g_{1T}^{\perp a}\,D_1^a$,
see Eq.~(\ref{Eq:FLT}). Assuming the Gaussian Ansatz, which gives a good 
approximation, see Sec.~\ref{Sec-5:pT-and-Gauss}, we have to model the 
prefactor $B_0^\prime$ in Eq.~(\ref{Eq:GaussFLT}).
For that let us rewrite that factor as
\be\label{Eq:B0prime-estimate}
   B_0^\prime = \frac{\sqrt{\pi}\,M}{\la p_T^2(f_1)\ra^{1/2}}\;
   \Biggl\{
   \frac{\la p_T^2(g_{1T}^\perp)\ra}{\la p_T^2(f_1)\ra}+
   \frac{\la K_T^2(D_1)\ra}{z^2\la p_T^2(f_1)\ra}\Biggr\}^{-1/2}\;.
\ee
For the first ratio in the curly brackets we use the model prediction
from the last column in Table~\ref{Table:pT-model}. For the second ratio 
in the curly brackets we use the numbers from Eq.~(\ref{Eq:fit-pT2-KT2}).

%------ BEGIN FIGURE 5: ALT -------------------------------------------
\begin{figure}[t!]
        \centering
        \vspace{-0.5cm}
        \includegraphics[width=13cm]{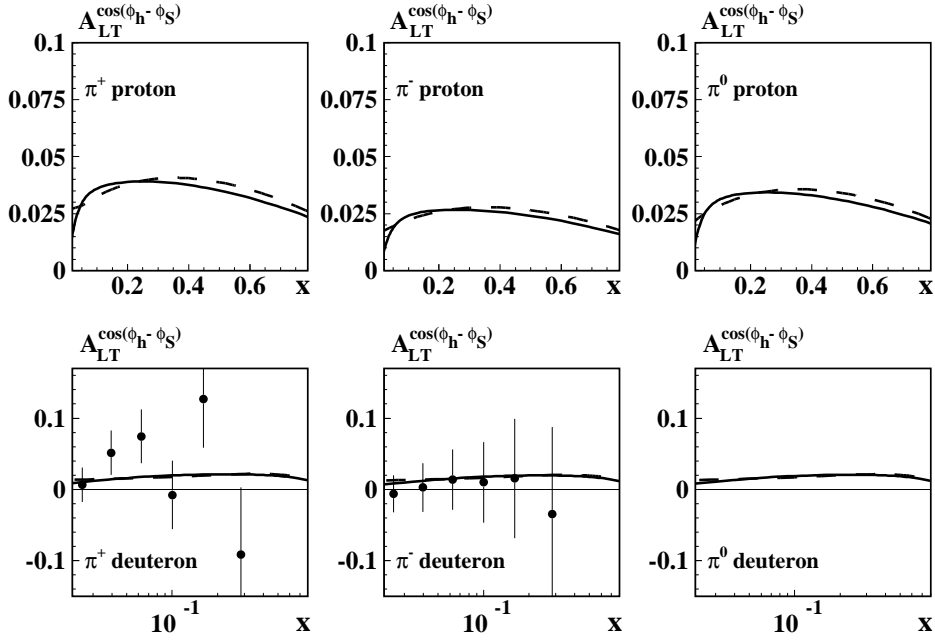}
        \vspace{0.3cm}
        \caption{\label{Fig-4:ALT}
	The double-spin asymmetry $A_{LT}^{\cos(\phi_h-\phi_S)}$ in DIS
	production of pions, as function~of~$x$, obtained using 
        $g_{1T}^{(1)a}(x)$ and $f_1^a(x)$ from the light-cone CQM 
        \cite{Pasquini:2008ax} in the following way:
        both functions are taken at the low scale of the model (dashed curves),
        and both are LO-evolved to $Q^2=2.5$ GeV$^2$ (solid curves).
        Hereby the scale dependence of $g_{1T}^{(1)a}(x)$ is 'simulated' 
        using the $g_1^a(x)$ evolution pattern, see text. 
 	The data points are preliminary COMPASS data
        for charged hadron production off deuteron
        \cite{Kotzinian:2007uv}.}
\end{figure}
%------ END FIGURE 5 --------------------------------------------------

Being interested in the $x$-dependence of the asymmetry, 
we further integrate over $z$
\be
   A_{LT}^{\cos(\phi_h -\phi_S)}(x) =
   \frac{\sum_a e_a^2 \,x\,g_{1T}^{\perp(1)a}(x)\,\la B_0^\prime D_1^a \ra}{
   \sum_a e_a^2 \,x\,f_1^a(x)\,\la D_1^a\ra}
\ee 
where $\la\dots\ra$ denotes the average over $z$ within the 
respective experimental cuts.
Here and in the following, we will consider the range $0.2\leq z \leq 0.7$
corresponding to the typical kinematics of HERMES. 
There is little difference if one uses the COMPASS cuts $0.2\leq z < 1$,
since the resulting $\la z\ra$ is similar. At JLab typically higher $\la z\ra$ 
are reached. For the present observable, however, this has little impact.
 
The results for $A_{LT}^{\cos(\phi_h-\phi_S)}$ in DIS-production 
of pions off different targets are shown in Fig.~\ref{Fig-4:ALT}.
For $g_{1T}^{\perp(1)a}(x)$ and $f_1^{a}(x)$ we take the 
results from the model \cite{Pasquini:2008ax}, and consider two options.
First, we take both functions at the low scale of the model (dashed curves).
Second, we consider both curves LO evolved to $Q^2=2.5\,{\rm GeV}^2$ 
(solid curves). 

Hereby, we use for 
$g_{1T}^{\perp(1)a}(x)$ the evolution equations for $g_1^{a}(x)$.
This is admittedly not the correct evolution pattern.
However, this is the evolution pattern of a chiral-even polarized
function, and the purpose of presenting it here is to shed some
light on the possible size of evolution effects.

Our crude estimate of evolution effects indicates, that the predictions 
for the asymmetries are presumably robust concerning scale dependence. 
The proton asymmetries reach 4$\,\%$ in the valence-$x$ 
region, which could be  measured --- especially at JLab.
The deuteron asymmetries are somewhat smaller.
For a deuteron target there also exist  preliminary data from the 
2002-2004 run of the COMPASS experiment~\cite{Kotzinian:2007uv}. 
As can be seen in Fig.~\ref{Fig-4:ALT}, our results are
compatible with these preliminary data.

Estimates for  $A_{LT}^{\cos(\phi_h-\phi_S)}$ were made 
also in \cite{Kotzinian:2006dw} on the basis of the approximation
\be\label{Eq:g1Tperp-in-WW-approx}
    g_{1T}^{\perp(1)a}(x,Q^2) \stackrel{\rm WW}{\approx} 
    x \int_x^1\frac{\di y}{y}\,g_1^a(y,Q^2)
\ee
using the parametrization \cite{Gluck:2000dy} for  $g_1^a(x)$. 
The approximation is 'justified' in QCD upon the neglect of pure twist-3
(quark-gluon) correlators and current quark mass terms
\cite{Avakian:2007mv,Metz:2008ib,Avakian:2009nj}. 
This is analog to the Wandzura-Wilczek (WW) 
approximation for the twist-3 parton distribution function $g_T^a(x)$ 
\cite{Wandzura:1977qf,Shuryak:1981pi,Jaffe:1989xx} --- hence the label
'WW' in (\ref{Eq:g1Tperp-in-WW-approx}). The WW-approximation for $g_T^a(x)$ 
is supported experimentally within the error bars of the present data
\cite{Zheng:2004ce,Amarian:2003jy,Anthony:2002hy}.
Whether the WW-type approximation (\ref{Eq:g1Tperp-in-WW-approx})
is supported by data equally well remains to be seen.

In~the light-cone CQM \cite{Pasquini:2008ax} the 'WW-type approximation',
Eq.~(\ref{Eq:g1Tperp-in-WW-approx}), is supported in the valence $x$-region 
with good accuracy. Furthermore our results support the findings
of \cite{Kotzinian:2006dw} also numerically.
Taking into account the different kinematical cuts applied in the calculation 
of~\cite{Kotzinian:2006dw}, we obtain  asymmetries of similar size, with 
a more flat $x$-dependence.

%===================  SECTION 7: A_UT^sin(phi+phiS) ==================
\section{\boldmath The single-spin asymmetry $A_{UT}^{\sin(\phi_h+\phi_S)}$}
\label{Sec-7:AUT-Coll}

Next we focus on the azimuthal SSA 
$A_{UT}^{\sin(\phi_h+\phi_S)}=F_{UT}^{\sin\left(\phi_h +\phi_S\right)}/F_{UU}$
due to transversity and the Collins function.
In the Gauss Ansatz (\ref{Eq:Gauss-ansatz}) the structure function in 
the numerator of this SSA is given by the expression
in Eq.~(\ref{Eq:GaussFUTCol}). 

For the Collins function, more precisely for $\la B_1
H_1^{\perp(1/2)a}\ra$ equal to $\la 2B_{\rm Gauss}
H_1^{\perp(1/2)a}\ra$ in the notation of \cite{Efremov:2006qm}, we~use
the results extracted in \cite{Efremov:2006qm} from the (preliminary)
HERMES data \cite{Diefenthaler:2005gx}.  Although meanwhile new data are 
available \cite{Diefenthaler:2007rj,Alekseev:2008dn,Abe:2005zx,Seidl:2008xc}
the results on $H_1^\perp$ from \cite{Efremov:2006qm} are still in
excellent agreement with updated extractions \cite{Anselmino:2008jk}.

When dealing with asymmetries due to chiral-odd TMDs, in our opinion a 
different approach is more appropriate as compared to the case of asymmetries
due to chiral-even TMDs. Let us explain this point in more detail.

When describing asymmetries due to chiral-even functions
in the previous Sections, we used model input for both, the numerator 
and the denominator of the asymmetries. In the model gluon (and sea quark) 
degrees of freedom are absent at the low scale, and generated by evolution 
at higher scales. Admittedly, in this way one cannot accurately describe 
absolute DIS cross section data. For that non-zero unpolarized gluon and sea 
quark distributions are needed already at low input scales \cite{Gluck:1998xa} 
(though the model scale is lower than the initial scale 
of the parametrizations~\cite{Gluck:1998xa}).
This 'shortcoming' of the model, however, affects similarly the numerator 
and the denominator of asymmetries due to chiral-even TMDs.
Indeed, we observed that these model uncertainties partly cancel in the 
ratio --- resulting in a useful description of (SI)DIS data on asymmetries
in the valence-$x$ region, see Sec.~\ref{Sec-4:A1-and-ALL}.

Can we expect a similarly good description of asymmetries, which are due 
to chiral-odd TMDs, using this~strategy? The answer is no, in our opinion.
Transversity has no gluon counterpart, in contrast with $f_1^a(x)$.
The absence of gluon degrees of freedom in an approach constitutes
therefore a 'lesser shortcoming' for $h_1^a(x)$ than for $f_1^a(x)$.
So one expects intuitively that in quark models transversity and other 
chiral-odd TMDs could be modeled more reliably than chiral-even ones,
though it is not clear how to put this expectation on a firm field
theoretical basis.

\begin{figure}[b!]

 \vspace{0.3cm}
 \hspace{-8mm}
 \includegraphics[height=4.2cm]{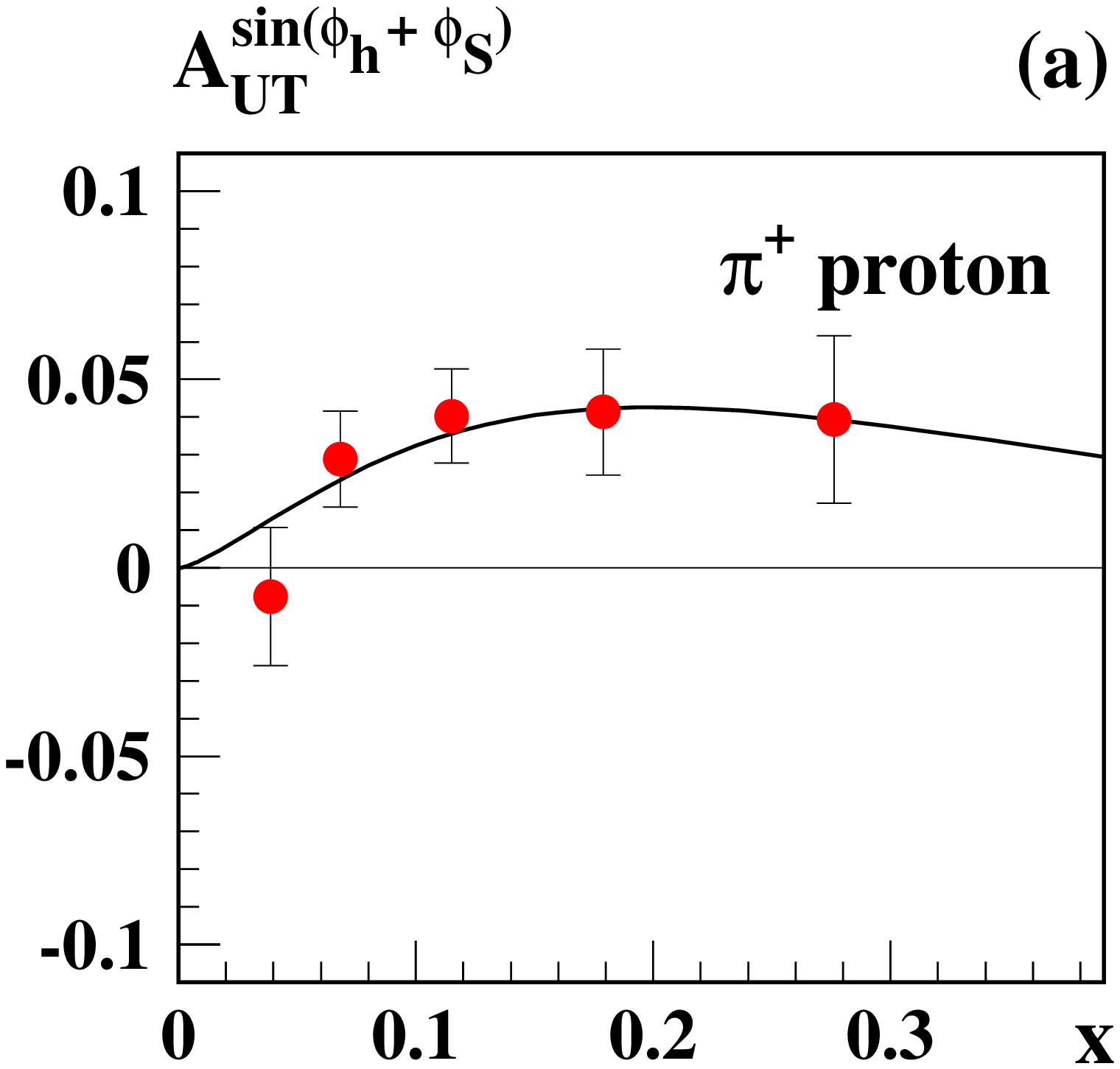}
 \hspace{-11mm}
 \includegraphics[height=4.2cm]{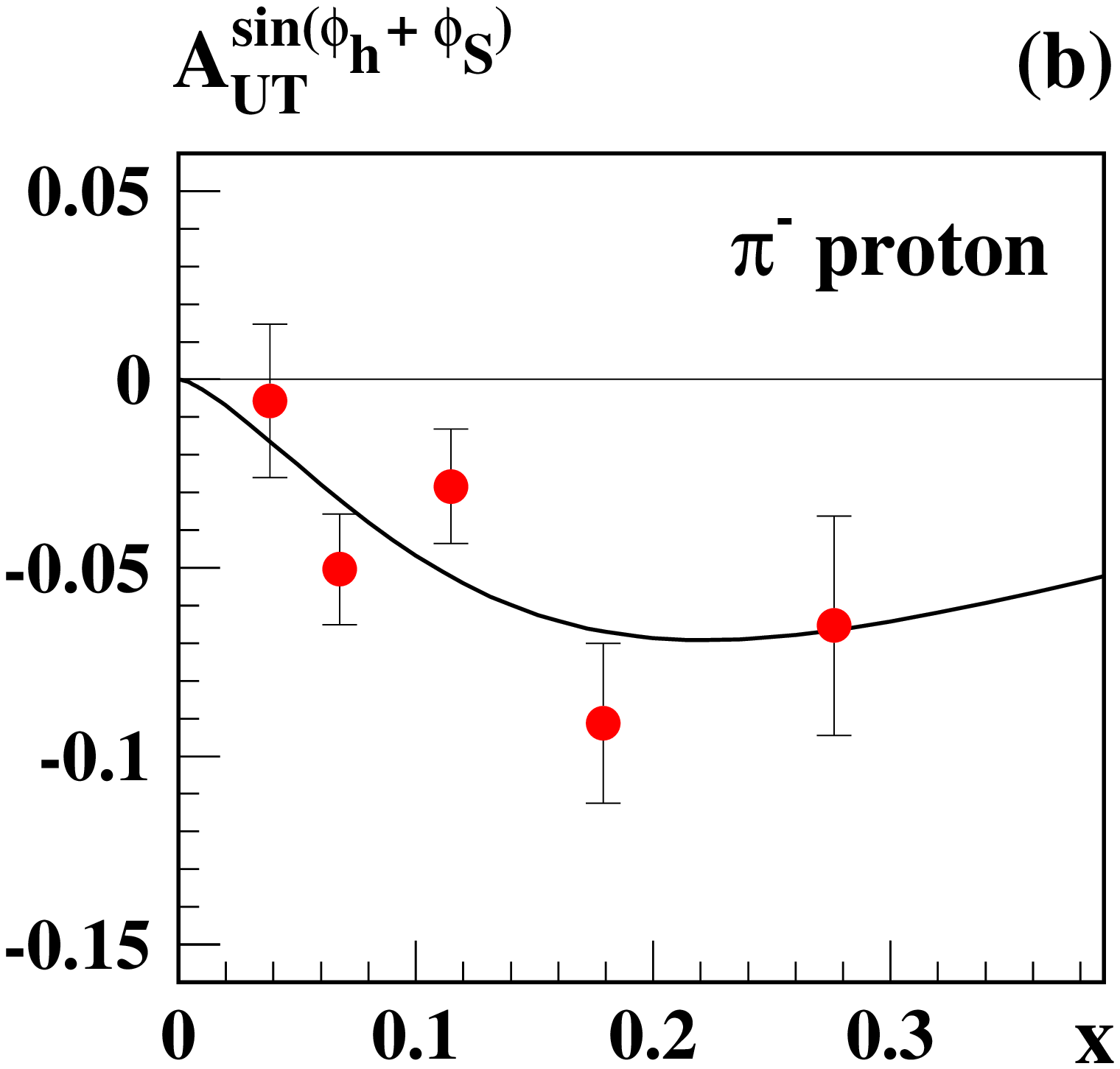}
 \hspace{-13mm}
 \includegraphics[height=4.2cm]{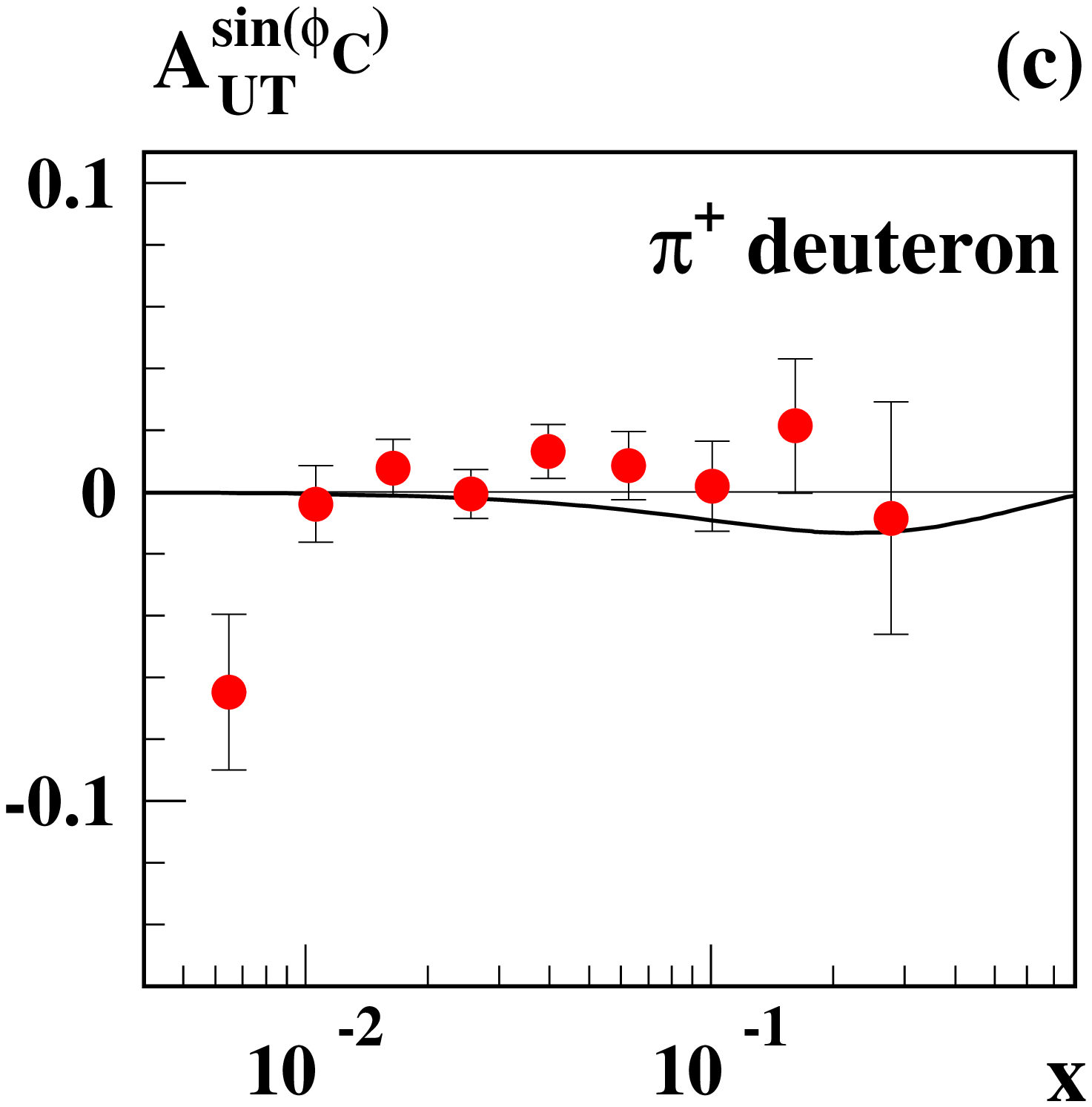}
 \hspace{-11mm}
 \includegraphics[height=4.2cm]{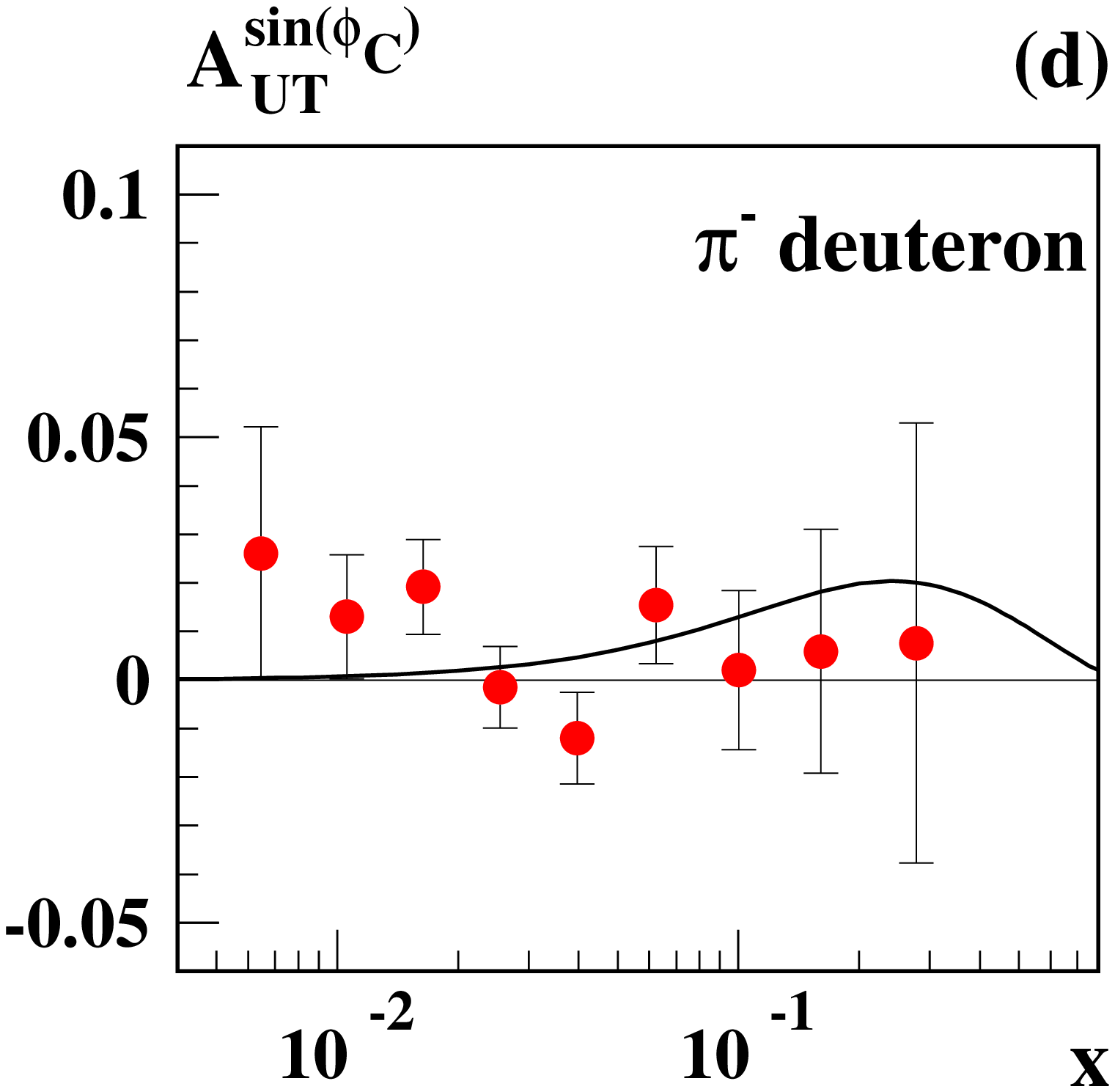}
 \hspace{-12mm}

 \vspace{-1.1cm}
	\caption{\label{Fig-5:AUT-Collins}
	The single-spin asymmetry 
        $A_{UT}^{\sin(\phi_h+\phi_S)}\equiv-A_{UT}^{\sin\phi_C}$ in DIS
	production of charged pions 
	off proton and deuteron targets, as function 
        of $x$. The theoretical curves are obtained on the basis of the 
        light-cone CQM predictions for $h_1^a(x,Q^2)$ from
	Ref.~\cite{Pasquini:2006iv,Pasquini:2008ax}, see text.
        The (preliminary) proton target data are 
        from HERMES \cite{Diefenthaler:2005gx}, 
	the deuteron target data are from COMPASS \cite{Alekseev:2008dn}.}
\end{figure}

Nevertheless, these considerations suggest to adopt the following strategy 
for the description of % asymmetries due to chiral-odd TMDs, 
the Collins SSA, namely to 
use $h_1^a(x)$ from the model LO-evolved~\cite{Pasquini:2006iv,Hirai:1997mm}
 to the experimental scale in 
the numerator of the SSA, and $f_1^a(x)$ from a parametrization, 
e.g.\ \cite{Gluck:1998xa}, taken at the corresponding scale.
In this way, the model uncertainty is limited to the numerator of the SSA 
only, while the denominator is described exactly.
We indeed observe that the above-described strategy yields by far
the best results in the case of the Collins SSA,
see Fig.~\ref{Fig-5:AUT-Collins}. 
Any other options, such as $h_1^a(x)$ and $f_1^a(x)$ from the model
at the low scale or $h_1^a(x)$ and $f_1^a(x)$ from the model LO-evolved,
gave unsatisfactory results.

Let us discuss in some more detail the results for 
$A_{UT}^{\sin(\phi_h+\phi_S)}$
in Fig.~\ref{Fig-5:AUT-Collins}, where for sake of clarity we refrain from
showing the error bands due to the statistical and systematic uncertainties 
of the extracted Collins function \cite{Efremov:2006qm}.
Figs.~\ref{Fig-5:AUT-Collins}a, b show the results for charged pion
production from a proton target in comparison to the preliminary HERMES 
data \cite{Diefenthaler:2005gx}. (It is consistent to compare to these
data, because the information on $H_1^\perp$ \cite{Efremov:2006qm}
was extracted from those data.)
The model results ideally describe these data --- including the 
small-$x$ region, see Figs.~\ref{Fig-5:AUT-Collins}a, b.
This is in line  with the favourable comparison between our model 
predictions~\cite{Pasquini:2005dk,Pasquini:2006iv} 
and the phenomenological extraction of the transversity and tensor charges in 
Ref.~\cite{Anselmino:2007fs,Anselmino:2008jk}.

%------ BEGIN FIGURE 6: A_UT^sin(phi+phiS) ----------------------------
%-------- END FIGURE 6 ------------------------------------------------

In Figs.~\ref{Fig-5:AUT-Collins}c, d we compare our results for
$A_{UT}^{\sin\phi_C}\equiv -A_{UT}^{\sin(\phi_h+\phi_S)}$
(since $\phi_C = \phi_h+\phi_S + \pi$) for charged pion production
from a deuterium target to the COMPASS data \cite{Alekseev:2008dn},
which extend down to much lower values of $x$.
Our results are compatible with the data also in this case, 
including again  the small-$x$ region.

On the basis of the presently available information on the Collins 
function extracted from SIDIS and $e^+e^-$ data,
we would predict $\pi^0$ SSAs compatible 
with zero within the uncertainties of the extractions 
\cite{Efremov:2006qm,Anselmino:2007fs,Anselmino:2008jk}.
However, this is a prediction due to our present understanding 
of the Collins effect, rather than due to the model for TMDs.
For this reason, here and in the following two Sections 
where we discuss further SSAs due to Collins effect,
we refrain from showing results for neutral pion production.

%===================  SECTION 8: A_UL^sin(2phi) ======================
\section{\boldmath The single-spin asymmetry $A_{UL}^{\sin(2\phi_h)}$}
\label{Sec-8:A_UL}

In this Section we discuss the azimuthal SSA 
$A_{UL}^{\sin(2\phi_h)}=F_{UT}^{\sin(2\phi_h)}/F_{UU}$
due to $h_{1L}^{\perp a}$ and the Collins function.
In the Gauss Ansatz (\ref{Eq:Gauss-ansatz}) the structure function in 
the numerator of this SSA is given by the expression
in Eq.~(\ref{Eq:GaussFULsin2phi}). Thus, in order to describe this 
SSA we need $\la B_2^\prime H_1^{\perp(1/2)a}\ra$ which we estimate 
on the basis of the Collins function extractions 
\cite{Collins:2005ie,Anselmino:2005nn} precisely as described 
in Ref.~\cite{Avakian:2007mv}.

The problem we face in the context of $A_{UL}^{\sin(2\phi_h)}$ concerns
the question how to evolve correctly $h_{1L}^{\perp(1)a}(x)$ from the
low initial scale of the model to the relevant experimental scale.
In contrast with transversity, exact evolution equations are not available
in this case. 

In our study of $A_{UT}^{\sin(\phi_h+\phi_S)}$ we learned that other 
strategies, such as leaving transversity at the low scale of the model
(and taking $f_1^a(x)$ in the denominator from the model or from
parametrization, at the low scale or evolved) resulted in
unfavourable descriptions of data, and we were able to understand
qualitatively why. 
Of course, this is here
a different observable. But the experience with the Collins SSA
does not encourage any other strategy than that adopted in that
case, in Sec.~\ref{Sec-7:AUT-Coll}, namely to evolve the chiral-odd 
TMD from the model, and use parametrizations for the denominator
of the SSA.

Not being able to evolve  $h_{1L}^{\perp(1)a}(x)$ correctly we use 
instead the $h_1^a(x)$-evolution pattern to evolve it 'approximately'. 
Since both functions are chiral-odd, the simulation of evolution effects 
in this way can be expected to be more promising than using any other 
evolution pattern.

%------ BEGIN FIGURE 7: A_UL^sin(2phi)  -------------------------------
%
\begin{figure}[b]
\vspace{10mm}

 \hspace{-8mm}
 \includegraphics[height=4.2cm]{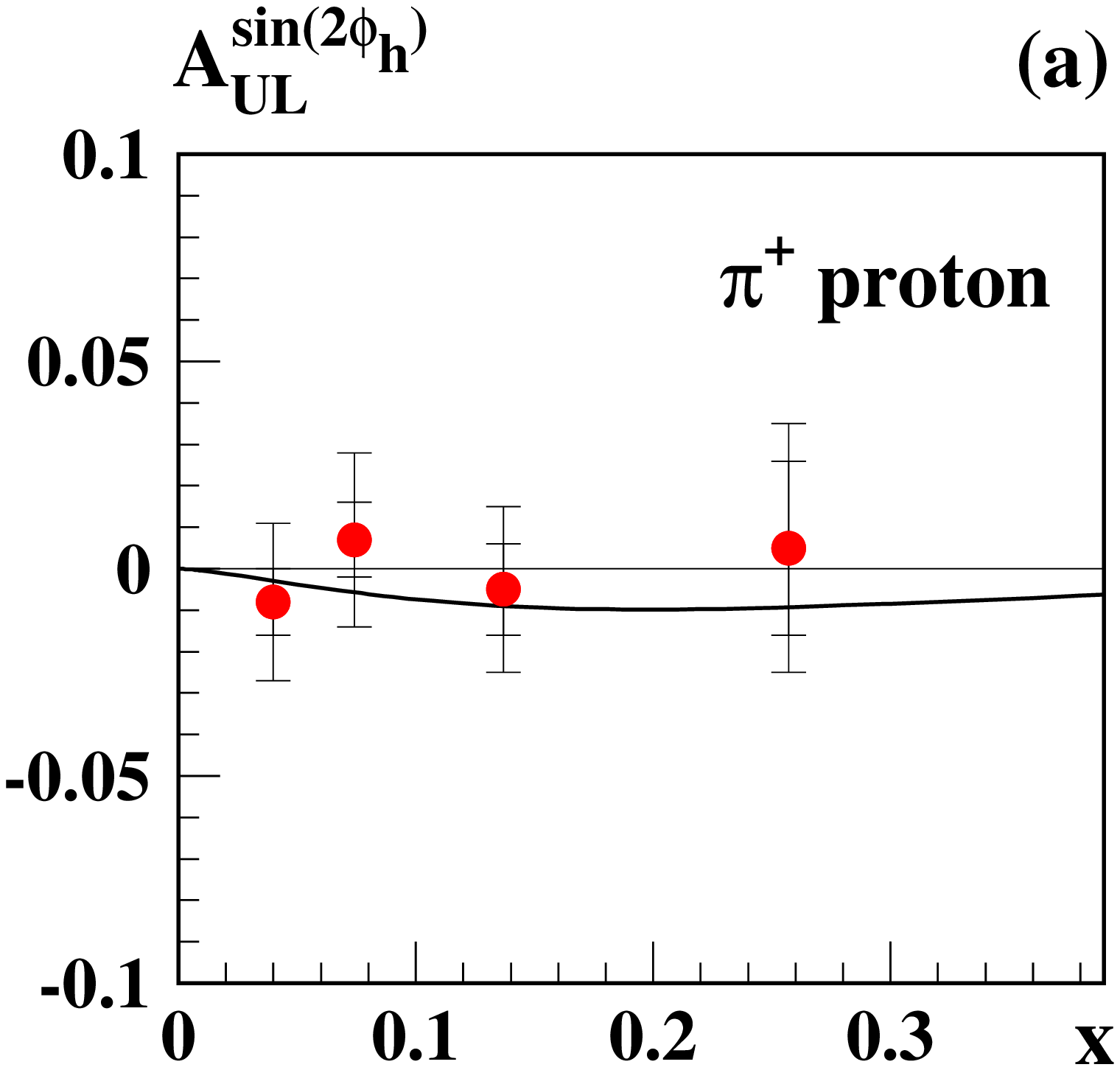}
 \hspace{-11mm}
 \includegraphics[height=4.2cm]{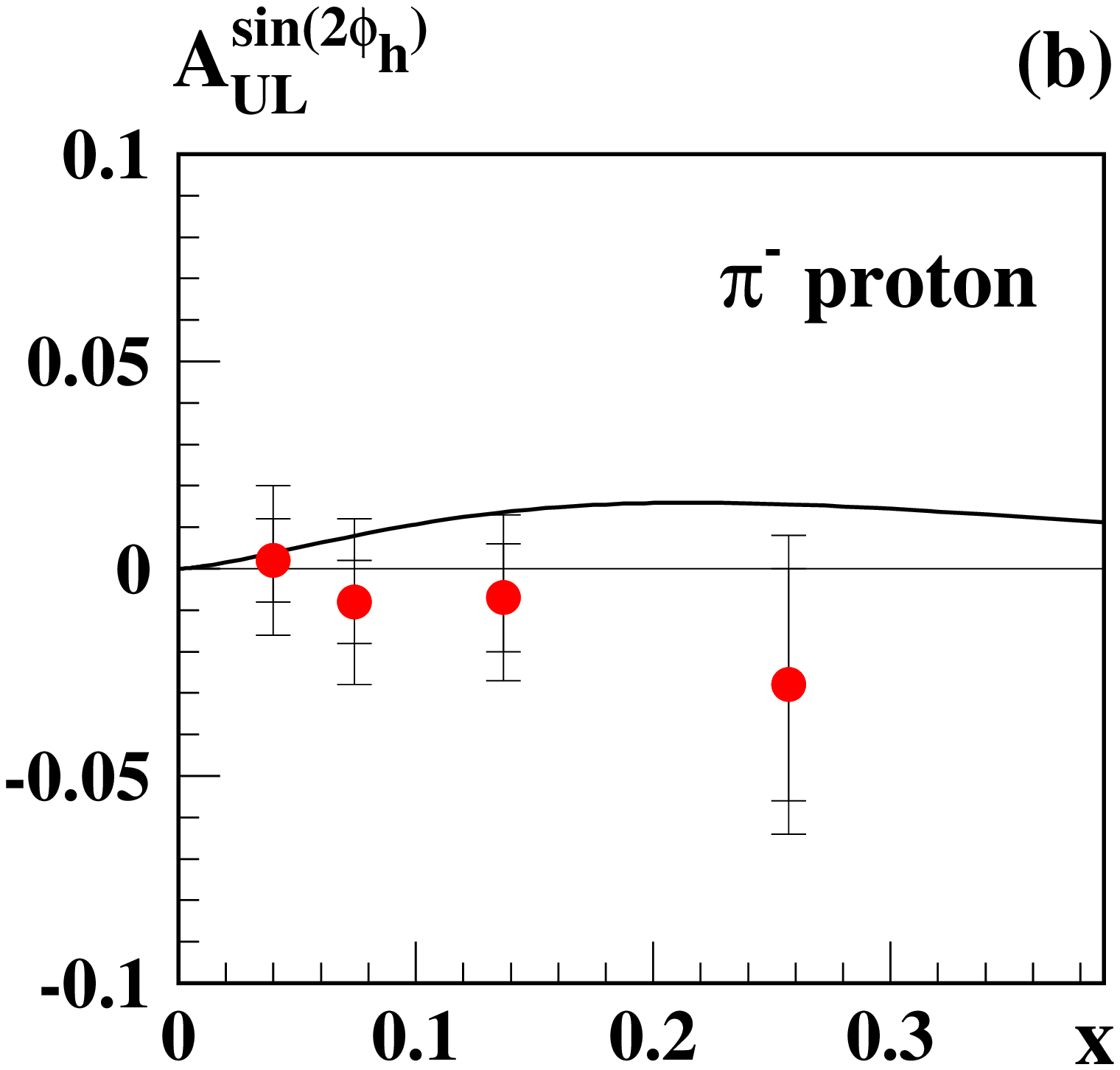}
 \hspace{-13mm}
 \includegraphics[height=4.2cm]{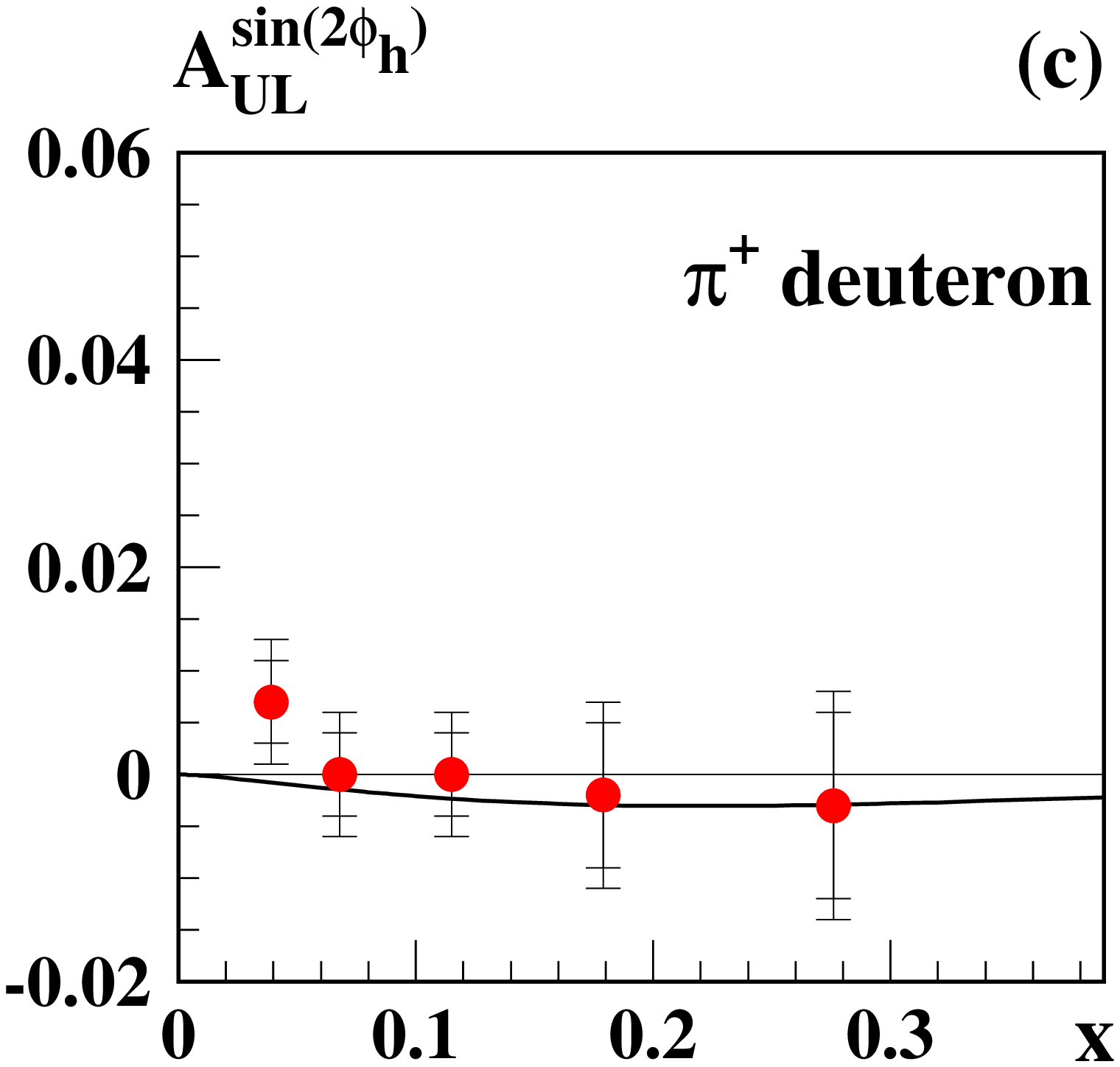}
 \hspace{-11mm}
 \includegraphics[height=4.2cm]{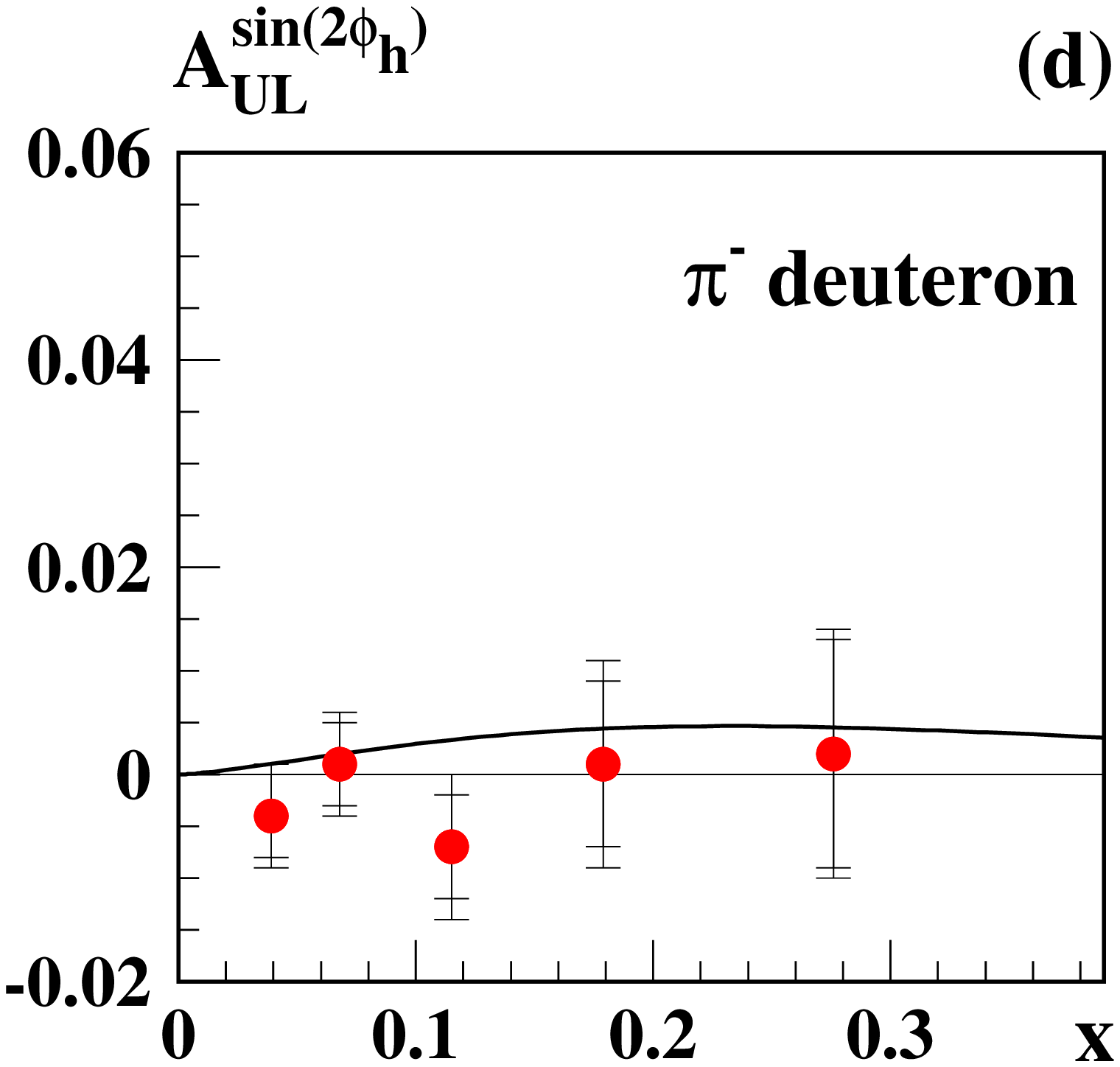}
 \hspace{-12mm}
 \vspace{-5mm}

    \caption{\label{Fig08:AULsin2phi}
	The single-spin asymmetry $A_{UL}^{\sin(2\phi_h)}$ in DIS
	production of charged pions off  proton and deuteron targets, 
	as function of $x$.
	The theoretical curves are obtained by evolving
	the light-cone CQM predictions for $h_{1L}^{\perp(1)a}$ of
	Ref.~\cite{Pasquini:2008ax} to $Q^2=2.5$ GeV$^2$, using
	the $h_1^a$ evolution pattern, see text.
	The data points are from 
	HERMES~\cite{Airapetian:1999tv,Airapetian:2002mf}. 
	The inner error bars are the statistical errors, 
	the outer error bars are the systematic errors.}
\end{figure}
%
%------ END FIGURE 7---------------------------------------------------

In Fig.~\ref{Fig08:AULsin2phi} we compare the results obtained in this way
to the HERMES data from proton and deuteron targets 
\cite{Airapetian:1999tv,Airapetian:2002mf}. We observe that our estimates
are well compatible with the data --- including again the small-$x$ region.
For comparison, in Appendix and Ref.~\cite{ProcSPIN08} we  also 
show the results obtained with the same ingredients as in 
Fig.~\ref{Fig08:AULsin2phi} but without approximate evolution of $h_{1L}^a(x)$.
This  approach yields a somewhat larger SSA, especially at large $x$, but it 
is similarly compatible with the data.

In Ref.~\cite{Avakian:2007mv} predictions for the $A_{UL}^{\sin(2\phi_h)}$ SSA
were made on the basis of the WW-type approximation 
\be\label{Eq:h1Lperp-in-WW-approx}
        h_{1L}^{\perp(1)a}(x) \stackrel{\rm WW}{\approx} 
        -x^2 \int_x^1\frac{\di y}{y^2}\,h_1^a(y)\;,
\ee
and model predictions for transversity from \cite{Schweitzer:2001sr}.
Eq.~(\ref{Eq:h1Lperp-in-WW-approx}) is analog to the approximation
(\ref{Eq:g1Tperp-in-WW-approx}), i.e.\ it also arises when certain 
quark-gluon correlator and current quark-mass terms are neglected. 
Interestingly, the light-cone CQM supports the approximation
(\ref{Eq:h1Lperp-in-WW-approx}) within a reasonable accuracy
\cite{Pasquini:2008ax}. Also the numerical results for the SSA 
obtained here and in \cite{Avakian:2007mv} agree well
qualitatively.

It is, of course, an important question how to quantify the theoretical
uncertainty we introduced in our study by employing the incorrect 
evolution pattern for $h_{1L}^{\perp(1)a}(x)$. 
Until exact evolution equations for this TMD will be available,
this question cannot be answered exactly. However, one may suspect that 
the uncertainties due to evolution are less dominating than other 
uncertainties within the model. The current HERMES data do not contradict
this expectation, see Fig.~\ref{Fig08:AULsin2phi}. 
We remark that there are also preliminary CLAS data \cite{Avakian:2005ps}. 
Our approach is compatible with the results for $\pi^+$ and $\pi^0$ but 
cannot explain the trend of the $\pi^-$ SSA,
similarly to Ref.~\cite{Avakian:2007mv}.
The situation will be further clarified in future experiments 
at JLab~\cite{Avakian-clas6,Avakian-clas12}, and COMPASS.

%===================  SECTION 9: PRETZELOSITY IN SIDIS ===============
\section{\boldmath The single-spin asymmetry $A_{UT}^{\sin(3\phi_h-\phi_S)}$}
\label{Sec-9:pretzelosity}

Finally we study the azimuthal SSA 
$A_{UT}^{\sin(3\phi_h-\phi_S)}=F_{UT}^{\sin\left(3\phi_h +\phi_S\right)}/
F_{UU}$ due to pretzelosity and the Collins function.
In the Gauss Ansatz (\ref{Eq:Gauss-ansatz}) the structure function in 
the numerator of this SSA is given by the expression
in Eq.~(\ref{Eq:GaussFUTpretzel}). 
The factor $\la B_3 H_1^{\perp(1/2)a}\ra$ we evaluate exactly as done in 
Ref.~\cite{Avakian:2008dz}.

Also in the context of the asymmetry $A_{UT}^{\sin(3\phi_h-\phi_S)}$
we face the question how to evolve $h_{1T}^{\perp(1)a}(x)$ from the
low initial scale of the model to the relevant experimental scale.
Exact evolution equations are not available in this case, either.
We follow here the approach developed in the previous Section,
and 'simulate' the evolution of $h_{1T}^{\perp(1)a}(x)$ by evolving 
it according to the transversity-evolution pattern.
Again, since pretzelosity and transversity are chiral odd, 
this way might be a useful estimate of evolution effects.

\begin{figure}[t!]
\vspace{1 truecm}
 \hspace{-8mm}
 \includegraphics[height=4.2cm]{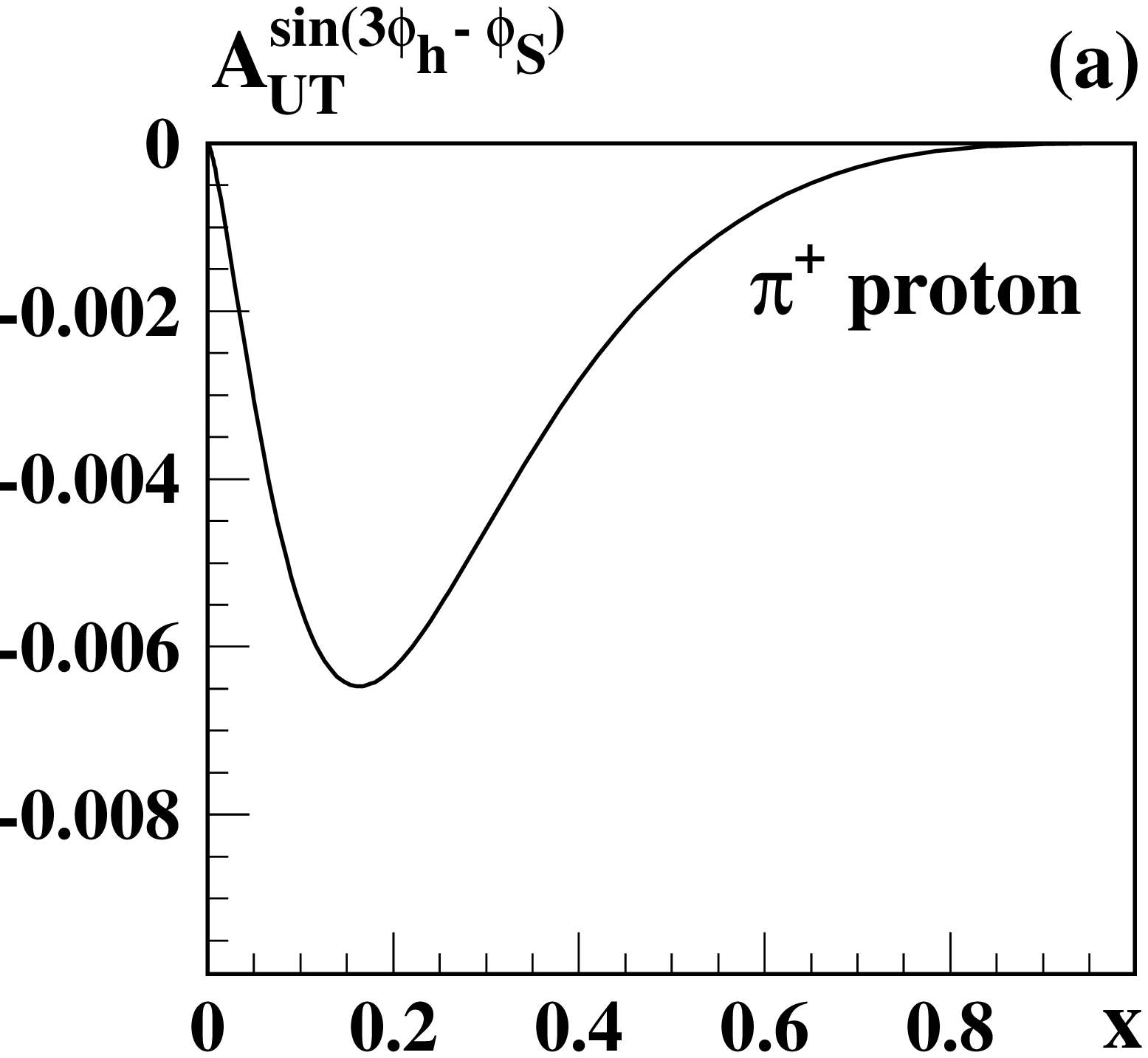}
 \hspace{-11mm}
 \includegraphics[height=4.2cm]{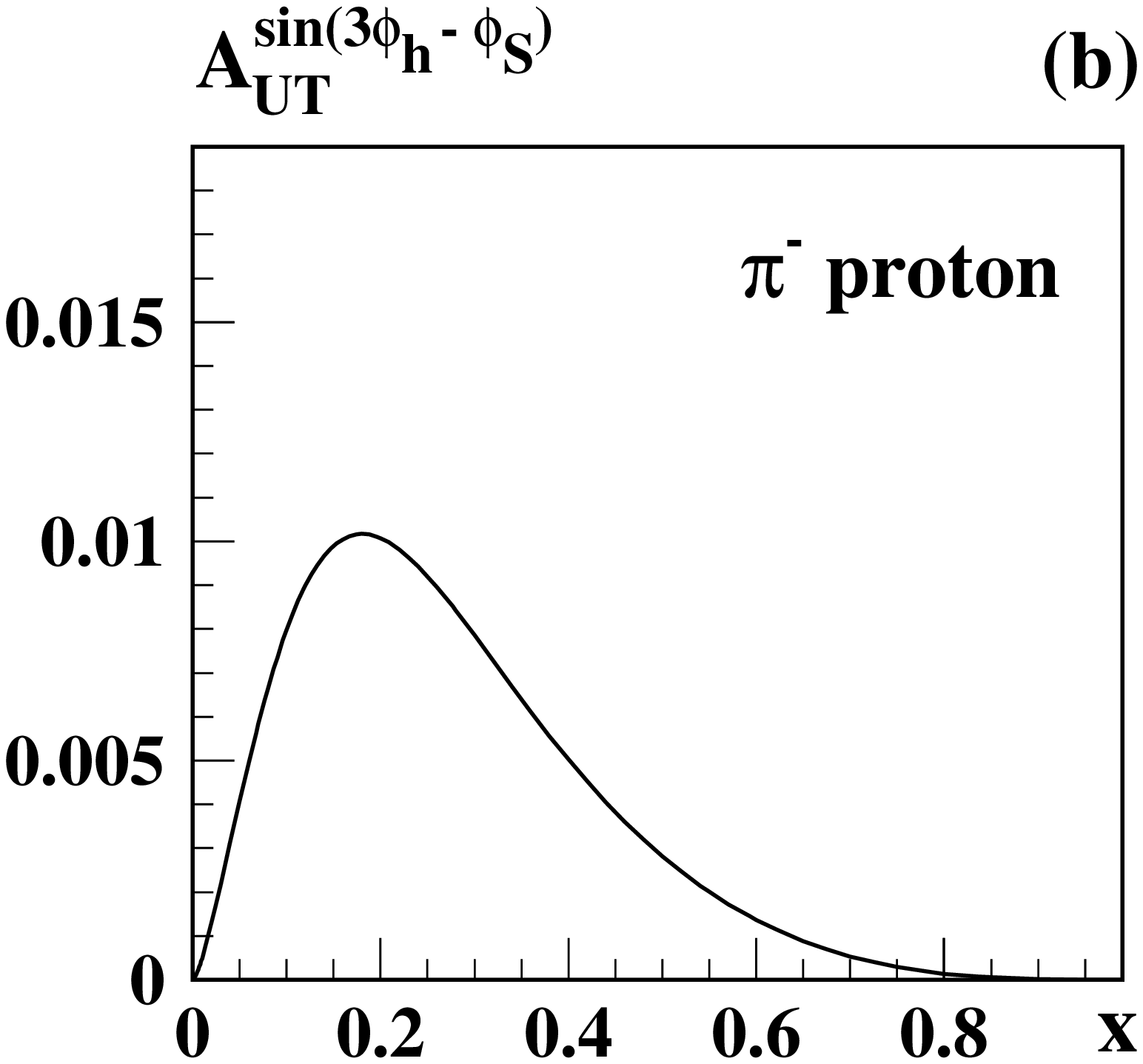}
 \hspace{-13mm}
 \includegraphics[height=4.2cm]{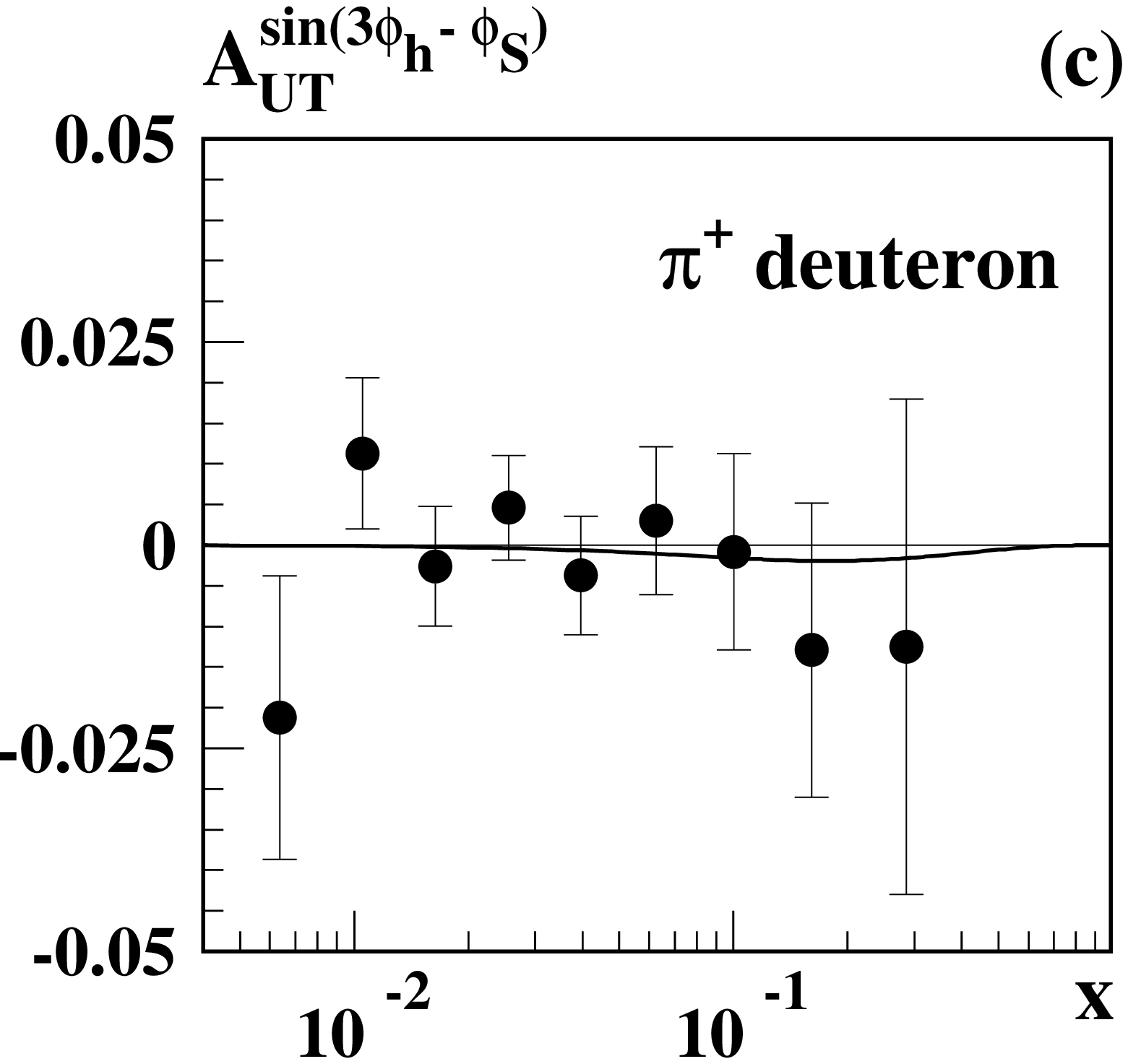}
 \hspace{-11mm}
 \includegraphics[height=4.2cm]{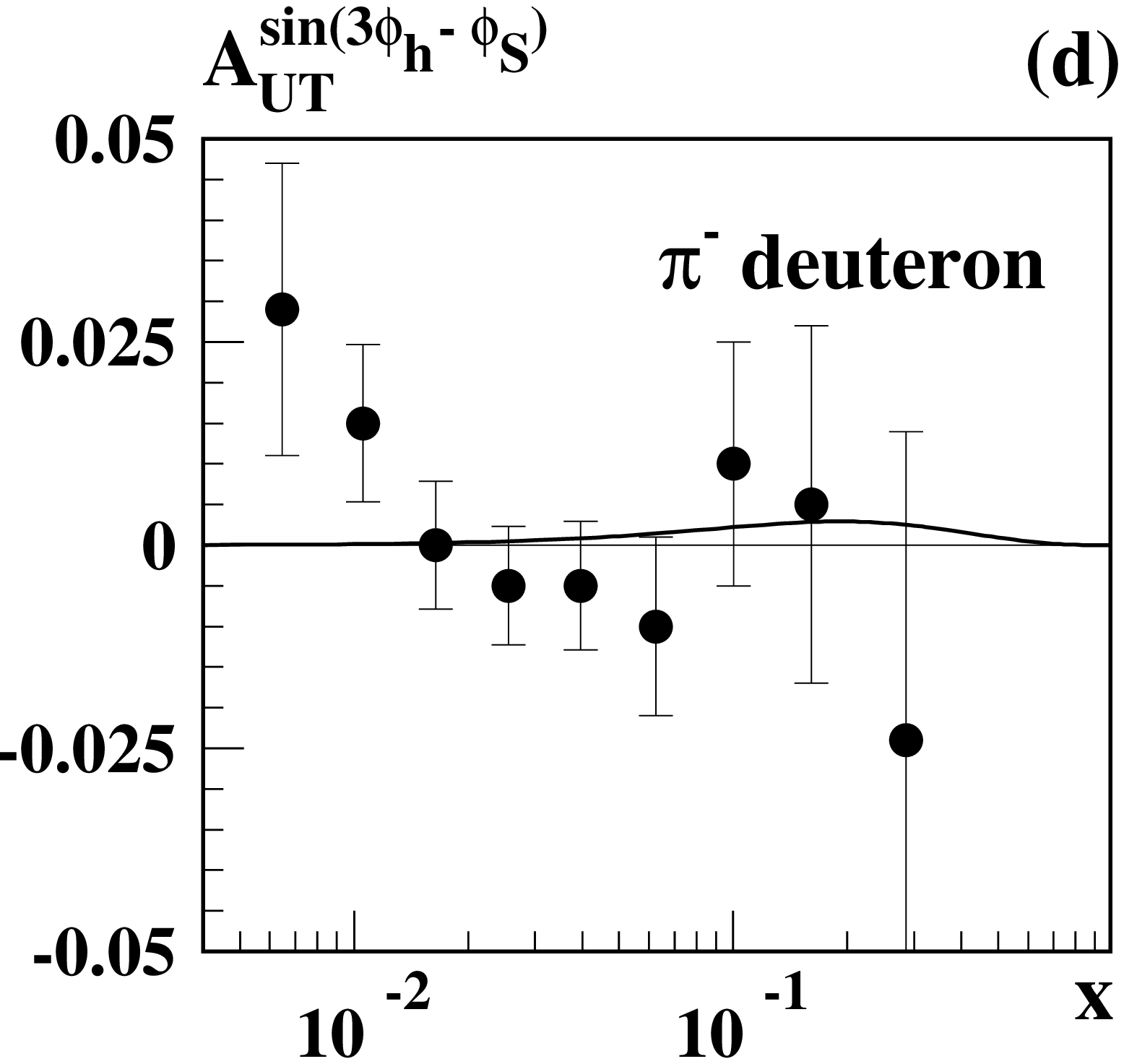}
 \hspace{-12mm}
{\vspace{-0.4 truecm}
    	\caption{\label{autbis}
	The single-spin asymmetry $A_{UT}^{\sin(3\phi_h-\phi_S)}$ in DIS
	production of charged pions off proton and deuteron targets, 
	as function of $x$.
	The theoretical curves are obtained by evolving
	the light-cone CQM predictions for $h_{1T}^{\perp(1)a}$ of
	Ref.~\cite{Pasquini:2008ax} to $Q^2=2.5$ GeV$^2$, using
	the $h_1^a$ evolution pattern, see text.
        The preliminary COMPASS data are from Ref.~\cite{Kotzinian:2007uv}.
}
}
\end{figure}

The results obtained in this way are shown in Fig.~\ref{autbis}.
We find the pretzelosity SSA rather small, about one percent in the
case of charged pions from a proton target, see Figs.~\ref{autbis}a, b. 
This makes it the probably most challenging asymmetry to be measured.
The deuteron SSAs are somewhat smaller, see Figs.~\ref{autbis}c, d
where we show for comparison the preliminary COMPASS data 
presented in Ref.~\cite{Kotzinian:2007uv}.
Our results are compatible with the data, and explain why
the effect was found consistent with zero within error bars at COMPASS.
The error bars of the preliminary data \cite{Kotzinian:2007uv} simply
do not allow to resolve an asymmetry smaller than one percent.

In Ref.~\cite{Avakian:2008dz} estimates for the asymmetry 
$A_{UT}^{\sin(3\phi_h-\phi_S)}$ were presented on the basis of the positivity 
bound $|h_{1T}^{\perp(1)a}(x)|\le \frac12\,(f_1^a(x)-g_1^a(x))$
\cite{Bacchetta:1999kz}, using the parametrizations \cite{Gluck:1998xa} 
for $f_1^a(x)$, $g_1^a(x)$. 
The results of the light-cone CQM for pretzelosity (as well as other TMDs),
of course, respect positivity bounds \cite{Pasquini:2008ax},
and the transverse moment of pretzelosity at the low scale 
of the model is not that small, see Fig.~\ref{Fig2:TMD}.
But after evolution (with the transversity evolution pattern) 
to a scale of $Q^2=2.5\,{\rm GeV}^2$, it is much smaller than its
bound constructed from parameterizations for $f_1^a(x)$, $g_1^a(x)$
at $Q^2=2.5\,{\rm GeV}^2$. Therefore, our estimates of the pretzelosity
SSA are significantly smaller than the maximal effect allowed by
positivity requirements \cite{Avakian:2008dz}.

Of course, we do not know to which extent our approach to estimate
the $h_{1T}^{\perp(1)}(x)$ evolution effects is really realistic.
For comparison, in Appendix and Ref.~\cite{ProcSPIN08} we 
also make predictions neglecting the evolution of pretzelosity.
In this way the results for the SSA are more sizable, and
the effects are larger especially in the region of 
intermediate and large $x$. The planned experiment at 
JLab  will allow us to discriminate 
among the different predictions~\cite{Avakian-LOI-CLAS12}.

%------ BEGIN FIGURE 8: A_UT^sin(3phi-phiS) ---------------------------
%
%---- END FIGURE 8 ----------------------------------------------------

%=================== SECTION 10: PT-DEPENDENCE =======================
\section{\boldmath $P_{h\perp}$-dependence of spin asymmetries}
\label{Sec-10:Ph-dep}

As discussed in Sec.~\ref{Sec-5:pT-and-Gauss}, care is required in order to
use the model results for the $p_T$-dependence of TMDs for phenomenological
applications. In this Section we shall exemplify how this can be done with a 
study of the $P_{h\perp}$-dependence of the double spin asymmetry $A_{LL}$.
In principle, we could discuss also other asymmetries, 
but $A_{LL}$ has the advantage that its $x$ and $z$ dependence is rather well 
known 
--- so we do not need the model input for that, and can focus on 
$P_{h\perp}$-dependence which is the only new concept in this case.
Would we discuss other (azimuthal) spin asymmetries, we would need to use
the model input also for the $x$-dependence of the novel TMD, and face the 
problems of how to evolve TMDs, make a meaningful estimate of Sudakov effects,
and deal with the small-$x$ region (see footnote~\ref{Footnote-1}). 
When dealing with $A_{LL}$ in the way described below, we avoid these problems.

Before discussing the $P_{h\perp}$-dependence of $A_{LL}$ in our approach,
let us remark that ideally a study of  $p_T$-effects should 
start with
absolute cross section data on the production of hadrons in unpolarized DIS.
Such data are difficult to produce, and 
experimentally it is preferable to study 
the $P_{h\perp}$-dependence of asymmetries, since detector acceptance 
effects in the numerator and denominator of the asymmetries (largely) 
cancel. Therefore, so far information from SIDIS on $p_T$-dependence of 
the unpolarized parton distribution and fragmentation functions, $f_1^a$ and 
$D_1^a$, has been obtained only indirectly, see Sec.~\ref{Sec-5:pT-and-Gauss}. 
It would be desirable to improve this situation.
Apart from the absolute cross section proportional to $F_{UU}$, the next 
'simplest' observable to learn about $p_T$-effects is probably 
$A_{LL}=F_{LL}/F_{UU}$.

In Sec.~\ref{Sec-5:pT-and-Gauss} we learned that the Gauss Ansatz is supported
within the model with reasonable accuracy. This justifies to make explicit 
use of it, also in this case. If we assume this Ansatz, then
\ba\label{Eq.XXX-10}
       F_{UU}(x,z,P_{h\perp}) 
       &=& \sum_a e_a^2\,xf_1^a(x)\,D_1^a(z)\;
       \frac{\exp(-P_{h\perp}^2/\la P_{h\perp}^{2,\rm unp}\ra)
       }{\pi\;\la P_{h\perp}^{2,\rm unp}\ra}
       \;,\;\;\;
       \la P_{h\perp}^{2,\rm unp}\ra = \la K_T^2\ra + z^2 \la p_T^2(f_1)\ra\;,\\
\label{Eq.XXX-11}
       F_{LL}(x,z,P_{h\perp}) 
       &=& \sum_a e_a^2\,xg_1^a(x)\,D_1^a(z)\;
       \frac{\exp(-P_{h\perp}^2/\la P_{h\perp}^{2,\rm pol}\ra)
       }{\pi\;\la P_{h\perp}^{2,\rm pol}\ra}        
       \;,\;\;\;
       \la P_{h\perp}^{2,\rm pol}\ra = \la K_T^2\ra + z^2 \la p_T^2(g_1)\ra\;.
\ea
If we assume that the widths are flavour and $x$- or $z$-independent,
then the $P_{h\perp}$-dependence of the double spin asymmetry is given by
\be\label{Eq.XXX-12}
      A_{LL}(P_{h\perp})  = \la A_{LL}\ra \;  
      \frac{\la P_{h\perp}^{2,\rm unp}\ra}{\la P_{h\perp}^{2,\rm pol}\ra}\;
      \exp\biggl[\frac{P_{h\perp}^2}{\la P_{h\perp}^{2,\rm unp}\ra}
      -\frac{P_{h\perp}^2}{\la P_{h\perp}^{2,\rm pol}\ra}\biggr]\;,
\ee
where $\la A_{LL}\ra$ denotes the spin asymmetry averaged over $x$ and $z$,
which is known in the experiment with precision.
(Now in Eq.~(\ref{Eq.XXX-12}) it is implied that ``$z^2$'' in 
(\ref{Eq.XXX-10}), (\ref{Eq.XXX-11}) is replaced by $\la z^2\ra$.
This is an approximation, and the treatment could be improved,
but we refrain from this in our illustrative study for sake of clarity.)

%------ BEGIN FIGURE 10: A_LL(PT)  ------------------------------------
	\begin{wrapfigure}[22]{RD}{10cm}
	\vspace{-0.7cm}
    \centering
\includegraphics[width=7.5cm]{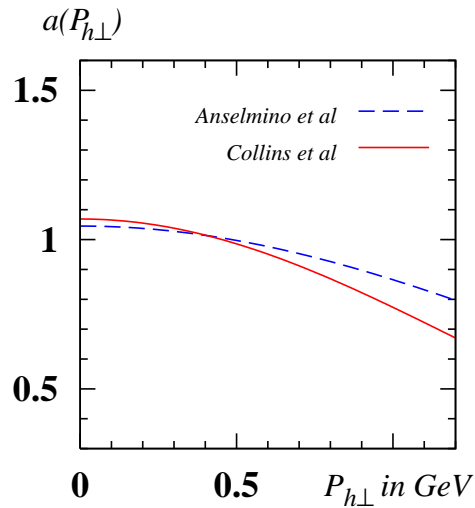}
        \vspace{-0.5cm}
        \caption{\label{Fig06:predictions}
        $a(P_{h\perp})\equiv A_{LL}(P_{h\perp})/\,\la A_{LL}\ra$
        vs.~$P_{h\perp}$ in SIDIS, for experiments with $\la z^2\ra=0.16$. 
        The results are obtained using the prediction
        (\ref{Eq:ZZZ}) from the model \cite{Pasquini:2008ax},
        and the Gauss model parameters in Eq.~(\ref{Eq:fit-pT2-KT2})
        from Ref.~\cite{Collins:2005ie} (solid curve) or the corresponding 
        parameters from Ref.~\cite{Anselmino:2005nn} (dashed curve).}
\end{wrapfigure}
%------ END FIGURE 10 -------------------------------------------------

We remark that positivity, i.e.\ $A_{LL}\le 1$ $\forall\;x$
and $P_{h\perp}$, dictates
\be\label{Eq:YYY}
       \frac{\la p_T^2(g_1)\ra}{\la p_T^2(f_1)\ra} \le
       \biggl|\frac{g_1^a(x)}{f_1^a(x)}\biggr|\;.
\ee
This implies that (in the Gauss Ansatz) the widths of helicity and 
the unpolarized distribution could be equal, if and only if the
equality $f_1^a(x)=|g_1^a(x)|$ were true.

Now, let us discuss how to use the model results in order to predict 
the $P_{h\perp}$-dependence of $A_{LL}$. From Table~\ref{Table:pT-model}
we know that
\be\label{Eq:ZZZ}
         \frac{\la p_T^2(g_1)\ra}{\la p_T^2(f_1)\ra} = 0.74\;.
\ee
If we take this ratio for granted, and assume for $\la p_T^2(f_1)\ra$
the result from \cite{Collins:2005ie}, Eq.~(\ref{Eq:fit-pT2-KT2}),
then we obtain for $a(P_{h\perp})\equiv A_{LL}(P_{h\perp})/\,\la A_{LL}\ra$ 
the results shown in Fig.~\ref{Fig06:predictions}.
We include in Fig.~\ref{Fig06:predictions} also predictions based on using the
results for $\la K_T^2\ra$, $\la p_T^2(f_1)\ra$ from 
\cite{Anselmino:2005nn}.
We observe a rather stable prediction which depends little on the choice of 
parameters \cite{Collins:2005ie} vs.\ \cite{Anselmino:2005nn}.
The prediction in Fig.~\ref{Fig06:predictions} depends more strongly 
on the model prediction (\ref{Eq:ZZZ}).

This result is (in our approximations) the same for any target
and produced hadron. 
In fact, 
in the SU(6) symmetric light-cone CQM of Ref.~\cite{Pasquini:2008ax} 
the widths as defined in Eq.~(\ref{Eq:define-mean-pT}) are always flavour
independent. But we recall, that the entire Gauss Ansatz is in the light
of the results of Ref.~\cite{Pasquini:2008ax} merely an approximation.

It will be instructive to learn to which extent
our predictions will be confirmed by experiment.
As mentioned, we could similarly discuss predictions from the model 
for azimuthal asymmetries, too. But those predictions would presumably 
have larger theoretical uncertainties, such that we shall content ourselves
here with the study of the $P_{h\perp}$-dependence of $A_{LL}$.

%=================== SECTION 11: CONCLUSIONS =========================

\section{Conclusions}
\label{sect:conclusions}

In this work we have studied all leading-twist azimuthal 
spin asymmetries in SIDIS due to T-even TMDs on the basis of predictions within one and the same model, i.e. the light-cone CQM of Ref.~\cite{Pasquini:2008ax}. 

By studying first the well known double spin asymmetries $A_1$ in DIS and $A_{LL}$ in SIDIS, we demonstrated that the approach is capable of describing the data on these asymmetries in the valence-$x$ region with an accuracy of ${\cal O}(20-30)\,\%$. The comparison with results from other constituent models has shown this to be a typical accuracy to which the constituent quark model scenario can be expected to work.

We paid particular attention to the question, how to apply the model results for TMDs obtained at a very low hadronic scale to the description of data referring to high scales of typically several ${\rm GeV}^2$. We made a test for the double spin asymmetries $A_1$ in DIS and $A_{LL}$ in SIDIS where the evolution equations involving the parton density $f_1^a(x)$ and the helicity distribution $g_1^a(x)$ are exactly known. In these cases we have been able to demonstrate the stability under evolution of our results in the valence-$x$ region.

For TMDs entering the description of azimuthal asymmetries, however, not only the evolution with renormalization scale has to be taken into account, but also Sudakov effects which broaden the $p_T$-distribution of the TMDs. We tackled this issue in two steps. First, we observed that  the light-cone CQM~\cite{Pasquini:2008ax} do not show a Gaussian $p_T$-dependence. Nevertheless, their $p_T$-dependence entering the azimuthal asymmetries is integrated over in certain convolution integrals, so that we have found that within the accuracy of our approach the effect of the true $p_T$-dependence in the TMDs can be approximated by a Gaussian dependence. We have therefore explicitly employed the Gaussian Ansatz, which allows to express azimuthal asymmetries  in terms of parton distribution functions or transverse moments of TMDs. In the second step, we used evolution equations to evolve the respective parton distribution functions or transverse moments of TMDs to the experimental scales. 

We have been able to do this {\sl exactly}, strictly speaking, only in the case of transversity $h_1^a(x)$.
In the other cases, the evolution of the transverse moments of TMDs was estimated by employing those evolution equations, which seem most promising to be able to simulate the correct evolution, which is
presently not available. For example, we evolved $g_{1T}^{\perp(1)a}(x)$ by means of the evolution pattern of the (also chiral-even) $g_1^a(x)$, while for $h_{1L}^{\perp(1)a}(x)$, $h_{1T}^{\perp(1)a}(x)$ we used the evolution pattern of the chiral-odd $h_1^a(x)$. The theoretical uncertainties due to these approximate treatment of the scale dependence is presumably not larger than the accuracy of the model.

Among the leading-twist azimuthal spin asymmetries due to T-even TMDs, the Collins SSA $A_{UT}^{\sin(\phi_h+\phi_S)}$ is the only non-zero one within the present day error bars. We observe a very good agreement of our results for the $x$-dependence of this SSA with the HERMES proton 
\cite{Airapetian:2004tw,Diefenthaler:2005gx}, as well as with the COMPASS deuteron target data
\cite{Alexakhin:2005iw,Ageev:2006da}.

The presently available final data on $A_{UL}^{\sin(2\phi_h)}$ \cite{Airapetian:1999tv,Airapetian:2002mf} or preliminary data on $A_{LT}^{\cos(\phi_h-\phi_S)}$ and $A_{UT}^{\sin(3\phi_h-\phi_S)}$ \cite{Kotzinian:2007uv} show results compatible with zero within error bars. Our results are compatible with these first or preliminary data. In future, our predictions of these azimuthal spin asymmetries
could be tested by more precise data --- especially from COMPASS and JLab.

In an exploratory study of the double spin asymmetry $A_{LL}$ we have shown how model results for TMDs obtained at very low scale could be applied for studies of the $P_{h\perp}$-dependence of spin asymmetries. We have chosen this observable, because here the $P_{h\perp}$-dependence is the only new aspect, the $x$- and $z$-dependence being known experimentally with good precision. For that we explored again the fact that the light-cone CQM \cite{Pasquini:2008ax} supports the Gaussian Ansatz for TMDs within a reasonable accuracy, and used as model input only the prediction for the ratio of the mean transverse-momentum squares of $g_1^a$ and $f_1^a$. This ratio is expected to be little affected by Sudakov effects in a first approximation. We made predictions for $A_{LL}(P_{h\perp})$ which could be tested soon, for example, at JLab \cite{Avakian-LOI-CLAS12}.

The advantage of our study is that the same model input has been used to describe all leading-twist spin asymmetries due to T-even TMDs. Wherever the data allow to draw definite conclusions, we observed a good agreement with the experiment in the range of applicability of the approach. It remains to be seen whether also our predictions for the other azimuthal spin asymmetries will be similarly confirmed 
by future data. If so, our approach will provide interesting insights in the spin and orbital angular momentum structure of the nucleon, which --- though being model dependent --- are of interest by themselves, as it is exposed in the Appendix.
                                                               
%===================  ACKNOWLEDGMENTS ================================
 \vspace{0.5cm}

 \noindent{\bf Acknowledgements.}
 We are grateful to Klaus Goeke for discussions and to the 
 Institute for Theoretical Physics II at the University of Bochum
 where this work was initiated for hospitality .  
 The work is partially supported by BMBF (Verbundforschung),
 and is part of the European Integrated Infrastructure Initiative Hadron
 Physics project under contract number RII3-CT-2004-506078.
A.I.E. is also supported by the Grants RFBR 09-02-01149 and 07-02-91557, RF
MSE RNP 2.1.1/2512 (MIREA) and by the Heisenberg-Landau Program of JINR.

\ \\

%===================  APPENDIX =======================================
\clearpage
\appendix
\section{Angular momentum decomposition of spin asymmetries}
\label{appendix}

In this Appendix we discuss the contribution from the different angular 
momentum components of the nucleon wave function to the spin asymmetries.
To this aim, we calculate the numerator of the 
asymmetries using the results of
the TMDs at the hadronic scale of the model, and separating them
into partial wave contributions according 
to the decomposition shown in Fig.~\ref{Fig2:TMD} of Sect.~\ref{Sec-3:model}.
In order to discuss how this decomposition behaves under evolution to 
higher scale, one would need to know the evolution equations for the 
different angular momentum components of the nucleon wave function separately, 
but, to our knowledge, this problem has never been  addressed so far
and  is beyond the scope of our work. 
Although this decomposition is model dependent and it is not possible
to  extract experimentally the absolute strength of the different 
partial waves, it is instructive to visualize how the
angular momentum content of the TMDs affects the spin asymmetries.
In particular,  the combined analysis
of different spin asymmetries  can give insights 
 about the relative strength of the  different partial waves,
and therefore can be useful in modeling 
the light-cone wave function of the nucleon.

The results presented in this Appendix, corresponds to an
alternative approach concerning the question how to use the model results
referring to a low hadronic scale for phenomenology at experimentally 
relevant scales. Namely, here we use the model at the low scale
only as input for the part which is responsible for the spin effects.
For the well known denominator of the spin asymmetries we use standard 
parametrizations at the experimental scale.
In this way, the model uncertainty is only in the numerator. 
The comparison of these results and those presented in 
Secs.~\ref{Sec-6:A_LT},~\ref{Sec-8:A_UL},~\ref{Sec-9:pretzelosity},
where the attempts were made to approximate evolution effects of the TMDs,
with forthcoming experimental data 
~\cite{Avakian-clas6,Avakian-clas12,Avakian-LOI-CLAS12}
may give us interesting information about the scale dependence of these 
observables.

\begin{figure}[h!]
 \includegraphics[width=9cm]{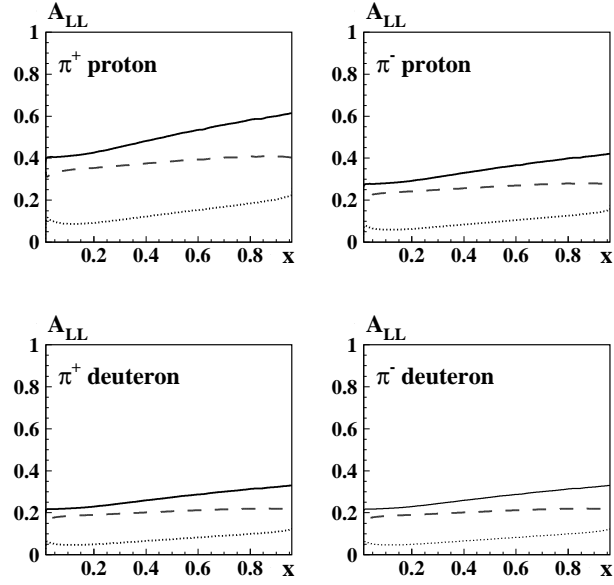}
	\caption{\label{Fig-9:ALL}
        The double-spin asymmetry $A_{LL}$ in DIS
	production of charged pions off proton and deuteron targets, 
	as function~of~$x$.
	The results are obtained using 
        $g_{1T}^{\perp(1)a}(x)$ and $f_1^a(x)$ from the light-cone CQM 
        \cite{Pasquini:2008ax} at the hadronic scale, and
         decomposing $g_{1T}^{\perp(1)a}$ into different partial wave 
	 contributions:
	 the dashed curves correspond to the contribution from S waves,
	 the dotted curves are the results for the P-wave contribution, and
	 the solid curves are the total results, sum of the S- and P-wave
	 contributions.}
\end{figure}

In Fig.~\ref{Fig-9:ALL}, we show the results for the $A_{LL}$ asymmetry, 
obtained by using both the unpolarized distribution function $f_1$ and the 
helicity distribution function
$g_{1}$  from the light-cone CQM at low scale.
The total results are further split into the contributions to $g_{1}$ 
from the S- (dashed curve) and P-wave (dotted curve)  components, 
while the D-wave contribution is not shown because it is negligible.
These separate terms reflect the dominance of the S-wave component 
with respect to the  P wave in $g_{1}$, 
as already observed  in Fig.~\ref{Fig2:TMD} of Sect.~\ref{Sec-3:model}.
Furthermore, the S-wave term is practically constant in the full $x$ range,
while the P-wave contribution is slowly increasing at larger value of $x$, 
reaching
a maximum of about 30$\%$ of the total result. 

\begin{figure}
    	\centering
        \includegraphics[width=12cm]{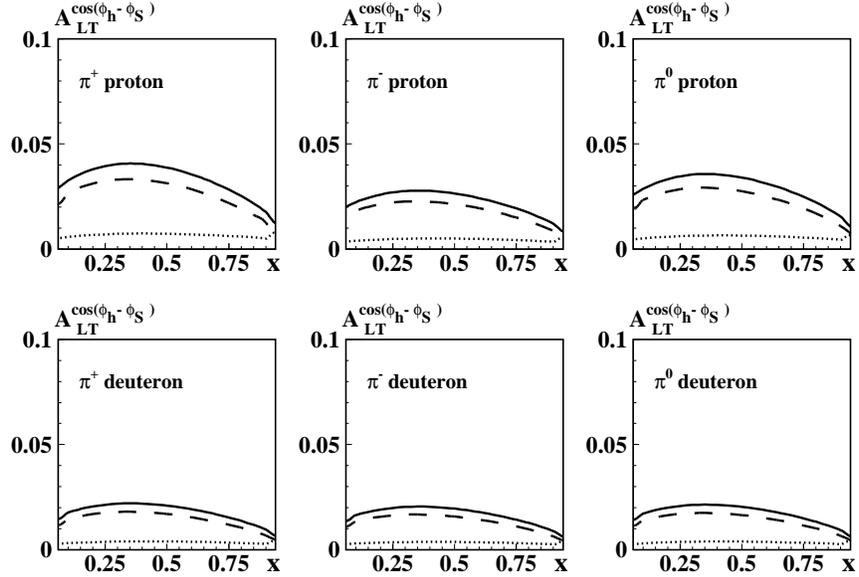}
        \caption{\label{Fig-10:ALT}
	The double-spin asymmetry $A_{LT}^{\cos(\phi_h-\phi_S)}$ in DIS
	production of pions off proton and deuteron targets, 
	as function~of~$x$.
	The results are obtained using 
        $g_{1T}^{\perp(1)a}(x)$ and $f_1^a(x)$ from the light-cone CQM 
        \cite{Pasquini:2008ax} at the hadronic scale, and
        decomposing $g_{1T}^{\perp(1)a}$ into different partial wave 
	contributions: the dashed curves correspond to the contribution from 
	the S- and P-wave interference,
	the dotted curves are the results for the P- and D-wave interference 
	term, and the solid curves are the total results.}
\end{figure}

The $A_{LT}^{\cos(\phi_h-\phi_S)}$ asymmetry shown in
Fig.~\ref{Fig-10:ALT} is calculated with both the unpolarized distribution 
function $f_1$ and the 
helicity distribution function
$g_{1T}^{(1)\perp}$ from the light-cone CQM at low scale.
Here we  separate the contribution to $g_{1T}^{(1)\perp}$ 
from the interference of S and P waves (dashed curves) and 
from the interference of P and D waves (dotted curves).
We see that the S- and P-wave interference term  governs
both the size and the shape in $x$ of the total results, while
the contribution of the P- and D-wave interference 
is rather small and constant in the full $x$ range.

\begin{figure}[t!]

\vspace{1cm}

 \hspace{-8mm}  \includegraphics[height=4.2cm]{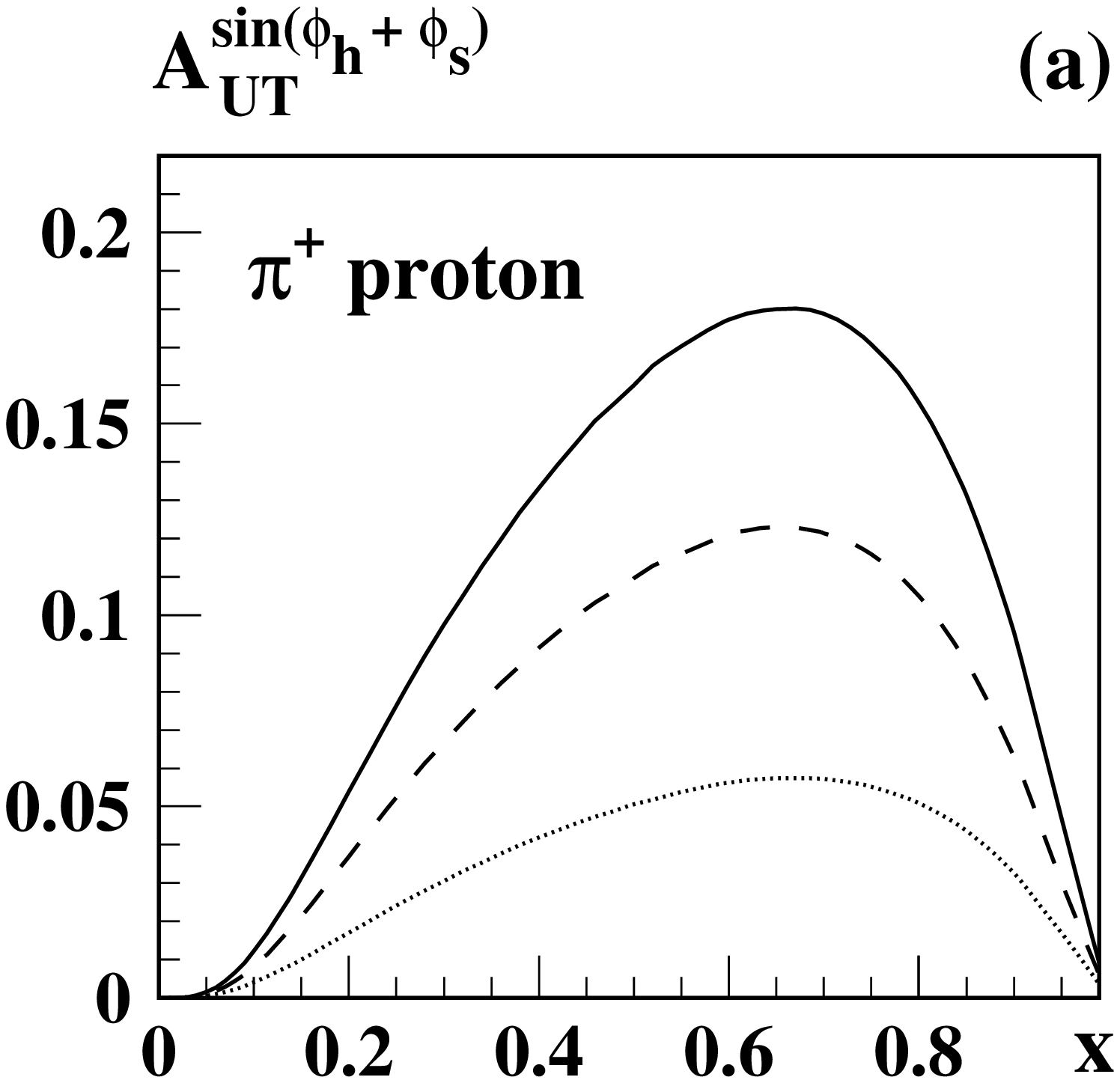}
 \hspace{-11mm} \includegraphics[height=4.2cm]{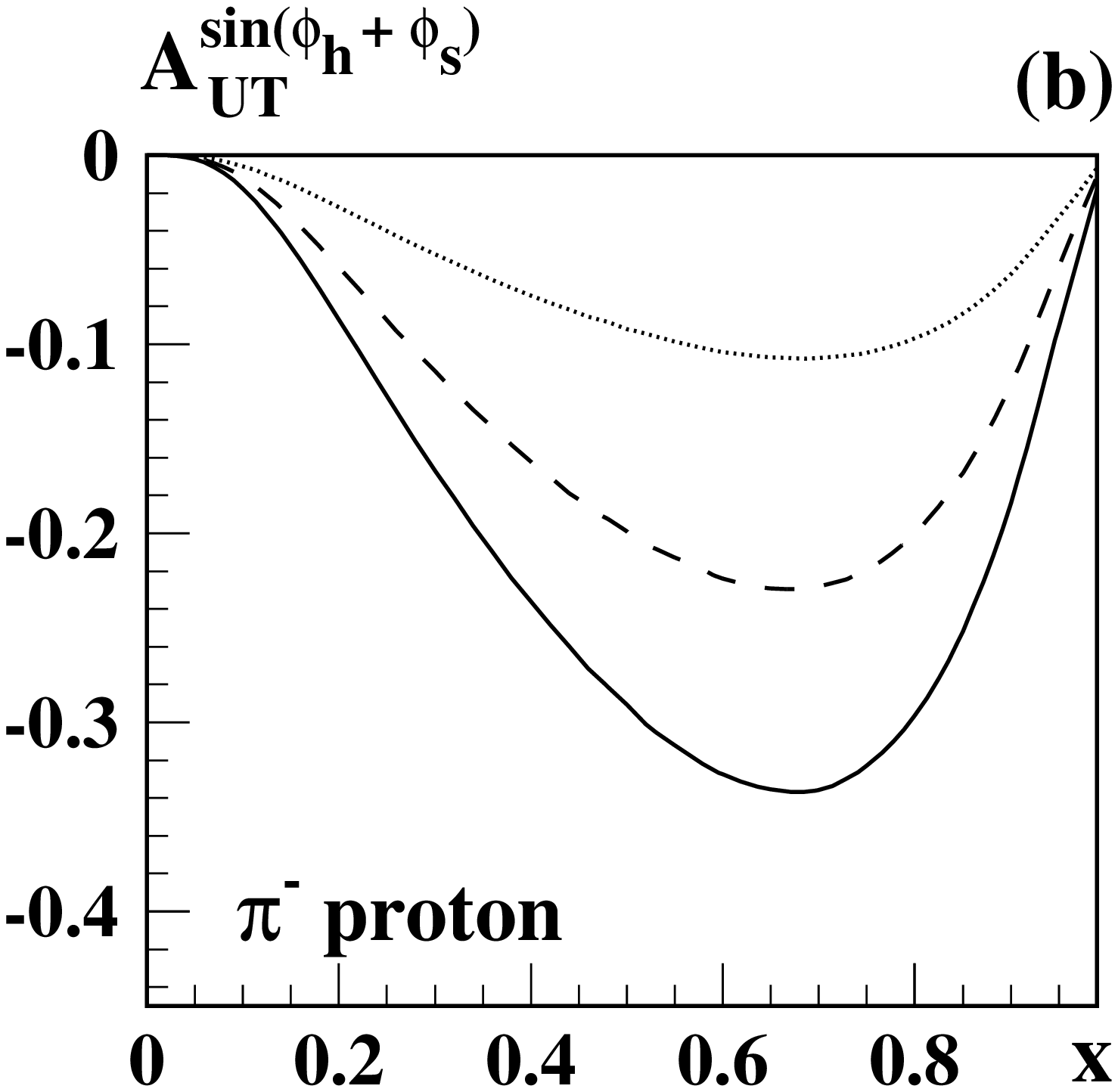}
 \hspace{-13mm} \includegraphics[height=4.2cm]{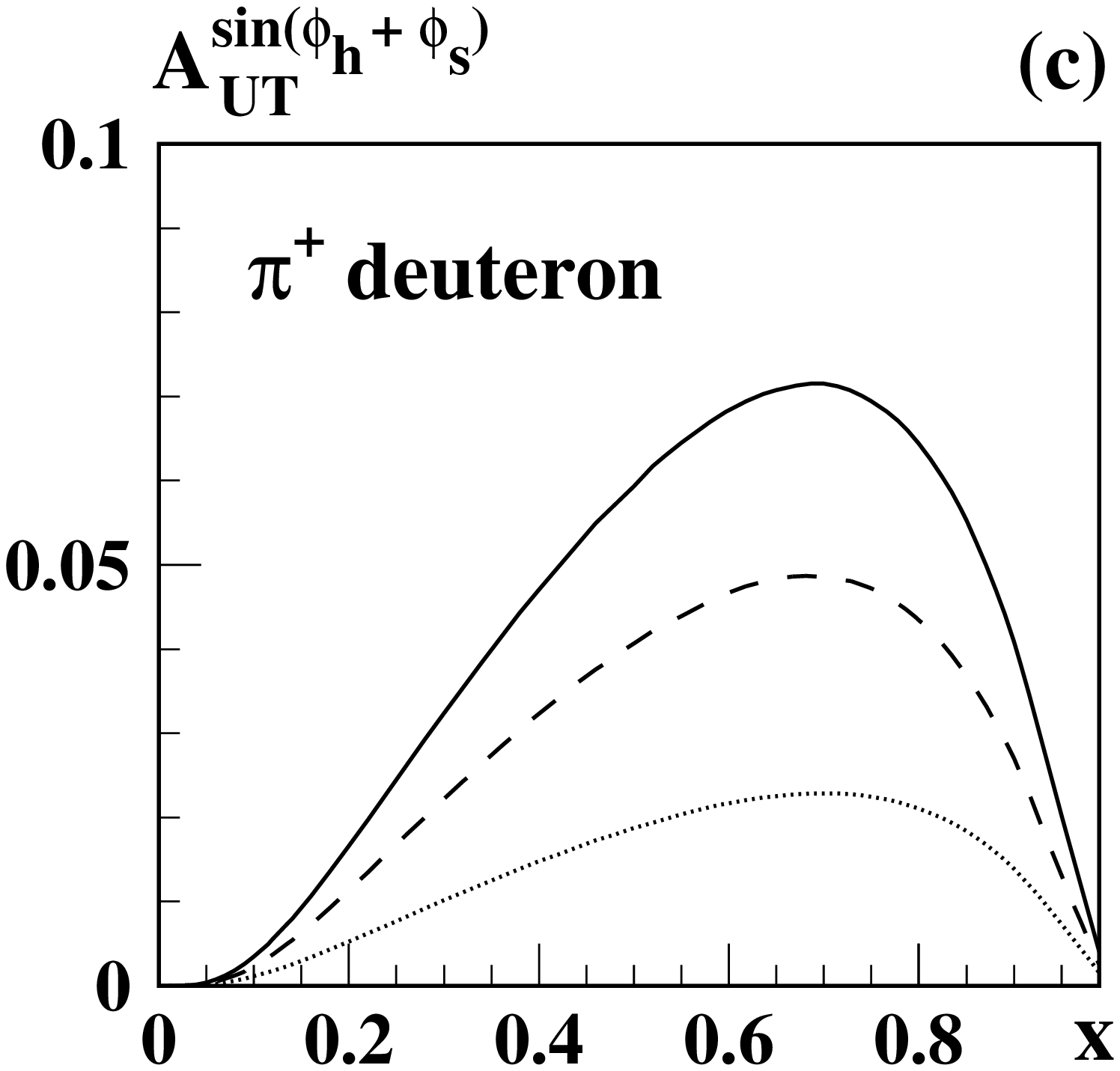}
 \hspace{-11mm} \includegraphics[height=4.2cm]{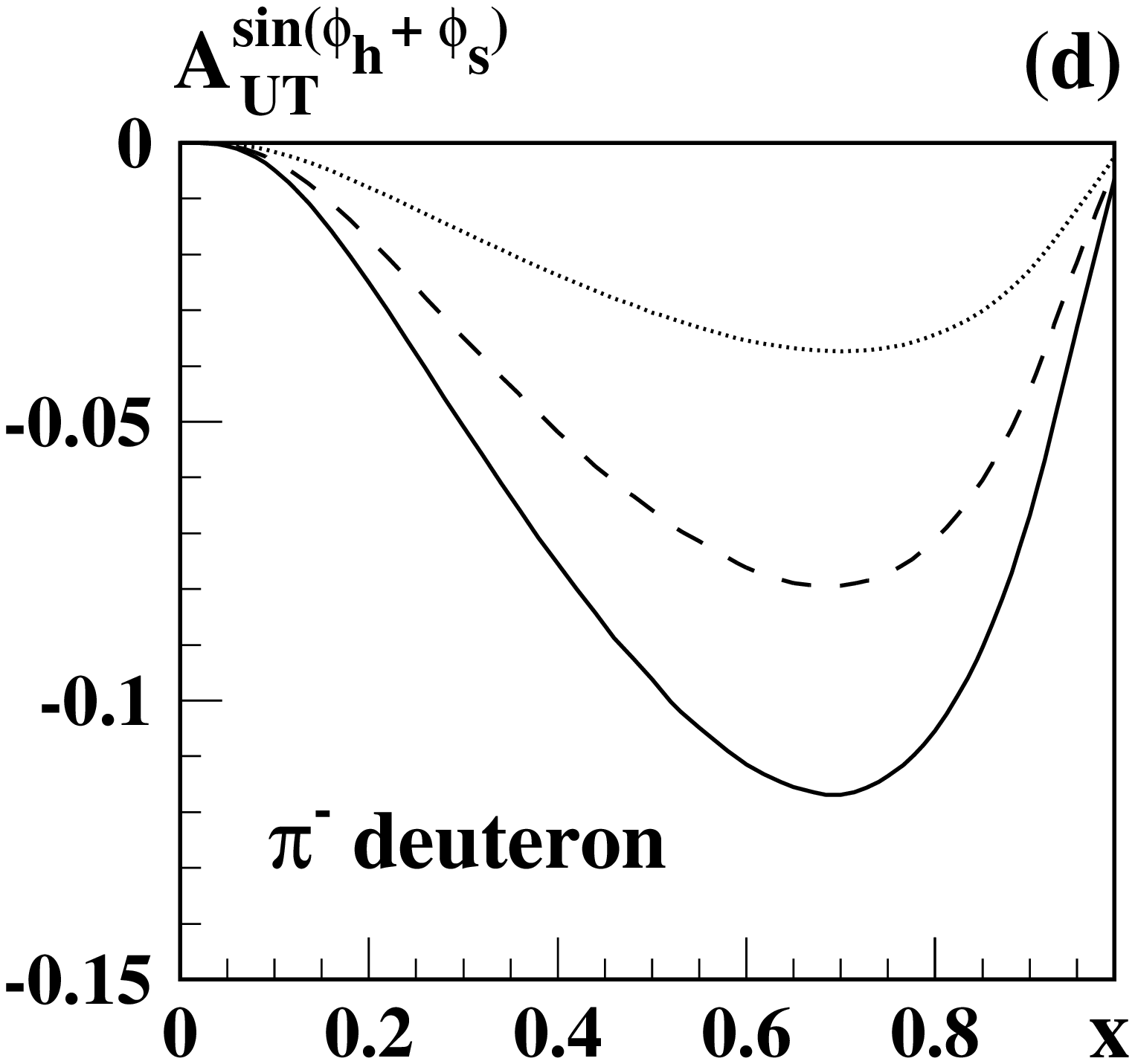}
 \hspace{-12mm}
{\vspace{-0.7 truecm}
\caption{
\label{Fig-10:AUT-Collins}
	The single-spin asymmetry $A_{UT}^{\sin(\phi_h+\phi_S)}$ in DIS
	production of charged pions 
	off proton and deuteron targets, as function 
        of $x$. The results are obtained using the parametrization 
	of Ref.~\cite{Gluck:1998xa} for $f_1^a(x)$ at $Q^2=2.5$ GeV$^2$, and
        the light-cone CQM predictions for $h_1^a(x)$ at the hadronic scale 
	from Ref.~\cite{Pasquini:2006iv,Pasquini:2008ax}.
	The dashed and dotted curves are obtained separating
	the contributions to  $h_1^a(x)$ from S and P waves, respectively.
	The solid curves show the total results, sum of the S- and P-wave
	contributions.}}

\vspace{15mm}
%------ BEGIN FIGURE 13: Angular momentum decomposition of AUL

 \hspace{-8mm}  \includegraphics[height=4.2cm]{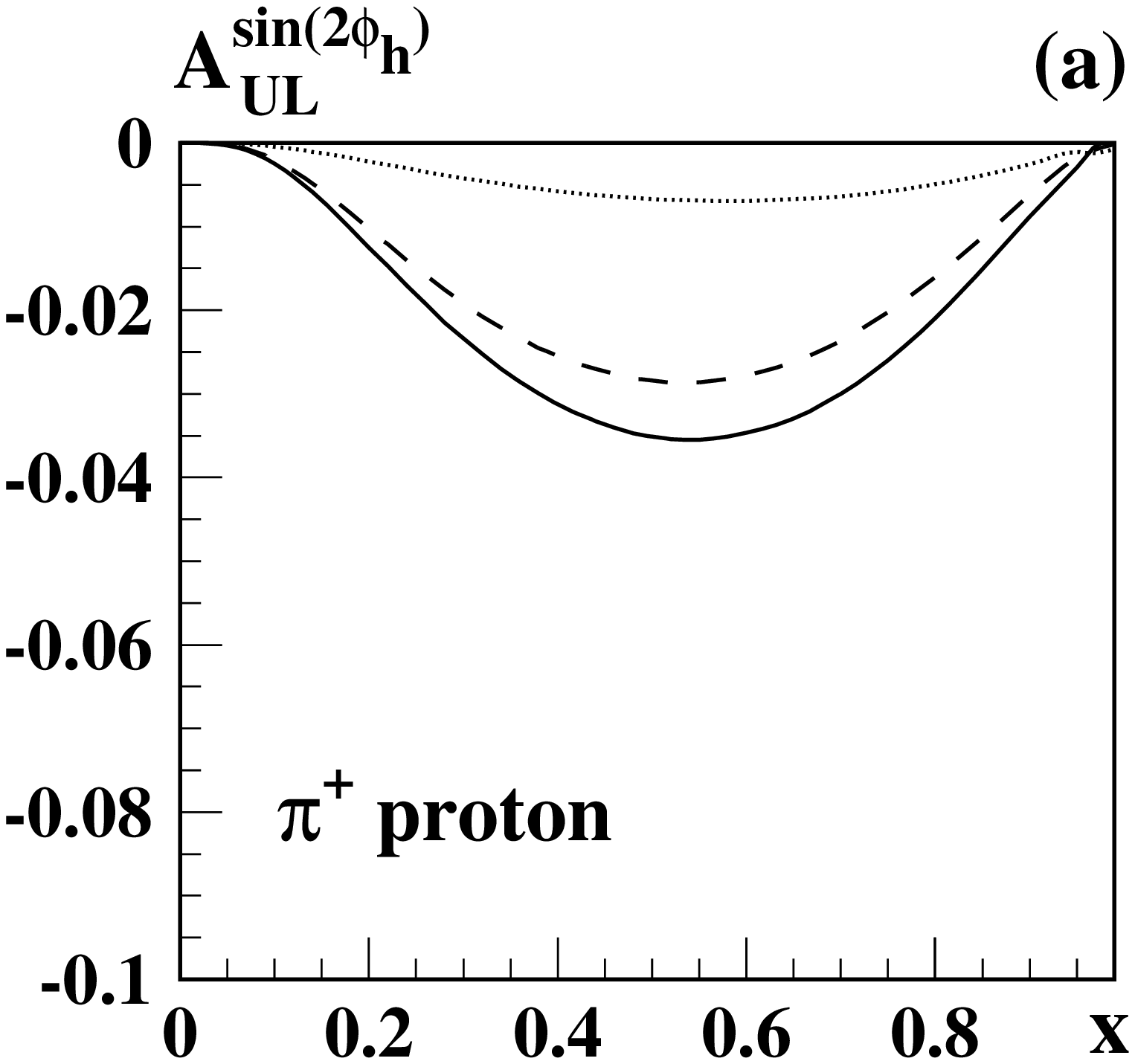}
 \hspace{-11mm} \includegraphics[height=4.2cm]{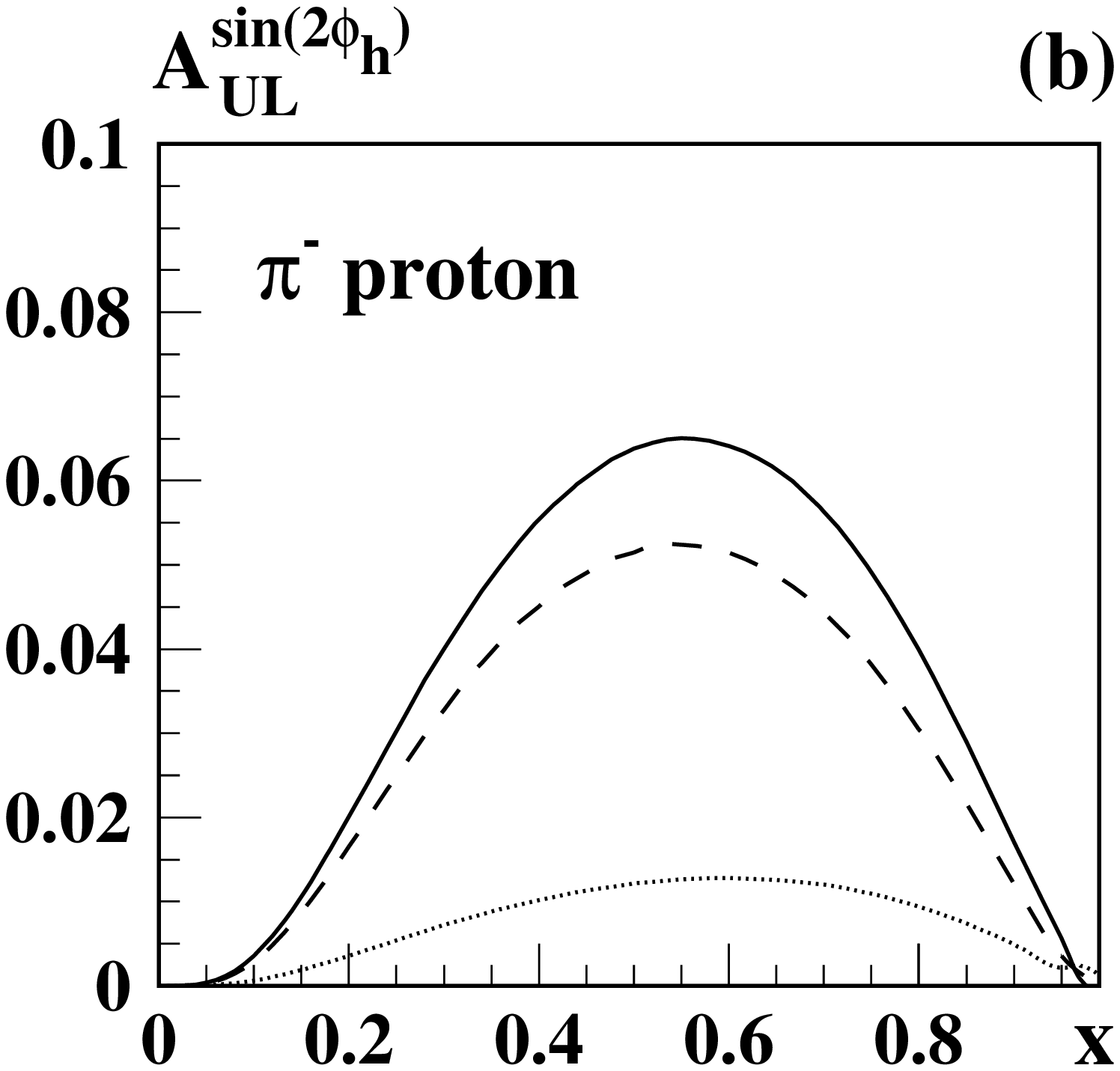}
 \hspace{-13mm} \includegraphics[height=4.2cm]{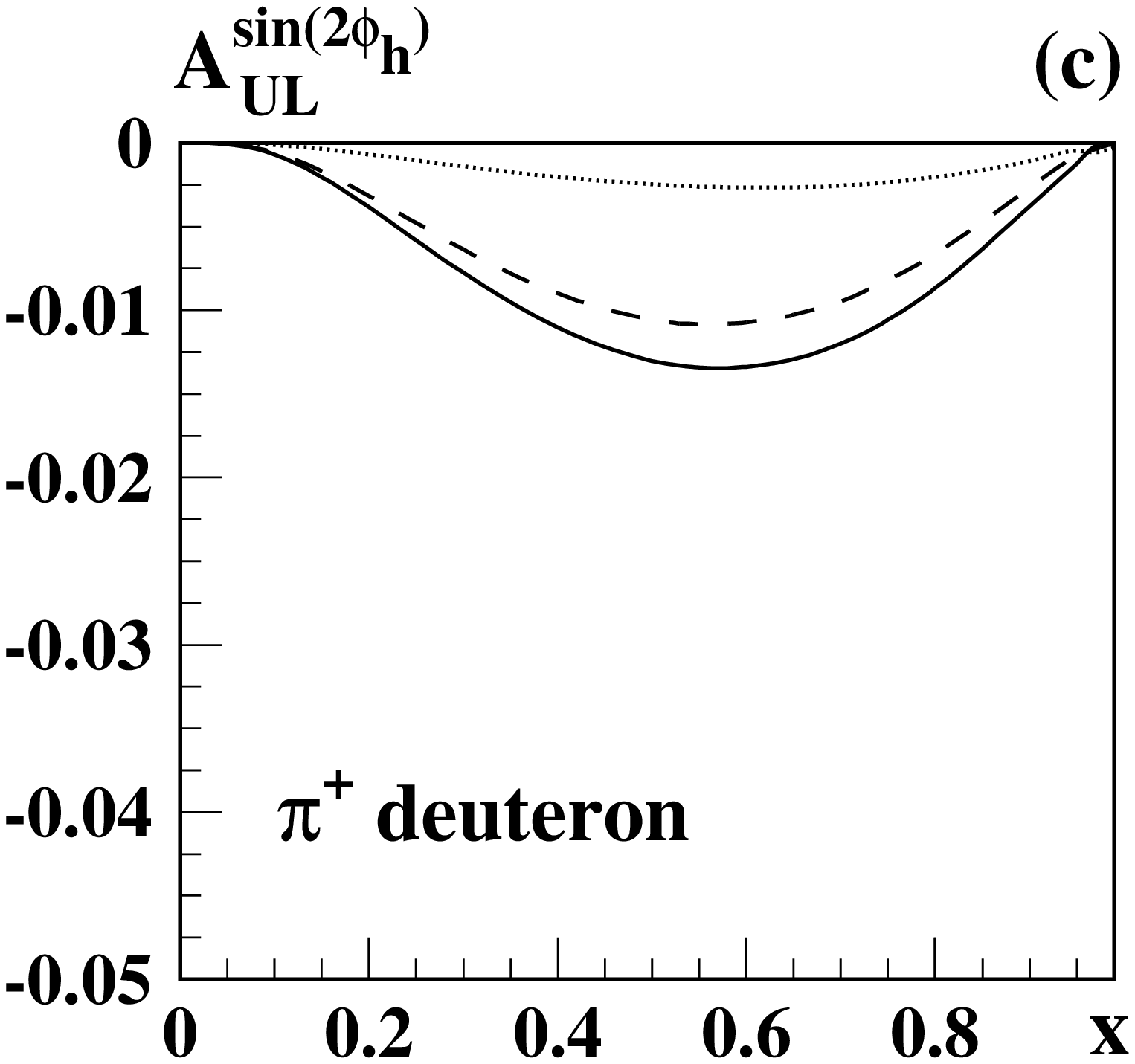}
 \hspace{-11mm} \includegraphics[height=4.2cm]{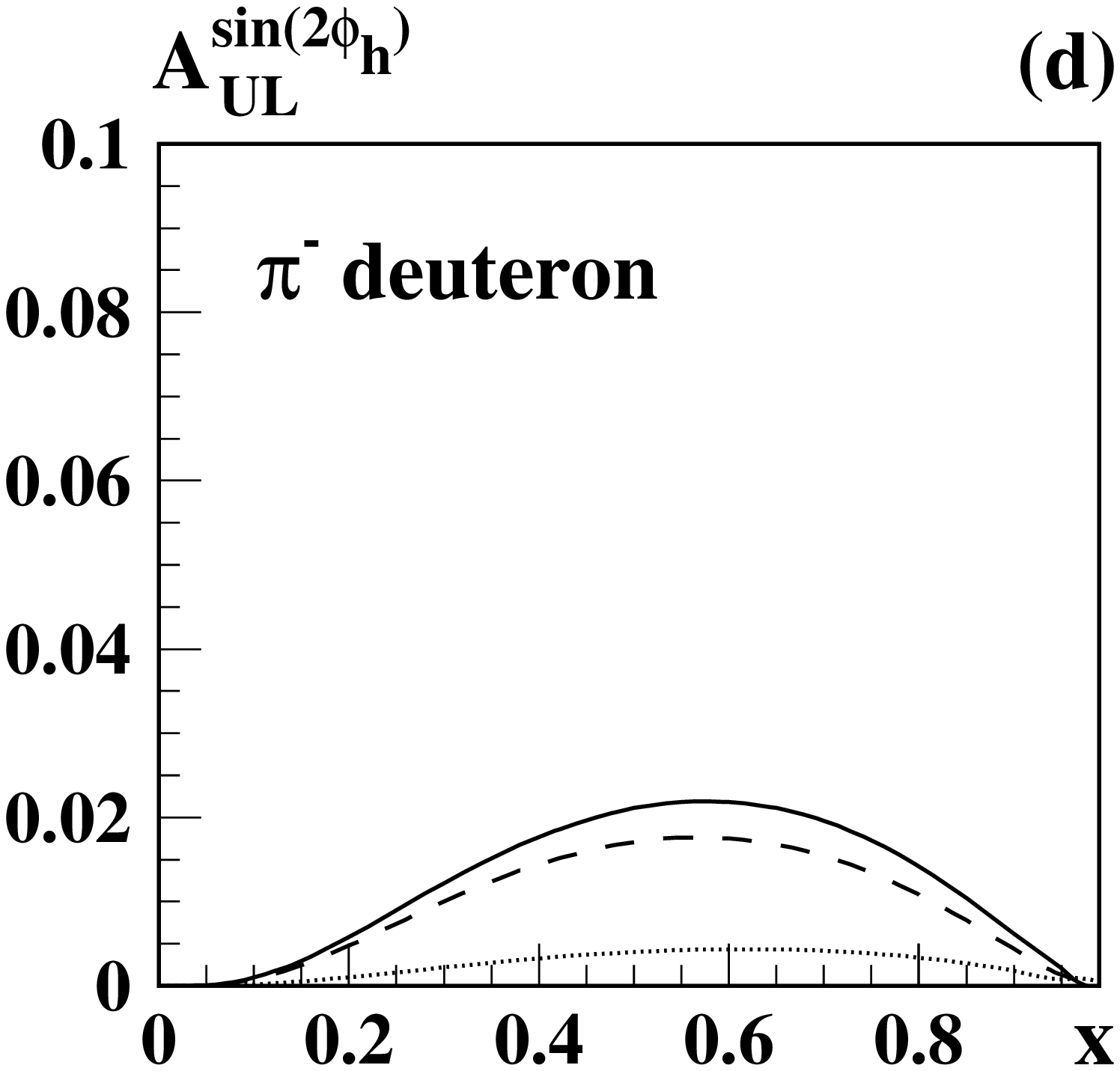}
 \hspace{-12mm}
{ \vspace{-0.7 truecm}
  \caption{\label{aul_ang_mom}
	The contribution from different angular momentum components to the 
	single-spin asymmetry $A_{UL}^{\sin(2\phi_h)}$ 
	in DIS production of charged pions off proton and deuteron targets, 
	as function of $x$. The results are obtained using the 
        light-cone CQM predictions for $h_{1L}^{(1)\perp a}(x)$ 
	at the hadronic scale and the parametrization 
	of Ref.~\cite{Gluck:1998xa} for $f_1^a(x)$ at $Q^2=2.5$ GeV$^2$.
	The dashed and dotted
	curves are obtained separating 
	the contributions to  $h_{1L}^{(1)\perp a}(x)$ 
	from  the interference of S-  and P-wave 
	components and from the interference of P- and D-wave
	components, respectively.
	The solid curves correspond to the total results.}}

\vspace{15mm}
%------ BEGIN FIGURE 14: Angular momentum decomposition of Pretzel asymmetry

 \hspace{-8mm}  \includegraphics[height=4.2cm]{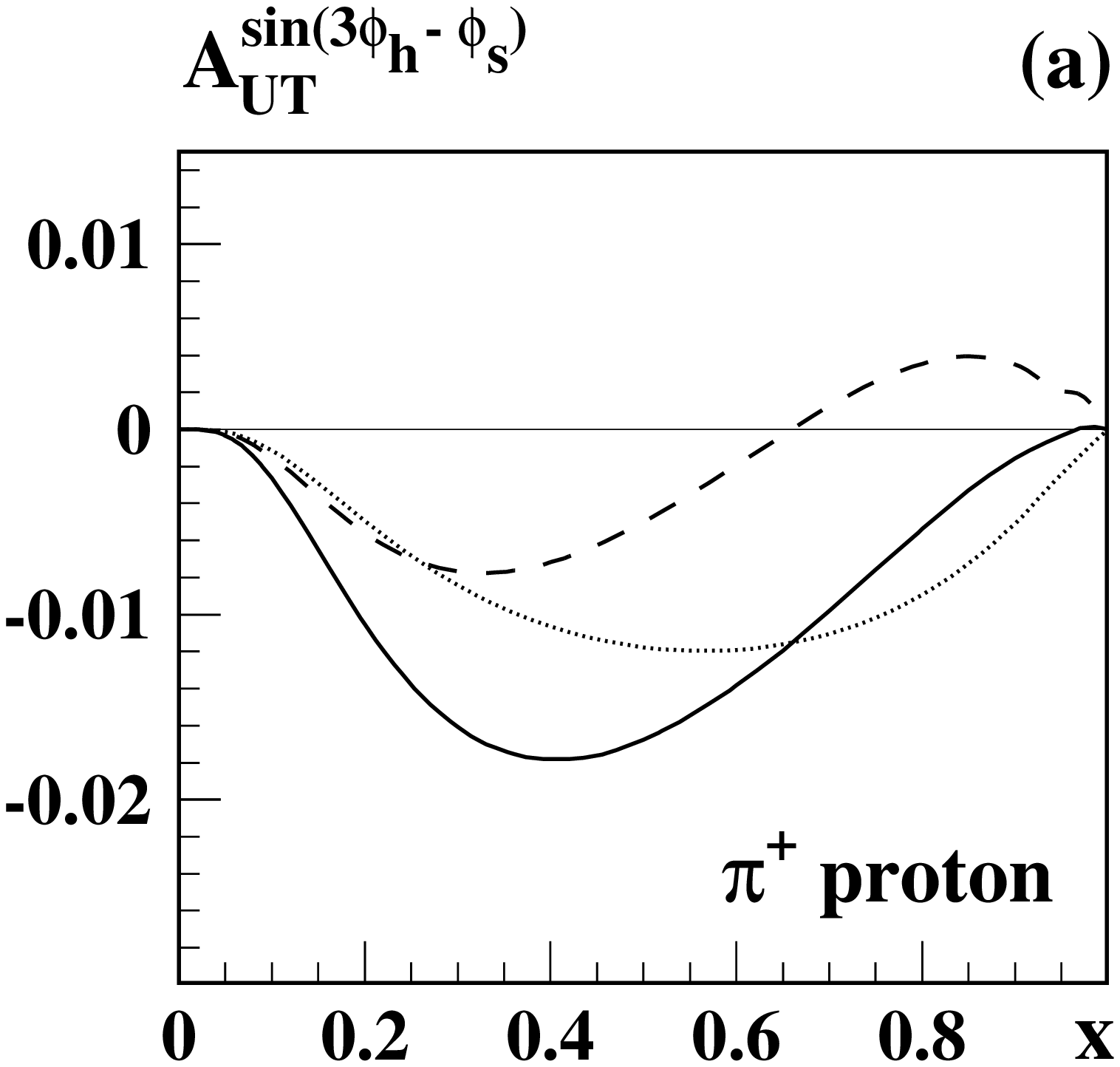}
 \hspace{-11mm} \includegraphics[height=4.2cm]{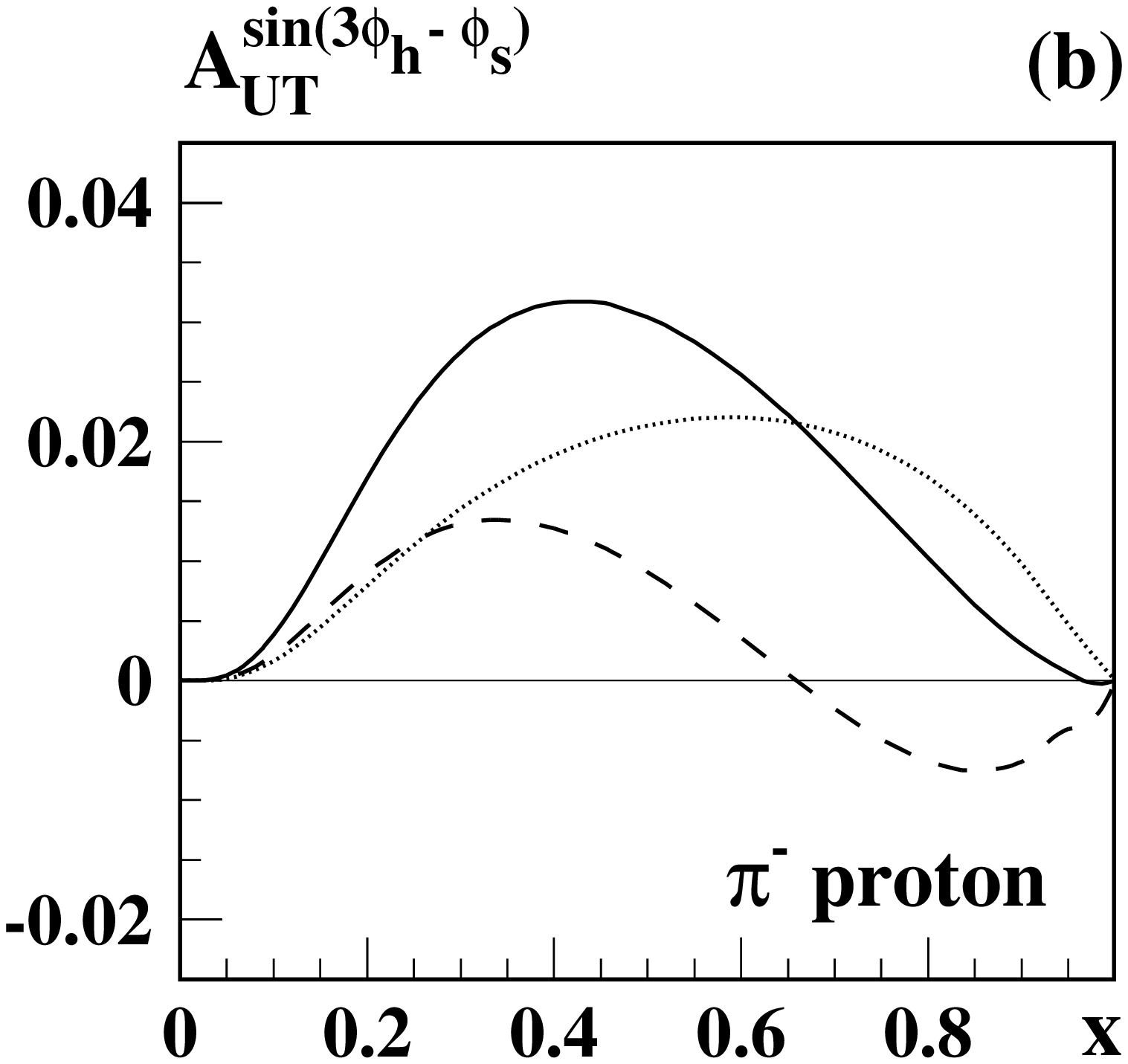}
 \hspace{-13mm} \includegraphics[height=4.2cm]{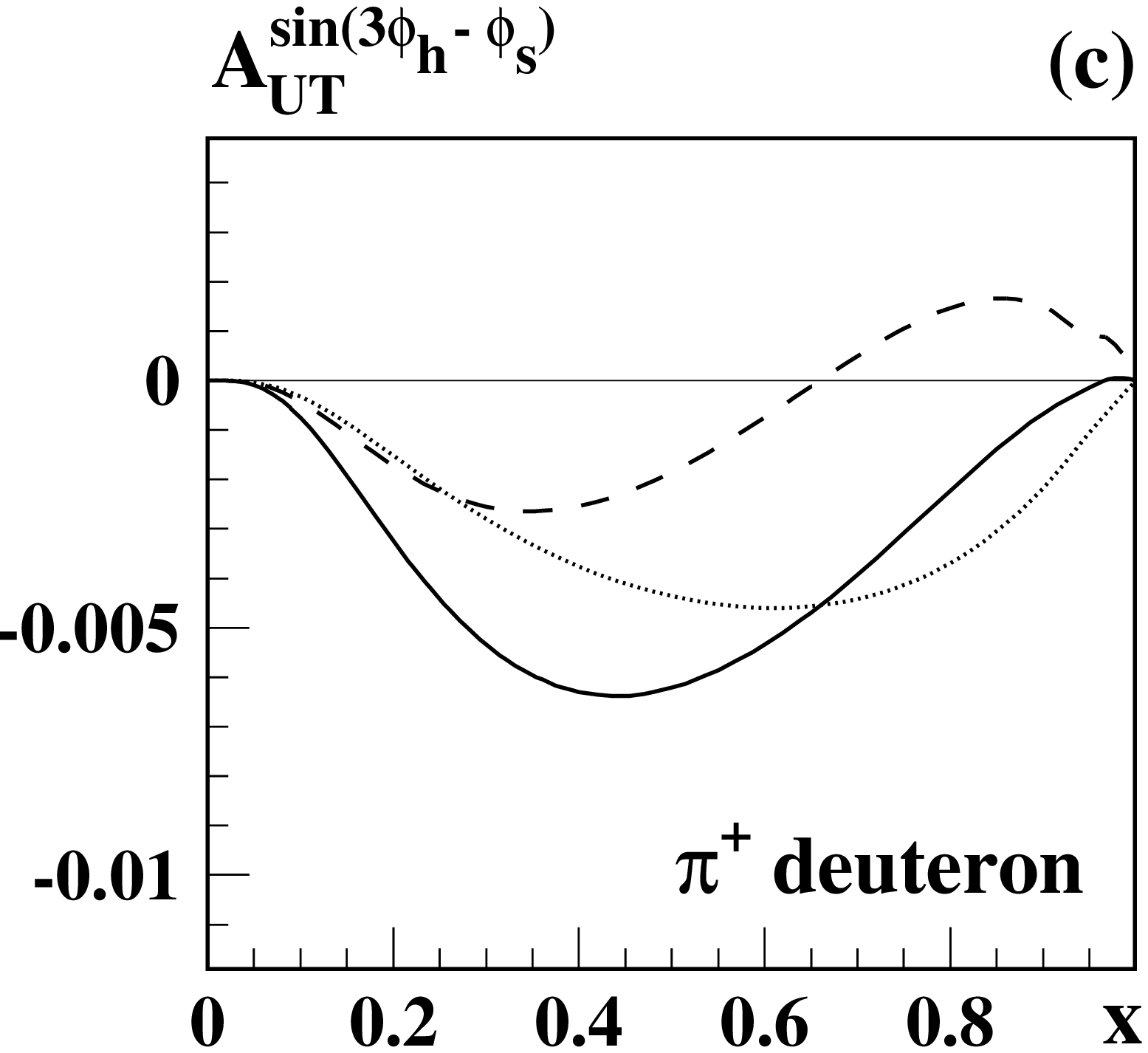}
 \hspace{-11mm} \includegraphics[height=4.2cm]{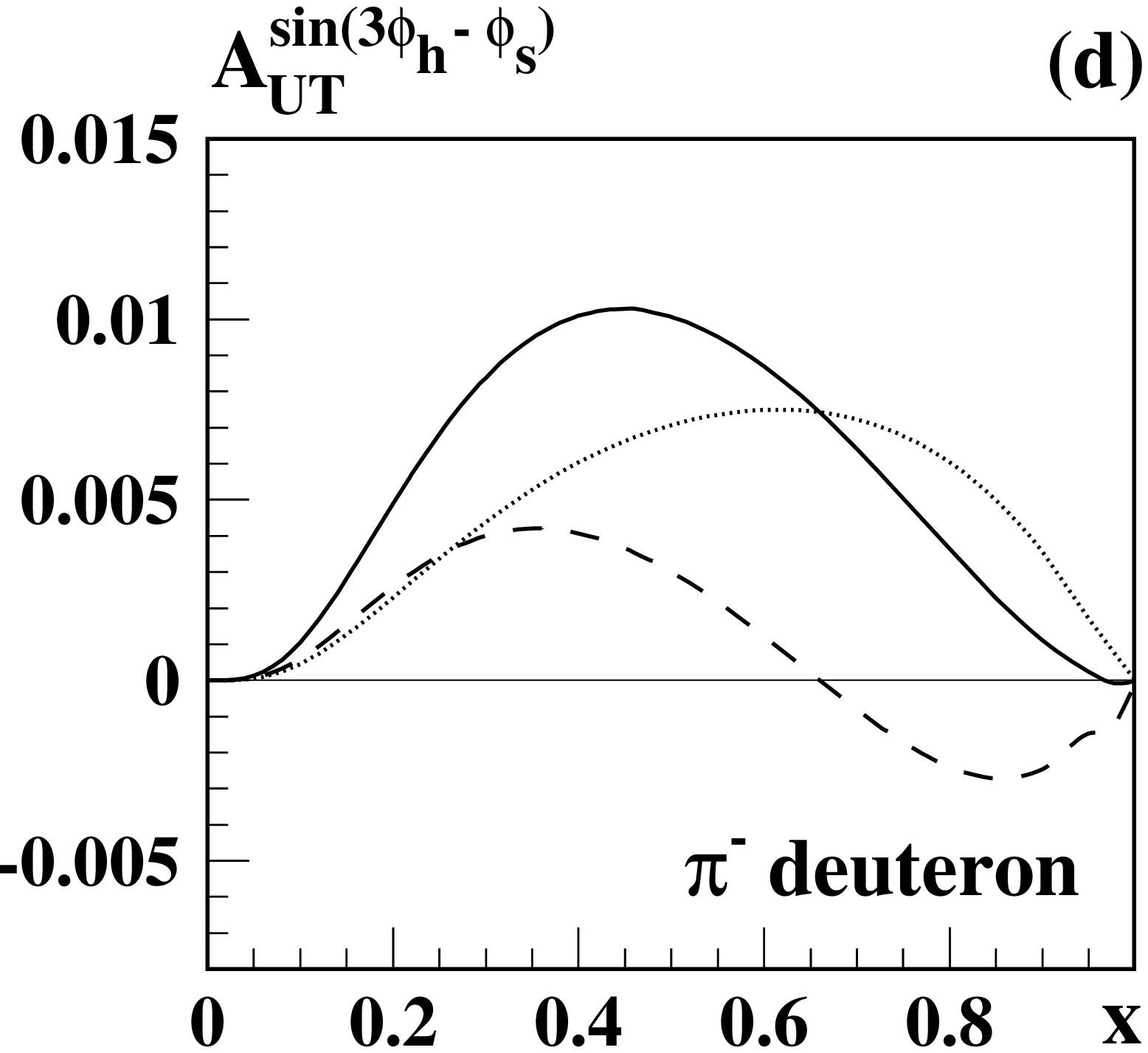}
 \hspace{-12mm}
{\vspace{-0.7 truecm}
    \caption{\label{pretzel_ang_mom}
	The contribution from different angular momentum components to the 
	single-spin asymmetry $A_{UT}^{\sin(3\phi_h-\phi_S)}$ 
	in DIS production of charged pions off proton and deuteron targets, 
	as function of $x$. The results are obtained using the 
        light-cone CQM predictions for $h_{1T}^{(1)\perp a}(x)$ 
	at the hadronic scale and the parametrization 
	of Ref.~\cite{Gluck:1998xa} for $f_1^a(x)$ at $Q^2=2.5$ GeV$^2$.
	The dashed and dotted curves are obtained separating
	the contributions to  $h_{1T}^{(1)\perp a}(x)$ 
	from the interference of $L_z=+1$ and  
	$L_z=-1$ angular momentum components, and
	from the interference of $L_z=0$ and  
	$L_z=+2$ angular momentum components, respectively.
	Solid curves: total results.}
}
\end{figure}

We now pass to consider single spin asymmetries
involving chiral-odd TMDs. In the following, we will use 
for $f_1$ the parametrization from Ref.~\cite{Gluck:1998xa} 
at $Q^2=2.5$ GeV$^2$, while for the  $h_1$, $h_{1T}^{(1)\perp}$, and
  $h_{1L}^{(1)\perp}$ TMDs we will use the results from the light-cone CQM at 
low scale.
The $A_{UT}^{\sin(\phi_h+\phi_S)}$ asymmetry is shown in 
Fig.~\ref{Fig-10:AUT-Collins}, with the separate contribution to $h_1$ from
the S- (dashed curves) and P-wave (dotted curves) components.
The contribution from the P waves is within 30$\%$ of the total results, and,
at variance with the double spin asymmetries discussed above, 
the two partial-wave contributions 
have very similar $x$ dependence, with a maximum at $x\simeq 0.7$. 

In Fig. ~\ref{aul_ang_mom} we show the results
for the $A_{UL}^{\sin(2\phi_h)}$ asymmetry, due to the to  the $h_{1L}^{\perp}$
 TMD.
Since in our model $h_{1L}^{\perp}=-g_{1T}^{\perp}$, the 
relative strength of the contributions from the
S- and P-wave interference (dashed curves) and the
P and D-wave interference (dotted curves) 
is the same as for the corresponding contributions
in $A_{LT}^{\cos(\phi_h-\phi_S)}$.

Finally, in Fig.~\ref{pretzel_ang_mom}, we show the results for the 
$A_{UT}^{\sin(3\phi_h+\phi_S)}$ asymmetry, separating
the contributions to $h_{1T}^{(1)\perp}$ from the interference 
of P waves (dashed curves) and S-D waves (dotted curves).
This is the only case where we can exploit the interference with
the large S-wave contribution to amplify the effects due to the small D wave.
The two interference terms  have a quite different shape 
as function of $x$: in the case of the P-wave interference,
 we have an oscillating 
behaviour, with a sign change at $x\simeq 0.7$, while 
the S-D wave interference term is similar to 
a bell-shaped curve with the maximum at  $x\simeq 0.7$.
The sum of these two contributions gives a total result which
is peaked at $x\simeq 0.4$.  At larger $x$, the S-D wave interference term 
gives the main contribution, while at smaller $x$ 
the P wave and the S-D wave interference terms contribute with the 
same strength.

%------ BEGIN FIGURE 11: Angular momentum decomposition of ALL
%
%
%------ BEGIN FIGURE 12: Angular momentum decomposition of ALT 
%
%------ END FIGURES 11 & 12 ----------------------------------------

%------ BEGIN FIGURE 13: Angular momentum decomposition of Collins Asymmetry
%------ END FIGURE 12 & 13 & 14 --------------------------------------	
\newpage
 \clearpage

%===================  REFERENCES =====================================

\end{document}